\documentclass[draft]{agujournal2019}
\usepackage{url} %this package should fix any errors with URLs in refs.
\usepackage{soul}

\usepackage{placeins}
\usepackage{ragged2e}
\usepackage{amsfonts,amsmath,amssymb} % math stuff
\justifying
\usepackage[export]{adjustbox}
\usepackage{subcaption}

\draftfalse

\journalname{Water Resources Research}

\begin{document}

\title{Two-phase pore-network model for evaporation-driven salt precipitation - representation and analysis of pore-scale processes}

%%%%%%%%%%%%%%%%%%%%%%%%%%%%%%%%%%%%%%%%%%%%%%%
%
%  AUTHORS AND AFFILIATIONS
%
%%%%%%%%%%%%%%%%%%%%%%%%%%%%%%%%%%%%%%%%%%%%%%%

\authors{Theresa Schollenberger\affil{1}, Christian Rohde\affil{2}, Rainer Helmig\affil{1}}

\affiliation{1}{Institute for Modelling Hydraulic and Environmental Systems, University of Stuttgart, Pfaffenwaldring 61, Stuttgart, 70569, Germany.}
\affiliation{2}{Institute of Applied Analysis and Numerical Simulation, University of Stuttgart, Pfaffenwaldring 57, Stuttgart, 70569, Germany.}

% Corresponding author mailing address and e-mail address:

% (include name and email addresses of the corresponding author.  More
% than one corresponding author is allowed in this LaTeX file and for
% publication; but only one corresponding author is allowed in our
% editorial system.)

% Example: \correspondingauthor{First and Last Name}{email@address.edu}

\correspondingauthor{Theresa Schollenberger}{theresa.schollenberger@iws.uni-stuttgart.de}

%%%%%%%%%%%%%%%%%%%%%%%%%%%%%%%%%%%%%%%%%%%%%%%
% KEY POINTS
%%%%%%%%%%%%%%%%%%%%%%%%%%%%%%%%%%%%%%%%%%%%%%%
%  List up to three key points (at least one is required)
%  Key Points summarize the main points and conclusions of the article
%  Each must be 140 characters or fewer with no special characters or punctuation and must be complete sentences

% Example:
% \begin{keypoints}
% \item	List up to three key points (at least one is required)
% \item	Key Points summarize the main points and conclusions of the article
% \item	Each must be 140 characters or fewer with no special characters or punctuation and must be complete sentences
% \end{keypoints}

\begin{keypoints}
\item A dynamic, two-phase, non-isothermal pore-network model is presented for evaporation-driven salt-precipitation processes
\item The model includes the mutual influence of pore-space alteration, interface location and local heterogeneities on flow and transport
\item This processes act on the pore-scale but influence the global behavior of the porous medium
\end{keypoints}

\begin{abstract}
Evaporation-driven salt precipitation occurs in different contexts and leads to challenges in case of e.g. soil salinization or stress-introducing precipitation in building material. During evaporation, brine in porous media gets concentrated due to the loss of water until the solubility limit is reached and salt precipitates. Different models on the REV-scale are available, which describe the effective behavior of the system. But several studies indicate that the controlling processes of evaporation-driven salt precipitation act on the pore scale. Thus, pore-scale models are necessary for a detailed understanding.
In this paper, we present a dynamic, two-phase, non-isothermal pore-network model for evaporation-driven salt precipitation. Salt precipitation is modeled with a kinetic reaction and the resulting pore-space alteration is considered. Further, the model includes influence of the salt concentration on the fluid properties and liquid corner films.
The capabilities of the model are demonstrated in several numerical evaporation experiments. The model is able to represent the effects of the decrease of pore space due to precipitation on the flow, phase displacement and salt transport. Further, the influence of the location of the brine-air interface and its dependency on the presence of liquid corner flow is described. Additionally, pore-scale heterogeneities and their effects like capillary pumping can be modeled. 
All these pore-scale effects have an influence on the salt precipitation process and the global behavior of the porous system. The presented model provides a general tool for further investigations of salt precipitation on the pore scale and transfer to the REV scale.
\end{abstract}

%%%%%%%%%%%%%%%%%%%%%%%%%%%%%%%%%%%%%%%%%%%%%%%
%
%  BODY TEXT
%
%%%%%%%%%%%%%%%%%%%%%%%%%%%%%%%%%%%%%%%%%%%%%%%

\newpage
\section{Introduction} \label{sec:Introduction}
%
%Why is this topic important
    %A considerable amount of research has been devoted to
Salt precipitation in porous media is an important process in many technical and environmental applications. Soil salinization, for example, describes the precipitation of salts in or on top of soils and occurs mainly in arid regions with high evaporation rates, like Australia \cite{Ondrasek2011, Rengasamy2006}, the Mediterranean coastline \cite{Daliakopoulos2016}, Tunisia \cite{Mejri2020} or Israel \cite{Nachshon2018a}. It leads to land degradation and reduces or hinders plant growth and agricultural production \cite{Singh2015, Qadir2014, Wicke2011}. Further, salt precipitation occurs in building material, leading to weathering and damage to the material~\cite{Espinosa2010, Scherer2004}, or at gas-injection wells for underground gas storage and geothermal wells, leading to blockage of the pore space and thus hindering the injection~\cite{Gholami2023, Miri2015}. 
\\
%Definition of key concept: salt precipitation
In all these applications, salt precipitation occurs in the context of evaporation. At the brine-gas interfaces in the porous medium, water evaporates, but the dissolved salt stays behind in the solution. 
This leads to an accumulation of salt at the interface and consequently an increase in salt concentration. Two counteracting processes influence the salt concentration at the interface: diffusive fluxes work on spreading the accumulated salt, whereas the advective fluxes due to evaporation and capillary forces transport the salt towards the interface. 
If the salt concentration exceeds the solubility limit, salt starts to precipitate as a solid and thus transforms the pore space of the porous medium, which again influences the flow and transport in the porous medium. 
\\
\\
%Brief literature review: 2-4 topics from broad to special
To understand and analyze the processes of salt precipitation, the development of numerical models is helpful. A model can be used to investigate the influence of different parameters on the salt precipitation and to identify important, controlling processes.
    %Salt precipitation models on the REV scale
        %REV-scale Jambhekar, Meijri, Bringedal
Numerous numerical models for salt precipitation have been developed on the REV-scale, in which the porous-medium and fluid properties are averaged over a representative elementary volume (REV)~\cite{Jambhekar2015, Mejri2017, Roy2022, Nicolai2007, Espinosa2010, Derluyn2014, Koniorczyk2012}. REV-scale models do not resolve the pore space or fluid distribution within the REV, but describe the averaged behavior of the system. These models are computationally effective, so problems on the field scale can be modeled. 
%\\
    %Experiments on salt precip: REV and pore scale
        %experiments reveal that pore-scale effects have a huge influence on precipitation process
        %REV: Shokri - wet patches, porous salt
        %Pore-scale: Naillon, Wu, Fehki
However, different studies on salt precipitation reveal that controlling processes of salt precipitation act on the pore scale \cite{Wu2023,Dong2021, Rad2015, Shokri2014, Camassel2005}. These pore-scale processes influence REV-scale flow, transport and precipitation but are not directly represented on the REV-scale. We identified three pore-scale processes and properties which have major effects: pore-scale, pore-space alterations, the location of brine-air interfaces and liquid films and pore-scale heterogeneities. We will discuss these processes and their limited representation on the REV-scale in the following.
\\
Pore-scale, pore-space alterations influence the global flow and transport in porous media. The narrowing of the pore space due to precipitation of salt influences the hydraulic properties of the porous medium and consequently the liquid and gas flow as well as the gas-liquid displacement \cite{Wu2023, Dong2021}.
On the REV scale, the influence of pore-space alterations on the permeability is often represented with a relationship of the permeability to the porosity. Different relations are presented in literature, e.g., in \citeA{Verma1988}, \citeA{Bernabe2003} and \citeA{Hommel2018}, but most commonly used is the Kozeny-Carman relation~\cite{Kozeny1927, Carman1937}. Further, the influence of pore-space alterations on the capillary pressure-saturation relationship can be considered with Leverett scaling~\cite{Leverett1941, Jambhekar2015}, which shifts the capillary pressure-saturation relationship based on the altered permeability and porosity.
\\
However, different studies state difficulties to predict the correct change in permeability and capillary pressure based on these relationships.
In \citeA{Jambhekar2015}, for example, the Kozeny-Carman relation and Leverett scaling is used to model precipitation of sodium-chloride.
The model has difficulties to predict the evaporation rate at the start of precipitation and in the period in which the hydraulic connectivity of the liquid to the porous-medium surface decreases. \citeA{Jambhekar2015} assume that this is caused partly by the inability of the Kozeny-Carman relationship to capture the porosity-permeability relation correctly for the case of salinization and the underprediction of the capillary pressure reduction by the Leverett scaling. The use of different permeability-porosity relationships do not show significant difference. % in the evaporation rate and cumulative salt precipitate.
\\
The porosity-permeability relationship depends not only on porosity, but also on further pore-scale properties like pore-space geometry, including pore-size distribution, pore shapes, connectivity, and distribution of precipitate \cite{Xu2004}. Depending on the location of the precipitation, a small reduction in porosity can lead to either large or small reduction in permeability. Large permeability reductions are caused by precipitation in the small throats causing clogging of the flow paths and disconnected void spaces, whereas the precipitation in larger pore spaces has minor influence on the permeability. 
\\
Further, the porosity-permeability relationships are specific to the physical processes as they describe the averaged behavior of the system, which depends on process-specific pore-scale processes.
%The averaged behaviour of the system is described which depends on process-specific, pore-scale processes not represented.
In~\citeA{Xu2004}, the Kozeny-Carman relation underestimate the loss in injectivity due to reduced permeability caused by mineral precipitation at a hot-brine geothermal injection well. For the scenario a different fitted relationship, presented in \cite{Verma1988}, is able to represent the processes better.
This emphasizes the importance of pore-scale considerations of pore-space alterations to represent the influence on the permeability and consequently flow and transport correctly.
\\
We also identified the position of the brine-air interfaces as another controlling process on the pore scale, as the salt precipitates predominantly near the interface \cite{Rad2015}. However, the interfaces can not be resolved in REV-scale models. To take interface locations into account pore-scale investigations are necessary. 
\\
Further, liquid films at the pore walls and in the corners lead to transport of salt and the presence of brine-air interfaces in air-invaded regions of the porous medium and so have a huge influence on the precipitation location \cite{Wu2023}. 
Through corner flow, additionally, the liquid connection to the porous surface could be maintained \cite{ Camassel2005, Fekih2024}. 
Consequently, in systems with and without corner flow totally different evaporation and precipitation dynamics develop \cite{Camassel2005}. \citeA{Wu2023} state that, "depiction of salt precipitation and its interplay with gas–liquid displacement and the corner liquid film flow is key to the development of the theory for evaporation in porous media with salt precipitation". On the REV-scale these liquid films can not be represented.
\\
Further, pore-scale heterogeneities are identified to have a huge influence on the global flow behavior.  
The liquid saturation distribution and evaporation stages are e.g. highly dependent on the grain size distribution, the degree of clustering of pore sizes and anisotropy of the porous medium \cite{Dashtian2018, Rad2015}.
\\
In \citeA{Rad2015} the influence of different grain sizes and distributions on evaporation-driven salt precipitation was investigated in a series of column experiments. Preferential evaporation and precipitation spots on the porous medium surface were found for larger grain sizes, whereas for smaller grain sizes a more homogeneous distribution develops. This is caused by local heterogeneities in the pore sizes. During evaporation from porous media, first the large pores are invaded whereas in the thin pores capillary-driven flow develops towards the porous medium surface from which water evaporates. Less thin pores are available for larger grain sizes.
With the REV-scale advection-diffusion equation, \citeA{Rad2013} were not able to represent the dynamics of salt precipitation and \citeA{Shokri2014} were not able to represent the salt concentration profiles due to pore-scale heterogeneities. \citeA{Shokri2014} concludes “that the pore-scale understanding of salt transport in drying porous media is required to describe the macroscopic profiles and that one must carry out full 3D pore network simulations with distributed heterogeneity at larger scales to account for the preferential phenomena occurring during solute transport in drying porous media.”
\\
\\
All these studies highlight the importance of pore-scale considerations to represent all relevant processes of evaporation-driven salt precipitation.
Hence, the development of numerical models on the pore scale is necessary. 
Different types of pore-scales models are available. 
Direct pore-scale models resolve the pore
space in detail, like e.g. Lattice Boltzmann models \cite{Yang2023, Kang2003, Yoon2015}, computational fluid dynamics models \cite{Molins2015, Molins2012, Zaretskiy2010} or phase-field models \cite{vonWolff2022, Rohde2021, Bringedal2020}.
Direct pore-scale modeling, however, becomes extremely complex and
the simulations of large domains are computational expensive, if not unfeasible. 
\\
\begin{figure}[t]%
\centering
\includegraphics[width=0.99\textwidth]{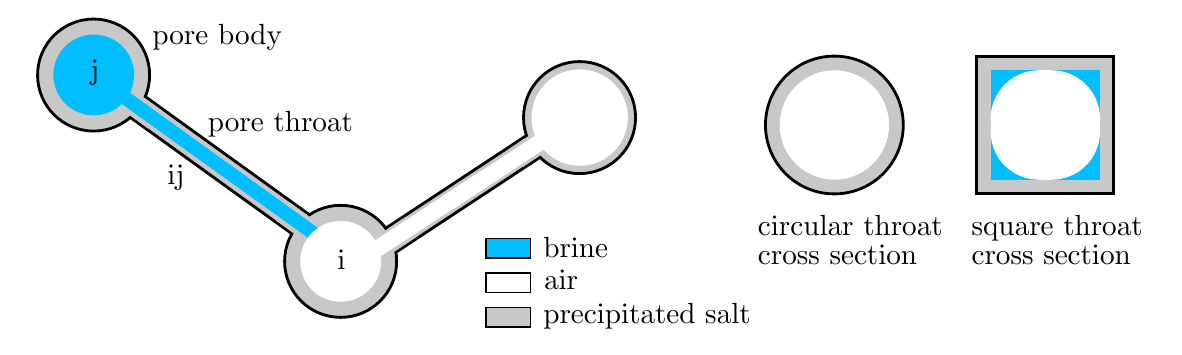}
\caption{Concept of the pore-network model}
\label{fig:PNMScheme_WaterBrine}
\end{figure}
Pore-network models, however, are able to incorporate pore-scale processes, but are at the same time capable to model a larger domain in the size of several REVs using a simplified pore-space representation. 
A pore-network model represents the pore space of a porous medium with a network of pore bodies and pore throats, see Figure~\ref{fig:PNMScheme_WaterBrine}. The pore bodies represent larger pore spaces. In every pore body, the balance equations are solved and one set of primary variables is stored. 
The fluid is assumed to be perfectly mixed within each pore body, leading to homogeneous fluid properties.
Fluxes between the pore bodies are calculated in the pore throats. The volume of throats is assumed to be neglectable. For a detailed introduction to pore-network models, see~\citeA{Blunt2017}. 
\\
Some pore-network models for precipitation are already presented in literature. 
In \citeA{Dong2021} a pore-network model is used to characterize the pore space and hydraulic properties before and after precipitation. The model does not describe the process of precipitation.
In \citeA{Nogues2013} a dynamic, one-phase, isothermal pore-network model is presented for the precipitation and dissolution of different carbonate species in the context of geological CO2-sequestration. Volume alterations of pore bodies and throats are included. Due to the different application, the model only considers one-phase flow and no evaporative processes.
In \citeA{Dashtian2018} a pore-network model for salt precipitation and evaporation is presented. For the saturation distribution, a static pore-network model using invasion percolation is used, and the salt transport is modeled using the advection-diffusion equation. Further, the model is isothermal and includes vapor diffusion, capillary effects, salt precipitation and pore-space alterations. However, the saturation distribution and the salt transport and precipitation are not directly coupled, as they are calculated sequentially. Further, the influence of liquid films, non-isothermal effects and salt concentration on the density and viscosity are neglected.
These models, however, are all very specific to their setups and applications.
\\
%Research gap   
    %the lack of research 
    %limitations in previous studies
    %practical problems that need solving
    %Numerous studies on have ben conducted. However the is a lack of research on ... In addition the previous studies are limited because
%Aim of the aim: 1-2 sentences
In this paper we present a dynamic pore-network model for non-isothermal, two-phase flow including salt transport and precipitation and the related pore-space alterations as well as transport through corner flow. The model is implemented in the open-source simulator DuMu$^\mathrm{X}$ and aims to be a general tool where further physical processes can be included, like the influence of free-flow on the evaporation. 
Further, we show in numerical experiments, that the presented model allows to investigate the influence of pore-scale, pore-space alterations, the bine-air interface location and local heterogeneities in the porous medium on flow, salt transport, precipitation and evaporation.
\\
%Structure of the paper
The model is presented in Section~\ref{sec:Model} of the paper, followed by the description of the setups of the numerical experiments in Section~\ref{sec:Setup} and the results and discussions in Section~\ref{sec:Results}. The main findings of the paper are summarized in the conclusion in Section~\ref{sec:conclusion}.
\section{Model Concept} \label{sec:Model}
Two fluid phases $\alpha$ are present in the pore-network model: a gaseous, non-wetting phase and a liquid, wetting phase (denoted with the subscript $\mathrm{g}$ and $\mathrm{l}$). The gaseous phase consists mainly out of the component air, but also contains water vapor. For the liquid phase either pure water, consisting out of the components water and air, or brine, which additionally includes the salt sodium chloride, is considered (denoted with the superscripts $\mathrm{a}$, $\mathrm{w}$ and $\mathrm{NaCl}$). In case of brine, additionally, a solid salt phase is present.
\\
The pore-network model in this paper is based on the model presented in~\cite{Weishaupt2022, Wu2024}. The fluid properties of brine are based on the work of~\cite{Jambhekar2016}.
Different relations are used for fluid properties to consider their dependency on e.g. salt concentration, pressure and temperature. An overview of the used relations is given in Table~\ref{tab:FluidProperties}.
\begin{sidewaystable}
\caption{Overview of relations for the brine and air fluid properties after Jambhekar et al. (2016) and as implemented in DuMu$^\mathrm{X}$}
\label{tab:FluidProperties}
\centering
\begin{tabular}{l r l l }
\hline
fluid properties & & relation & source and variable specification\\
\hline
mass fraction & $X_\mathrm{l}^\mathrm{NaCl}$ & $M^\mathrm{NaCl}  x_\mathrm{l}^\mathrm{NaCl} / (x_\mathrm{l}^\mathrm{NaCl}  M^\mathrm{NaCl} + (1 - x_\mathrm{l}^\mathrm{NaCl} - x_\mathrm{l}^\mathrm{a})  M^\mathrm{w} + x_\mathrm{l}^\mathrm{a}  M^\mathrm{a})$ & $M^\kappa$: molar mass of $\kappa$ \\
\hline
average molar mass  & $M_\mathrm{avg,l}$ & $M^\mathrm{w}x_\mathrm{l}^\mathrm{w} + M^\mathrm{NaCl} x_\mathrm{l}^\mathrm{NaCl} + M_\mathrm{l}^\mathrm{a} x_\mathrm{l}^\mathrm{a}$ & \\
\hline
mass density  	&$\rho_\mathrm{l}$ &$ \rho_w + 1000 \cdot X_\mathrm{l}^\mathrm{NaCl} \cdot (0.668 + 0.44 \cdot X_\mathrm{l}^\mathrm{NaCl}$ & \citeA{Batzle1992}\\
& & $+10^{-6}(300 \cdot p -2400 \cdot p \cdot X_\mathrm{l}^\mathrm{NaCl}$  & $\rho_\mathrm{w}$: IAPWS$^{*}$, $p$ [MPa], $T$ [$^{\circ}$C]\\
  				& & $+T \cdot (80.0 +3 \cdot T -3300 \cdot X_\mathrm{l}^\mathrm{NaCl}- 13 \cdot p +47 \cdot p \cdot X_\mathrm{l}^\mathrm{NaCl})))$ & \\
  				&$\rho_\mathrm{g}$ & $\rho_g^\mathrm{w}(T, x_\mathrm{g}^\mathrm{w} p_\mathrm{g}) + \rho_g^a(T, x_\mathrm{g}^\mathrm{a} p_\mathrm{g}) $ & $\rho_\mathrm{g}^\mathrm{w}(T,p)$: IAPWS$^{*}$,  $\rho_\mathrm{g}^\mathrm{a}$: ideal gas law\\
\hline
molar density  	&$\rho_\mathrm{m,l}$ & $\frac{\rho_l}{ M_\mathrm{avg,l}}$ & \\
  				&$\rho_\mathrm{m,g}$ &$\rho_\mathrm{m,g}^\mathrm{w}(T, x_\mathrm{g}^\mathrm{w} p_\mathrm{g}) + \rho_\mathrm{m,g}^\mathrm{a}(T, x_\mathrm{g}^\mathrm{a} p_\mathrm{g}) $ &  $\rho_\mathrm{m,g}^\mathrm{w}(T,p)$: IAPWS$^{*}$,  $\rho_\mathrm{m,g}^\mathrm{a}$: ideal gas law\\
\hline
viscosity  		& $\mu_\mathrm{l}$& $0.1 + (0.333 \cdot X_\mathrm{l}^\mathrm{NaCl}) + (1.65+(91.9 \cdot (X_\mathrm{l}^\mathrm{NaCl})^3)) \cdot e^{-A}$ & \citeA{Batzle1992}\\ 
  				& &$A = ((0.42 \cdot ((X_\mathrm{l}^\mathrm{NaCl})^{0.8}-0.17)^ 2 + 0.045) \cdot T ^ {0.8} $ & \\
  				&$\mu_g$ & $1.496 \cdot 10^{-6}  \frac{\sqrt{T^3} }{T+ 120.0} \left(1.0 + p_\mathrm{corr}(\frac{p}{1.0 \cdot 10^{5}} - 1.0 ) \right)$& $p_\mathrm{corr} = 9.7115\cdot 10^{-9} (T - 273.15)^2 $\\
  				& & & $- 5.5 \cdot 10^{-6} (T-273.15) + 0.0010809$\\
\hline
vapor pressure	&$p_\mathrm{vapor}$ & $\frac{p_\mathrm{vapor}^\mathrm{w}(T)}{x_\mathrm{l}^\mathrm{w}} \mathrm{exp}\left( \frac{-p_\mathrm{c}}{ \rho_\mathrm{m,l} R T}\right)$& \citeA{Kelvin1871}\\
 & & & $p_\mathrm{vapor}^\mathrm{w}(T) $: IAPWS$^{*}$, $R$: ideal gas constant \\%Influence of salt and capillary pressure\\
\hline
diffusion coefficient  	& $D_\mathrm{l}^\mathrm{w, a}$ &$D_\mathrm{ref}~ T/T_\mathrm{ref}$ & \citeA{Ferrell1967} \\
 & & & $D_\mathrm{ref} = 2.01 \cdot 10^{-9} \frac{m^2}{s} $, $T_\mathrm{ref} = 298.15 K $ \\
  						& $D_\mathrm{l}^\mathrm{w, NaCl}$& $1.54 \cdot 10^{-9} \mathrm{\frac{m^2}{s}}$& \citeA{Rard1979}\\
  						& $D_\mathrm{g}^\mathrm{w, a}$&$D_\mathrm{ref} \frac{p_\mathrm{g,ref}}{p} \left( \frac{T}{T_\mathrm{ref}}\right)^\Theta $ & $\Theta =1.8$, $D_\mathrm{ref} =2.13 \cdot 10^{-5} \mathrm{\frac{m^2}{s}}$\\
  						& & & $p_\mathrm{g, ref} =1.0 \cdot 10^{5}$ Pa, $T_\mathrm{ref} =273.15$ K \\
  						%& $D_\mathrm{g}^\mathrm{H2O, NaCl}$ & $10^{-12}$ & is zero, numeric\\
\hline
enthalpy				& $h_{\mathrm{l}}$ & $(1-X_\mathrm{l}^\mathrm{NaCl}) \cdot h_\mathrm{l}^\mathrm{w} + X_\mathrm{l}^\mathrm{NaCl} \cdot h_\mathrm{l}^\mathrm{NaCl} + X_\mathrm{l}^\mathrm{NaCl} \cdot \Delta h  $&  $\Delta h(T,X_\mathrm{l}^\mathrm{NaCl})$ after \citeA{Michaelides1981} \\
  						& & & $h_\mathrm{l}^\mathrm{NaCl}(T)$ after \citeA{Daubert1989}\\
  						& & & $h_\mathrm{l}^\mathrm{w}$: IAPWS$^{*}$\\
  						& $h_\mathrm{g}$& $X^\mathrm{w}_\mathrm{g} h_\mathrm{g}^\mathrm{w} + X^\mathrm{a}_\mathrm{g} h_\mathrm{g}^\mathrm{a}$& $h_g^\mathrm{w}$: IAPWS$^{*}$\\
  						& & & $h_\mathrm{g}^\mathrm{a} = c_\mathrm{p} (T-273.15)$, $h_\mathrm{g}^\mathrm{NaCl} = 0.0$\\
\hline
thermal conductivity	& $\lambda_\mathrm{l}$& $c ~ \lambda_l^\mathrm{w}$& $c(T, x_\mathrm{l}^\mathrm{NaCl})$: \citeA{Yusufova1975}, $\lambda_l^\mathrm{w}:$ IAPWS$^{*}$ \\
  						& $\lambda_\mathrm{g}$ & $\lambda_\mathrm{g}^\mathrm{a} X_\mathrm{g}^\mathrm{a} + \lambda_\mathrm{g}^\mathrm{w} X_\mathrm{g}^\mathrm{w} $ & $\lambda_\mathrm{g}^\mathrm{w}$: IAPWS$^{*}$, $\lambda_\mathrm{g}^\mathrm{a} = 0.0255535~\mathrm{\frac{W}{m K}}$\\
\hline
heat capacity			&$c_\mathrm{p,l}$ & $\left(h_\mathrm{l}^\mathrm{NaCl}(p_\mathrm{l}, T+\Delta T, X_\mathrm{l}^\mathrm{NaCl}) - h_\mathrm{l}^\mathrm{NaCl}(p_\mathrm{l}, T, X_\mathrm{l}^\mathrm{NaCl})\right)/\Delta T$ & \\
  						& $c_\mathrm{p,g}$ & $c_\mathrm{p,g}^\mathrm{a} X_\mathrm{g}^\mathrm{a} + c_\mathrm{p,g}^\mathrm{w} X_\mathrm{g}^\mathrm{w}$ & $c_\mathrm{p,g}^\mathrm{w}$: IAPWS$^{*}$, $c_\mathrm{p,g}^\mathrm{a} = \lambda_\mathrm{g}^\mathrm{a}(T)$: tabulated\\
\hline
\multicolumn{4}{l}{$^{*}$ IAPWS: for water the relations from "The International Association for the Properties of Water and Steam (IAPWS)"}\\
\multicolumn{4}{l}{presented in \cite{IAPWS2008} are used.}
\end{tabular}
\end{sidewaystable}
\subsection{Balance Equations}\label{subsec:BalanceEquations}
\subsubsection{Component Mole Balance}
%Balance equation
For each component $\kappa$, a mole balance is solved in every pore body $i$:
\begin{linenomath*}
\begin{align}
V_{\mathrm{ini},i} \frac{\partial}{\partial t} \left( \sum_\alpha^{N_\alpha} \left( x_\alpha^\kappa \rho_\mathrm{m,\alpha} S_\alpha \phi \right)_i \right) + \sum_\alpha^{N_\alpha} \sum_j^{N_j} \left( \frac{x_\alpha^\kappa \rho_\mathrm{m,\alpha} }{\mu_\alpha}\right)_\mathrm{up} j_{\mathrm{adv,\alpha,}ij} + &\sum_\alpha^{N_\alpha} \sum_j^{N_j} j_{\mathrm{diff,\alpha,}ij}^\kappa = s_{i}^\kappa, \label{eq:MoleBalance}\\
&\alpha \in \{ \mathrm{l,g} \}, ~\kappa \in \{\mathrm{w,a,NaCl}\}. \nonumber
\end{align}
\end{linenomath*}
The first term in the equation is the storage term, with $V_{\mathrm{ini},i}$ as the initial pore-body volume, $N_\alpha$ as the number of fluid phases, $x_{\alpha,i}^\kappa$ as the mole fraction of component $\kappa$ in phase $\alpha$, $\rho_{\mathrm{m,\alpha},i}$ as the molar density of $\alpha$, $S_{\alpha,i}$ as the saturation of $\alpha$ and $\phi_i$ as the void volume fraction in pore body $i$. The second term describes the sum of the advective fluxes over the number of adjacent pore throats $N_j$, which are specified in Section~\ref{subsubsec:advectiveFluxes}. The respective upstream value of $ x_\alpha^\kappa$ and $\rho_\mathrm{m,\alpha}$ is used.
The third term describes the sum of diffusive fluxes through the adjacent pore throats $j$, which are described in more detail in Section~\ref{subsubsec:diffusiveFluxes}. The fourth term includes the precipitation and dissolution reaction of salt, which is specified for $\mathrm{NaCl}$ in Section~\ref{subsec:precipReaction} and is zero for the other components.
To close the system of equations, it is given that all saturations and all mole fractions sum up to one:
$\sum_\alpha^{N_\alpha} S_\alpha = 1 $ and
$\sum_\kappa^{N_\kappa} x_\alpha^\kappa = 1$.
\\
For the solid salt phase the mole balance reduces to the storage and reaction term:
\begin{linenomath*}
\begin{align}
V_{\mathrm{ini},i} \rho_\mathrm{m,s} \frac{\partial \phi_\mathrm{s}}{\partial t} &= s_\mathrm{s,i}.
%s_{\alpha,i}^\kappa &= - s_\mathrm{s,i}
\label{eq:solidSaltBalance}
\end{align}
\end{linenomath*}
Here $\rho_\mathrm{m,s} $ describes the molar density of the solid salt, $\phi_\mathrm{s} $ the volume fraction in pore body $i$ occupied by solid salt and $s_\mathrm{s,i} $ the reaction term for solid salt. The balance equations~\ref{eq:MoleBalance} and~\ref{eq:solidSaltBalance} are solved for the primary variables gas pressure $p_\mathrm{g}$, liquid saturation $S_l$, salt mole fraction $x_\mathrm{l}^\mathrm{NaCl}$ and the precipitated salt volume fraction $\phi_\mathrm{s}$.
\subsubsection{Energy Balance Equation}
Thermal equilibrium between the liquid and solid phases is assumed. This results in one energy balance equation and temperature for each pore body $i$:
\begin{linenomath*}
\begin{align} 
V_{\mathrm{ini},i} & \frac{\partial}{\partial t}  \left( \sum_\alpha^{N_\alpha} \left( \rho_\alpha u_\alpha S_\alpha \phi \right)_i + \left(T c_\mathrm{s} \rho_\mathrm{s} \phi_\mathrm{s} \right)_i\right) + \sum_\alpha^{N_\alpha} \sum_j^{N_j} \left( \frac{\rho_\alpha h_\alpha}{\mu_\alpha} \right)_\mathrm{up} j_{\mathrm{adv,\alpha,}ij} \\ \nonumber
&+ \sum_\alpha^{N_\alpha} \sum_j^{N_j} \sum_\kappa^{N_\kappa} j_{\mathrm{diff,\alpha},ij}^\kappa h_\mathrm{\alpha,up}^\kappa 
+ \sum_\alpha^{N_\alpha} \sum_j^{N_j} j_{\mathrm{cond},\alpha,ij} %\\ \nonumber
%\\
= q_{\mathrm{e},i} \nonumber
&\alpha \in \{ \mathrm{l,g} \}.\nonumber
\end{align}
\end{linenomath*}
The first term describes the energy storage in the fluid and solid salt phase. Here, $\rho_\alpha $ is the mass density, $u_\alpha $ the internal energy, $T$ the temperature, $c_\mathrm{s} $ the solid heat capacity and $\rho_\mathrm{s} $ the mass density of salt. The second term describes the transport of energy through advective fluxes and the third through diffusive fluxes, where $ h_\mathrm{\alpha}^\kappa $ is the specific component enthalpy and $ h_\mathrm{\alpha} $ the phase enthalpy. The fourth term  describes heat conduction, specified in Section~\ref{subsubsec:condFlux}, and the fifth term the energy source term. The energy balance equation is solved for the primary variable temperature $T$.
\subsection{Transport Processes}
\subsubsection{Advective Flow}\label{subsubsec:advectiveFluxes}
%Advective flux
The advective fluxes are driven by the phase pressure difference $p_\mathrm{\alpha,i} - p_\mathrm{\alpha,j}$ between pore body $i$ and $j$:
\begin{linenomath*}
\begin{align}
j_{\mathrm{adv,\alpha},ij} = g_\mathrm{ij,\alpha}  ( p_\mathrm{\alpha,i} - p_\mathrm{\alpha,j}) .
\end{align}
\end{linenomath*}
The transmissibility $g_\mathrm{ij,\alpha}$ of the phase $\alpha$ in pore throat $ij$ depends on throat geometry and phase distribution. For throats, which are completely filled with one phase, either liquid or gas, the transmissibility can be described by the relationship of~\citeA{Bruus2011}. In this paper, we consider square shaped cross sections with side length $w_\mathrm{ij}$ and circular shaped cross sections with radius $r_\mathrm{ij}$ and $l_{ij}$ as the throat length:
\begin{linenomath*}
\begin{align}
g_\mathrm{ij,1p,square} &= \frac{w_\mathrm{ij}^4}{28.4 l_\mathrm{ij}}, \label{eq:1pTransmissibilitySquare}\\
g_\mathrm{ij,1p,circle} &= \frac{\pi r_\mathrm{ij}^4}{8 l_\mathrm{ij}}.
\end{align}
\end{linenomath*}
In angular throats as the square shaped throat, two phases are present after invasion. The wetting phase is located in throat corners and liquid films establish whereas the non-wetting phase is in the throat middle.
In circular throats always only one phase is present as no liquid corner films can form.
The transmissibility of the wetting phase in a square shaped throat can be described by the relationship of \citeA{Ransohoff1988}:
\begin{linenomath*}
\begin{align}
g_\mathrm{ij,l} &= \sum_{c=0}^{N_\mathrm{c}} \left( \frac{A_{\mathrm{ij,l},c} r_\mathrm{curv}}{l_\mathrm{ij}R_{\mathrm{crevice},c}} \right).
\label{eq:liquidTransmissibility}
%r_\mathrm{curv} = \frac{\gamma}{ p_\mathrm{c}}
\end{align}
\end{linenomath*}
Here, $N_\mathrm{c}$ is the number of corners in the throat cross section,  $A_{\mathrm{ij,l},c} $ the liquid phase cross-sectional area, $r_\mathrm{curv}= \frac{\gamma}{ p_\mathrm{c}} $ the curvature radius of the liquid-gas interface of the corner film in a throat cross section dependent on the capillary pressure $p_\mathrm{c}$ and the surface tension $\gamma$. $R_{\mathrm{crevice},c} $ is a dimensionless resistance factor depending on the contact angle and the corner half angle (for corner angles in squares, right angles, it is $\frac{\pi}{4}$)~\cite{Zhou1997}.
\\
The transmissibility of the non-wetting phase in the square throat can be described by the relationship of \citeA{Bakke1997}:
\begin{linenomath*}
\begin{align}
g_\mathrm{ij,g} &= \frac{r_\mathrm{eff}^2 A_\mathrm{ij,g}}{8 l_\mathrm{ij}},\\
r_\mathrm{eff} &= 0.5 \left(\sqrt{\frac{A_\mathrm{ij,g}}{\pi}} + \frac{w_\mathrm{ij}}{2} \right).
\end{align}
\end{linenomath*}
Here, $r_\mathrm{eff} $ is the effective radius and $A_\mathrm{ij,g} $ the throat cross-sectional area of the gas phase, see Section~\ref{subsec:phaseAreas}.
\subsubsection{Diffusive Fluxes}\label{subsubsec:diffusiveFluxes}
%Diffusive fluxes
The diffusive flux is driven by the mass fraction difference $X_{\alpha,i}^\kappa - X_{\alpha,j}^\kappa$ as described by Ficks law:
\begin{linenomath*}
\begin{align}
j_{\mathrm{diff,\alpha},ij} = \frac{\bar{\rho}_{\mathrm{\alpha},ij}}{M^\kappa} \tilde{D}_{\alpha,ij}^\kappa A_{\alpha,ij} \frac{X_{\alpha,i}^\kappa - X_{\alpha,j}^\kappa}{l_{ij}}.
\end{align}
\end{linenomath*}
Here, the arithmetical mean of the mass density in pore body $i$ and $j$ is denoted as $\bar{\rho}_{\mathrm{\alpha},ij}$ and the harmonic mean of the molecular diffusion coefficient as $\tilde{D}_{\alpha,ij}^\kappa$. Further, ${M^\kappa}$ describes the molar mass and  $A_{\alpha,ij}$ the cross-sectional area of phase $\alpha$ in pore throat $ij$, see Section~\ref{subsec:phaseAreas}.
%and $l_{ij}$ the pore-throat length.  
%
%
\subsubsection{Conductive Fluxes}\label{subsubsec:condFlux}
The conductive flux is driven by the temperature difference $T_i - T_j$ between pore body $i$ and $j$. Here, $\tilde{\lambda}_{\alpha,ij}$ is the harmonic mean of the thermal conductivity of phase $\alpha$ of pore body $i$ and $j$:
\begin{linenomath*}
\begin{align}
j_{\mathrm{cond},\alpha,ij} = \tilde{\lambda}_{\alpha,ij} A_{\alpha,ij} \frac{T_i - T_j}{l_{ij}}.
\end{align}
\end{linenomath*}
\subsubsection{Phase Areas} \label{subsec:phaseAreas}
If only one phase is present in the pore throat the phase area corresponds to the throat cross-sectional area:
\begin{linenomath*}
\begin{align}
A_\mathrm{ij,square} &= w_\mathrm{ij}^2, \\
A_\mathrm{ij,circle} &= \pi r_\mathrm{ij}^2.
\end{align}
\end{linenomath*}
In case of two present phases, the wetting-film throat cross-sectional area can be calculated based on the curvature radius $r_\mathrm{curv}$, the contact angle $\theta$ and the corner half angle $\beta$:
\begin{linenomath*}
\begin{align}
A_\mathrm{ij,l} &= \sum_{c=0}^{N_\mathrm{c}} r_\mathrm{curv}^2 \left( \frac{\mathrm{cos}(\theta) \mathrm{cos}(\theta + \beta)}{\mathrm{sin}(\beta)} + \beta + \theta - \frac{\pi}{2} \right). 
\end{align}
\end{linenomath*}
The non-wetting throat cross-sectional area can be calculated as the difference of the wetting area to the total cross-sectional area:
\begin{linenomath*}
\begin{align}
A_\mathrm{ij,g} = A_\mathrm{ij} - A_\mathrm{ij,l}.
\end{align}
\end{linenomath*}
\subsection{Salt Precipitation}\label{subsec:SaltPrecip}
\subsubsection{Precipitation Reaction}\label{subsec:precipReaction}
The precipitation of salt is described by a kinetic reaction:
\begin{linenomath*}
\begin{align}
    s_i^\mathrm{NaCl} = - s_{\mathrm{s},i}^\mathrm{NaCl} = - k A_{\mathrm{react},i} \left( \Omega -1 \right). %mol/s
    \label{eq:sourceTerm}
\end{align} 
\end{linenomath*}
Here, the reaction constant $k  = 10^{-3} \mathrm{\frac{mol}{m^2 s}}$ is used. For the reactive surface $A_{\mathrm{react},i}$ the surface area of the pore body is used. $A_{\mathrm{react},i}$ is independent of $S_\mathrm{l}$ based on the assumption that the wetting brine phase covers the whole pore-body walls.
For cubic pore bodies, used in this paper, with side length $W_i$ it is $A_{\mathrm{react,cubic},i} = 6 (W_i)^2$. The saturation index $\Omega$ describes the degree of oversaturation ($\Omega > 1$) and undersaturation ($\Omega < 1$) of the solution:
\begin{linenomath*}
\begin{align}
\Omega &= \frac{[\mathrm{Na}^+][\mathrm{Cl}^-]}{K_\mathrm{eq}^\mathrm{NaCl}}.
\end{align}
\end{linenomath*}
Here, $[\mathrm{Na}^+][\mathrm{Cl}^-]$ is the product of the dissolved ion activities. It is assumed that $[\mathrm{Na}^+]=[\mathrm{Cl}^-]$ and so $\mathrm{Na}^+$ and $\mathrm{Cl}^-$ can be considered to be one component for the balance equations. The activities are calculated using a simplified Pitzer approach~\cite{Perez2002}, also presented in~\citeA{Schollenberger2024}. The equilibrium constant $K_\mathrm{eq}^\mathrm{NaCl} = 38.0464$ is calculated by applying the Pitzer approach to the solubility limit and corresponds to $x_\mathrm{l}^\mathrm{NaCl}=0.0996$. 
\\
\subsubsection{Pore-space Alterations}
Due to the precipitation of salt as solid phase the volume of the pore bodies and throats decreases. For the pore body the actual volume results from the salt volume fraction and initial volume:
\begin{linenomath*}
\begin{align}
V_i &= \left(1 - \phi_{\mathrm{s},i} \right) V_{\mathrm{ini},i}.
\label{eq:poreBodyVolumeAlterations}
\end{align}
\end{linenomath*}
It is assumed that the salt precipitates in a homogeneous layer at the pore-body walls, so for cubic pore bodies it is: $W_i =  V_i ^{1/3}$.
In the pore throats no balance equations is solved, consequently no value for the amount of salt in the pore throat is available. Different concepts for an estimation of the precipitated volume and the resulting pore throat volume decrease are presented and discussed in~\citeA{Schollenberger2024} for one-phase flow. The concepts either use the information from the adjacent pore bodies to estimate the amount of salt precipitation or solve balance equations in a discretized throat. In this paper, we concentrate on one concept, which uses the source terms in the adjacent pore bodies to calculate the amount of precipitated salt and resulting change in the pore-throat volume for the current time step $t$:
\begin{linenomath*}
\begin{align}
\Delta V_{ij}^t &= \frac{1}{2} \left( \frac{s_i^{\mathrm{NaCl},t}}{A_{\mathrm{react},i}^t} + \frac{s_j^{\mathrm{NaCl},t}}{A_{\mathrm{react},j}^t} \right) \frac{A_{\mathrm{react},ij}^t \Delta t}{\rho_\mathrm{m,s}}.
\label{eq:poreThroatVolumeAlterations}
\end{align}
\end{linenomath*}
The source terms, given in Equation~\ref{eq:sourceTerm}, are normalized by the reactive surface area of the respective pore body and averaged arithmetically. This reaction rate is then applied to the reactive surface of the pore throat $A_{\mathrm{react},ij}$. 
For the presence of two phases in the throats, it is assumed that the liquid phase covers the whole pore-throat walls. So, the reactive surface equals the throat wall surface: for square throat cross sections it is $A_{\mathrm{react,square},ij} = 4 w_{ij} l_{ij}$ and for circular throat cross sections it is $A_{\mathrm{react,circle},ij} = 2 \pi r_{ij} l_{ij}$. With the time step size $\Delta t$ and the molar density of solid salt $\rho_\mathrm{m,s}$ the change in pore-throat volume is calculated. The dimensions for the current time step $t$ are calculated based on the values of the previous time step. It follows for the actual width $w_{ij}^t = \sqrt{(V_{i}^{t-1}+\Delta V_{i}^{t-1})/l_{ij}}$ or radius $r_{ij}^t = \sqrt{(V_{i}^{t-1}+\Delta V_{i}^{t-1})/(l_{ij} \pi)}$.
\subsection{Relations for Phase Distribution}
\subsubsection{Pore Body}
For each pore body a local $p_\mathrm{c}$-$S_\mathrm{l}$-relationship is given with $p_{\mathrm{c},i} = p_{\mathrm{g},i} - p_{\mathrm{l},i}$. For cubic pore bodies it is \cite{Joekar2010}: 
\begin{linenomath*}
\begin{align}
p_{c,i} &= \frac{2 \gamma}{\frac{W_i}{2} \left(1-\mathrm{exp}(-6.83 ~S_{\mathrm{l},i})\right)}.
\end{align}
\end{linenomath*}
\subsubsection{Pore Throat}
A saturated pore throat gets invaded if the entry capillary pressure $p_{c,e}$ of the respective throat is reached by one of the adjacent pore bodies. For square shaped cross sections $p_{c,e}$ is \cite{Oren1998, Mason1991, Ma1996}:
\begin{linenomath*}
\begin{align}
p_{c,e} &= \frac{\gamma (1+2 \sqrt{\pi G}) \mathrm{cos}(\theta)F_d}{w_{ij}/2}, \label{eq:pceSquare}\\
F_d &= \frac{1+\sqrt{1+4G \frac{E}{\mathrm{cos}^2(\theta)}}}{1+2\sqrt{\pi G}},\label{eq:pceSquare1}\\
E &= \pi - 3\theta + 3 \mathrm{sin}(\theta) \mathrm{cos}(\theta)- \frac{\mathrm{cos}^2(\theta)}{4 G}.\label{eq:pceSquare2}
\end{align}
\end{linenomath*}
The shape parameter $G=\frac{A}{P^2}$ describes the throat geometry and is for square cross section $G_\mathrm{square}=\frac{1}{16}$. Further, the corner half angle for square cross sections is $\beta_\mathrm{square} = \frac{\pi}{4}$.
\\
For circular throat cross sections it is \cite{Blunt2017}:
\begin{linenomath*}
\begin{align}
p_{c,e} = \frac{2 \gamma ~\mathrm{cos}(\theta)}{r_{ij}}.
\end{align}
\end{linenomath*}
The process of an invaded throat getting completely filled with wetting phase again is called snappoff. It occurs if the capillary pressure falls below the so called snappoff capillary pressure. As in this paper no snappoff occurs in any scenario, we refer to \cite{Wu2024} for description.
\subsection{Numerical Model and Implementation}
The dynamic pore-network model is implemented in the open-source simulator $\mathrm{DuMu^X}$ \cite{Koch2020}, which is based on the numerical software Dune~\cite{Bastian2021}. A first-order backward Euler scheme is used for the temporal discretization. 
All balance equations are assembled in one system of equations and linearized using the Newton method. 
\\
Different regularizations are used to make the model more robust in terms of the invasion process, which are presented in~\cite{Wu2024}. 
\section{Setups} \label{sec:Setup}
\begin{figure}[t]%
\centering
\begin{subfigure}[t]{0.45\textwidth}
\includegraphics[width=\textwidth]{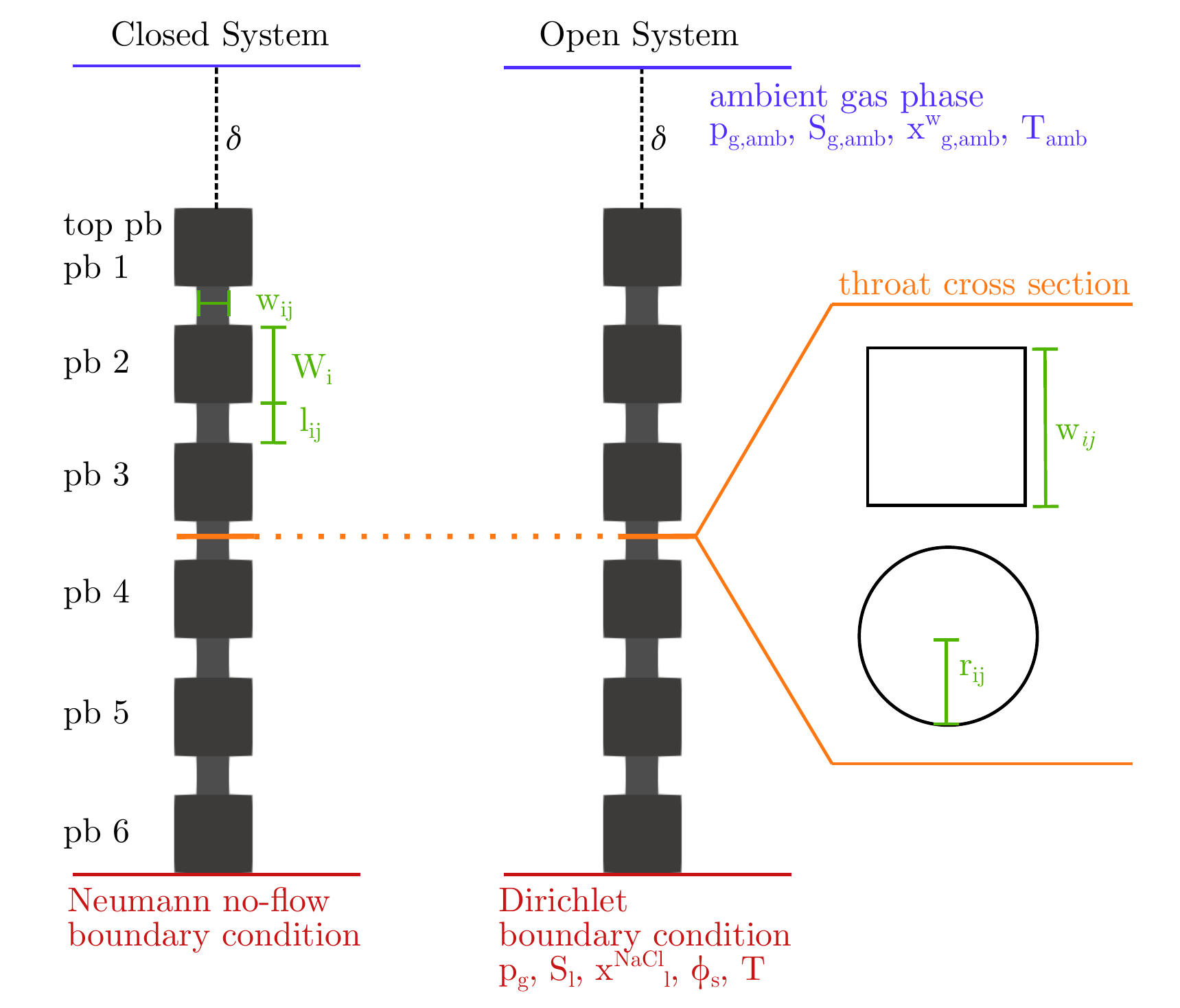}
\subcaption{One-dimensional setup}
\label{fig:1DSetup}
\end{subfigure}
\begin{subfigure}[t]{0.45\textwidth}
\includegraphics[width=\textwidth]{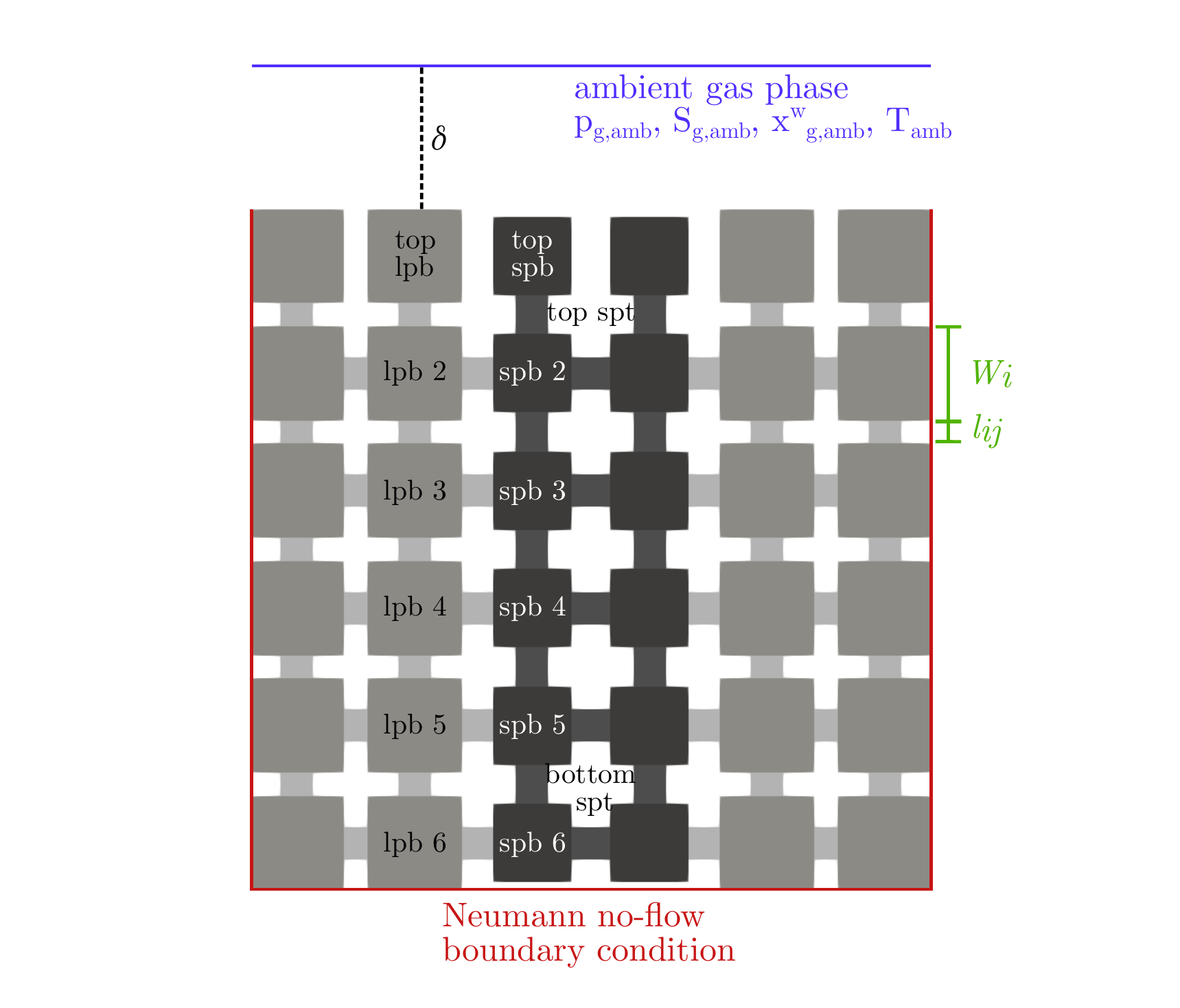}
\subcaption{Two-dimensional setup}
\label{fig:2DSetup}
\end{subfigure}
\caption{Geometry and boundary conditions of setups }
\label{fig:Setups}
\end{figure}

To demonstrate the characteristics of the pore-network model two different setups are considered: a simple, homogeneous, one-dimensional setup and a heterogeneous, two-dimensional setup, see Figure~\ref{fig:Setups}.
%\FloatBarrier
%
%
\subsection{One-dimensional Setup} \label{subsec:1DSetup}
The one-dimensional setup consists out of five equal sized, cubic pore bodies (pb1-pb5), which are stacked above each other and connected through equal sized pore throats (pt1-pt4) with square or circular shaped cross sections. The dimensions can be seen in Figure~\ref{fig:1DSetup} and Table~\ref{tab:Setups}.
\\
%Initial conditions
Initially, the system is nearly saturated, and no pore throat is invaded with gas phase. The system is either filled with pure water or a high concentrated brine. Further, the gas phase is at atmospheric pressure and no precipitated salt is present initially. The exact values for the initial conditions can be found in Table~\ref{tab:Setups}.
\\
%Boundary conditions
For the one-dimensional setup, two different systems are considered, which differ in the boundary condition at the bottom: an open or a closed system. The open system is considered to be connected to a reservoir at the bottom which has constant parameters. The system is open at the bottom and fluxes can enter or leave the system. This is realized with a Dirichlet boundary condition fixing the primary-variables in the bottom pore body (pb5) to the initial values. In the closed system, no fluxes can enter or leave at the bottom. Here, at the bottom, a Neumann no-flow boundary condition is used. In both systems, Neumann no-flow boundary conditions are also used at the left and right. 
\\
The top, in both systems, is open for gas phase and energy fluxes, including the evaporative flux of water vapor. Therefor, parameters ($p_\mathrm{g,amb}$, $S_\mathrm{g,amb}$, $x_\mathrm{g,amb}^\mathrm{w}$, $T_\mathrm{amb}$) are fixed in the ambient gas phase, see Table~\ref{tab:Setups}. Advective, diffusive and energy fluxes are calculated according to the balance equations described in Section~\ref{subsec:BalanceEquations} based on the gradients of $p_\mathrm{g}$, $x_\mathrm{g}^\mathrm{\kappa}$, $T$ arising between the constant parameters in the ambient gas phase and the values in the top pb over the boundary layer thickness $\delta = 1.6 \cdot 10^{-3}~\mathrm{m}$. It is assumed that no liquid phase can leave the system at the top. The exact equations of the top boundary condition can be found in \ref{app:TopBC}.
\begin{table}
\caption{Parameters of the one- and two-dimensional setup: geometry of the setups, initial conditions and values of the boundary condition in the ambient gas phase}
\label{tab:Setups}
\centering
\begin{tabular}{l r r r l l}
\hline
\textbf{parameter}  & {\textbf{1D}} & \multicolumn{2}{c}{\textbf{2D }} & \textbf{unit} & \textbf{description} \\
%\hline
%\cline{2-4}
  &  & \textbf{small} & \textbf{large} & &  \\
\hline
\multicolumn{6}{l}{\textbf{Geometry}}\\
\hline
  $W_\mathrm{i}$  & $4$ & $4$ & $4.8$ & $10^{-4} \mathrm{m}$ & initial pore-body radius \\
  $w_\mathrm{ij}$  & $2.2$ & $2.2$ & $2.64$ & $10^{-4} \mathrm{m}$ & initial pore-throat cross section \\
    &  & & & &  side length  \\
  $l_\mathrm{ij}$  & $2$ & $2$ & $1.2$ & $10^{-4} \mathrm{m}$ & pore-throat length\\
\hline
%\multicolumn{2}{l}{$^{a}$Footnote text here.}
%
%\hline
%parameter  & values  &  & & description\\
% & 1D & 2D & & \\
\multicolumn{6}{l}{\textbf{Initial condition}}\\
\hline
  $S_\mathrm{l}$  & $0.9$ & \multicolumn{2}{c}{$0.9$} & $-$& initial liquid saturation \\
  $p_\mathrm{g}$  & $1\cdot 10^{5}$ & \multicolumn{2}{c}{$1\cdot 10^{5}$} & $\mathrm{Pa}$ & initial gas pressure  \\
  $x_\mathrm{l}^\mathrm{NaCl}$  & $0.09$ & \multicolumn{2}{c}{$0.06$} & $-$& initial salt mole fraction \\
  $\phi_\mathrm{s}$  & $0.0$ & \multicolumn{2}{c}{$0.0$} & $-$& initial precipitated volume fraction \\
  $T$  & $293.15$ & \multicolumn{2}{c}{$293.15$} &$\mathrm{K}$ & initial temperature  \\
\hline
%\multicolumn{2}{l}{$^{a}$Footnote text here.}
%
%parameter  & value & & description \\
\multicolumn{6}{l}{\textbf{Boundary condition in the ambient gas phase}} \\
\hline
  $p_\mathrm{g,amb}$  & \multicolumn{3}{c}{$1\cdot 10^{5}$} & $\mathrm{Pa}$ & gas pressure of the ambient \\
  $S_\mathrm{g,amb}$  & \multicolumn{3}{c}{$1.0$} & $-$& gas saturation of the ambient \\
  $x_\mathrm{g,amb}^\mathrm{w}$  & \multicolumn{3}{c}{$4.7 \cdot 10^{-3}$} & $-$& water mole fraction in the ambient \\
  $T_\mathrm{amb}$  & \multicolumn{3}{c}{$293.15$} & $\mathrm{K}$ & temperature in the ambient \\
\hline
%\multicolumn{2}{l}{$^{a}$Footnote text here.}
\end{tabular}
\end{table}
\subsection{Two-dimensional Setup} \label{subsec:2DSetup}
The two-dimensional setup consists out of a network of six by six pore-bodies, connected by vertical and horizontal pore throats, see Figure~\ref{fig:2DSetup}. The system has a heterogeneous pore-size distribution. The two columns of pore bodies and throats in the middle have the same dimensions as in the one-dimensional setup. The remaining pore bodies and throats on the right and left have larger dimensions, see Table~\ref{tab:Setups}. 
\\
Initially, the system is nearly saturated with a salt solution with $x_\mathrm{l}^\mathrm{NaCl} = 0.06$. Further, the gas phase pressure is at atmospheric pressure and the precipitated volume fraction is zero. Initially, the pore throats in the region with smaller pore sizes are fully saturated with brine, whereas the larger throats are invaded by the gas phase. 
\\
For the two-dimensional system, we consider a closed system with Neumann no-flow boundary conditions at the bottom, left and right. At the top the same evaporation boundary condition is applied as for the one-dimensional setup, also described in \ref{app:TopBC}.
\section{Results and Discussions} \label{sec:Results}
In this section, we show and discuss processes during evaporation-driven salt precipitation. In this way, the capabilities of the previously introduced pore-network model are demonstrated. To analyze different process correlations separately, at the beginning, the simple, one-dimensional setup is used. 
First, general processes during evaporation and precipitation from a pore-network model and the differences between water and brine are shown in Section~\ref{subsec:WaterBrine}. Further, different pore-scale processes are analyzed in detail: the influence of pore-space alterations of pore bodies and pore throats in Section~\ref{subsec:PoreSpaceAlterations} and the influence of the location of interfaces on the evaporation and precipitation in Section~\ref{subsec:InterfaceLocation}. Finally, the interactions of these processes and the influence of local heterogeneities are considered in Section~\ref{subsec:LocalHeterogeneities} using the two-dimensional heterogeneous setup.
\\
To investigate the processes the capillary pressure, the liquid saturation, the evaporation rate, the salt mole fraction and the precipitated volume fraction of the top pb are evaluated, in some cases additionally also the parameters from pb2. 
\subsection{General Processes and Differences Between Water and Brine During Evaporation} \label{subsec:WaterBrine}
The open and closed one-dimensional setup with square-shaped throat cross section is used to show the processes during evaporation from the pore-network model. The evaporation of pure water and brine is considered. The volume alterations due to precipitation, in case of brine, are neglected, in this section.  
%\\
\begin{figure}[p]%
\centering
\includegraphics[width=0.47\textwidth]{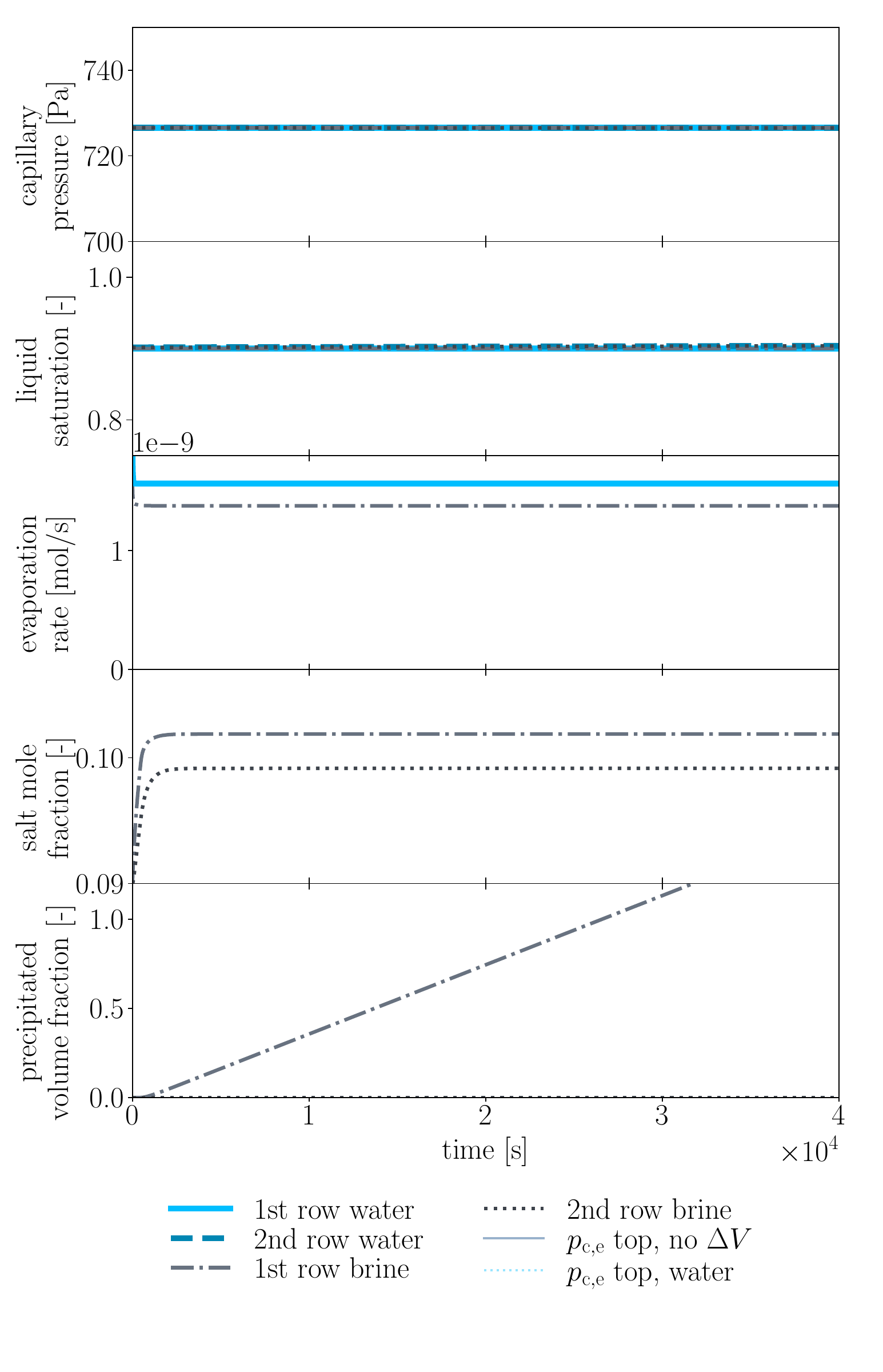}
\includegraphics[width=0.47\textwidth]{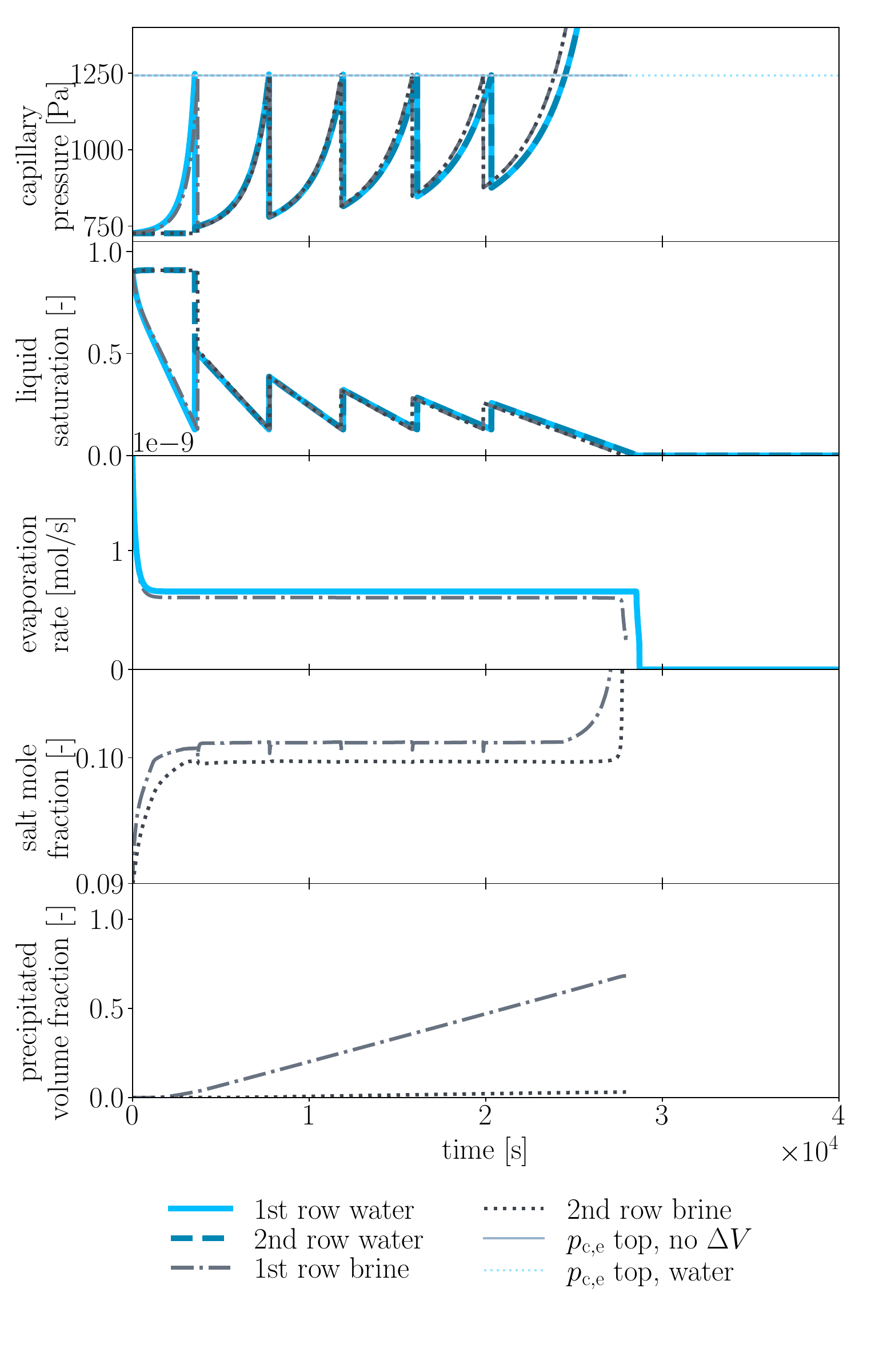}
\caption{Comparison of pure water and brine for square-shaped throat cross sections, considering an open (left) and closed system (right)}
\label{fig:WaterBrine_Square}
%\end{figure}
%\FloatBarrier
%
%
%
%\begin{figure}[h]%
\centering
\begin{subfigure}[t]{0.3\textwidth}
\includegraphics[width=\textwidth]{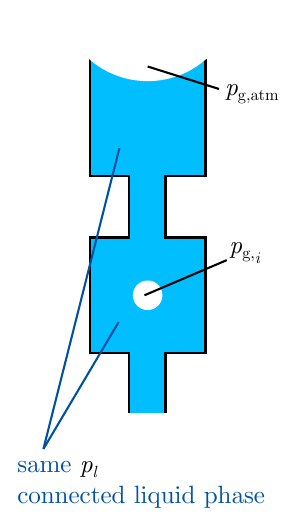}
\subcaption{Before invasion}
\label{fig:Invasion1}
\end{subfigure}
\begin{subfigure}[t]{0.3\textwidth}
\includegraphics[width=\textwidth]{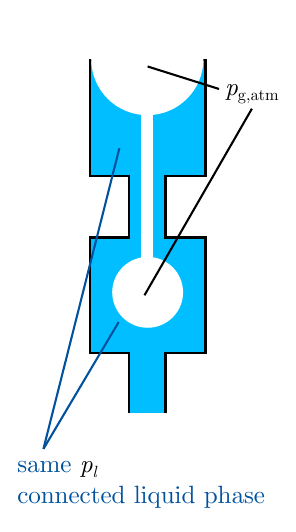}
\subcaption{Invasion of first throat}
\label{fig:Invasion2}
\end{subfigure}
\begin{subfigure}[t]{0.3\textwidth}
\includegraphics[width=\textwidth]{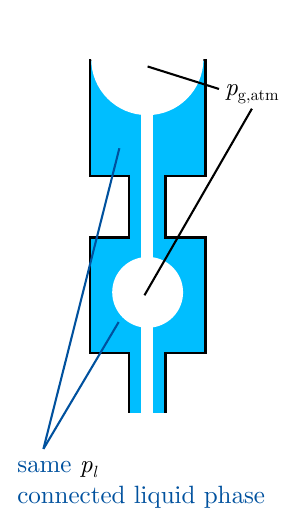}
\subcaption{Invasion of second throat}
\label{fig:Invasion3}
\end{subfigure}
\caption{Invasion process }
\label{fig:Invasion}
\end{figure}
%\FloatBarrier
%
\subsubsection{Open System} \label{subsubsec:WaterBrineOpen}
% Open system
In the open system, the capillary pressure stays constant over time and is the same for every pore body, as shown in Figure~\ref{fig:WaterBrine_Square} for the top pb and pb2. Based on the capillary pressure-saturation relationship, applied to each pore body, this results in a constant liquid saturation over time and the domain. The flux, which enters the system from the bottom in the open system, balances the loss of liquid phase due to evaporation of water which leads to this stationary saturation distribution. In this setup, no throat gets invaded by air. 
Pure water and brine differ in evaporation rate. The evaporation rate is constant over time but higher for water than for brine. This corresponds to the lower saturated vapor pressure of brine, presented in Table~\ref{tab:FluidProperties}.
In case of evaporating brine, the initial salt mole fraction increases over time. The upwards flux of brine transports salt towards the top pb. Here, the water evaporates, whereas the salt accumulates.
In the top pb the solubility limit is reached after $480$s and the salt mole fraction stays constant at $x_\mathrm{l}^\mathrm{NaCl} = 0.102$. An equilibrium establishes between transport of salt into the top pore body, the amount of salt which precipitates and diffusive fluxes, which transport salt downwards driven by the salt mass fraction gradient. In the top pore body salt starts to precipitate as soon as the precipitation limit is reached. The saturated volume fraction increases linearly and exceeds $1.0$. This high volume fraction is nonphysical but would be limited to $1.0$ if volume alterations were taken into account. In cases with such high precipitated volume fractions the volume change of the pore body is not neglectable anymore.
%\\
\subsubsection{Closed System} \label{subsubsec:WaterBrineClosed}
% Closed system
%Invasion process in terms of pc and Sl
In the closed system, the evaporated water from the liquid phase gets replaced by gas phase, which flows in at the top, as no fluid can enter from the bottom. First the saturation in the top pb decreases and the capillary pressure rises accordingly, see Figure~\ref{fig:WaterBrine_Square}. In pb2, the liquid saturation and the capillary pressure stay at the initial value, at first. The lower pore bodies (pb2-pb5) can not supply the top pb with liquid phase and contribute to the evaporation without a connection of the gas phase to the ambient gas phase, as there is no other fluid, which could replace out-flowing liquid phase. 
\\
In Figure~\ref{fig:Invasion} the state in the top pb and pb 2 are demonstrated schematically at different times. 
Initially, the liquid phase is connected in the whole domain, which leads to the same liquid pressure in the whole domain. The gaseous phase in the top pb is connected to the ambient air and so the gas pressure equals atmospheric pressure. However, in the lower bodies, the gas phase is not connected to the ambient air and consequently the air pressures are decoupled, see Figure~\ref{fig:Invasion1}. Based on that, different capillary pressures can develop. 
When the capillary pressure in the top pb reaches the entry capillary pressure of the top pt, the throat is invaded with gaseous phase from the top. Now gaseous phase from the top pb can enter pb2 and the gaseous phase of the two pore bodies is connected, see Figure\ref{fig:Invasion2}. Over the period of a few seconds, advective gas fluxes equalize the gas pressures, so in both pore bodies there is atmospheric pressure. Liquid phase is flowing from pb2 to the top pb. The out-flowing liquid of pb2 can now be replaced with gas phase from the ambient. This leads to jumps in saturation. After invasion, the two pore bodies have the same capillary pressure and same liquid saturation. 
These processes repeat at the next invasions of pt2, pt3 and pt4, which are indicated by the jumps in capillary pressure and liquid saturation in Figure~\ref{fig:WaterBrine_Square}.
\\
%Salt mole fraction and precipitation
The evaporation rate of water is higher as for brine as for the open system. This results in slightly faster decrease of the liquid saturation and earlier invasions in case of water. When the system is completely dried out, the evaporation rate drops to zero. In case of brine, the salt mole fraction increases and exceeds the solubility limit. As for the open system, the salt mole fraction gets constant due to an equilibrium of advective, diffusive fluxes and precipitation of salt. A drop in salt mole fraction is observed at every invasion as liquid from lower pore bodies with lower salt mole fractions enters the upper pore bodies. In the top pore body, the solubility limit is reached and salt precipitates. Also in pb2, the solubility limit is reached and a small amount of salt precipitates. In the lower pore bodies, small amounts of salt precipitate at the end when the system is nearly dried out. Then the solubility limit is also exceeded in lower pore bodies, as just a residual amount of water is left.
\subsection{Pore-Space Alterations due to Precipitation} \label{subsec:PoreSpaceAlterations}
\begin{figure}[t]%
\centering
\includegraphics[width=0.47\textwidth]{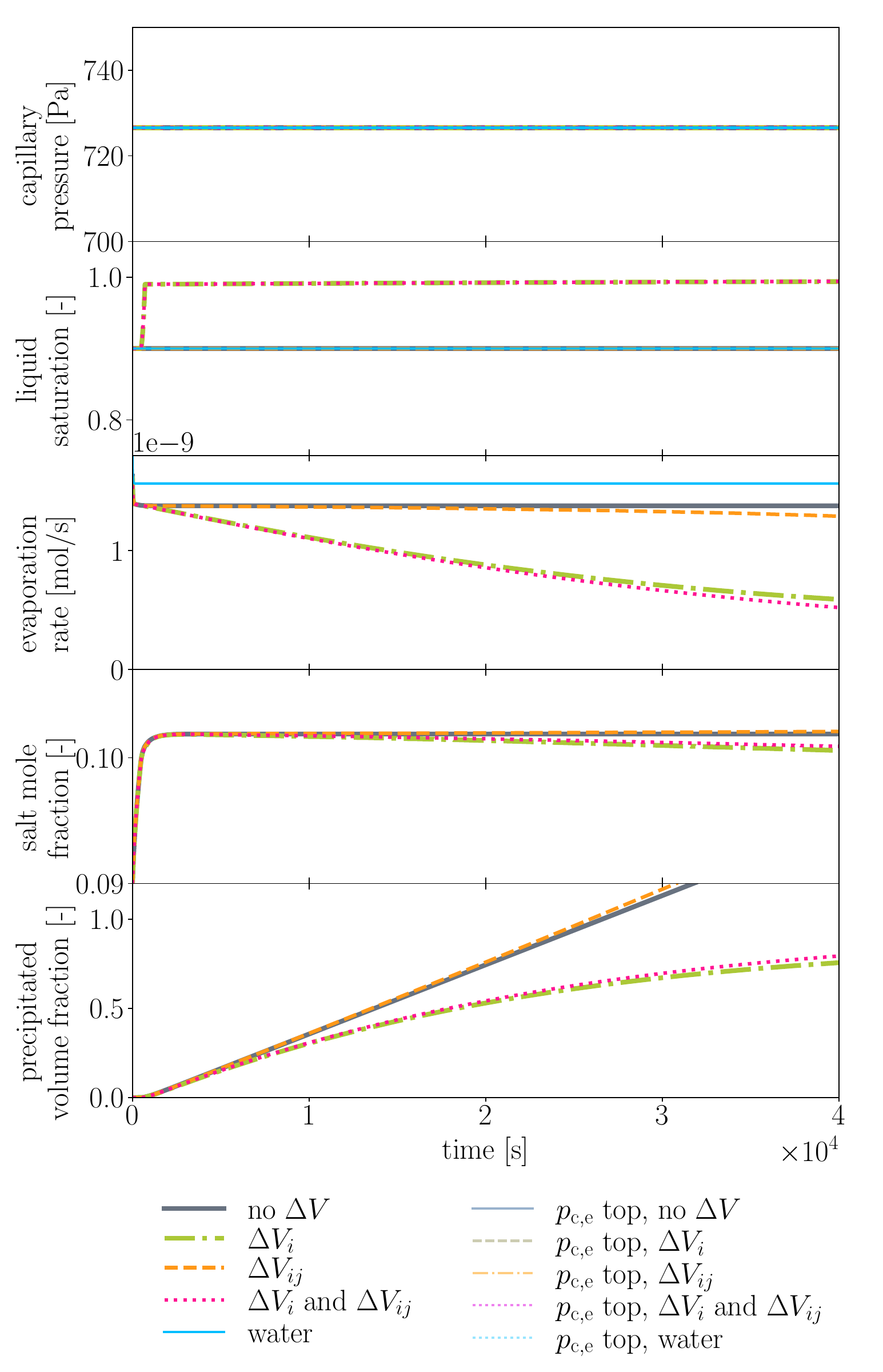}
\includegraphics[width=0.47\textwidth]{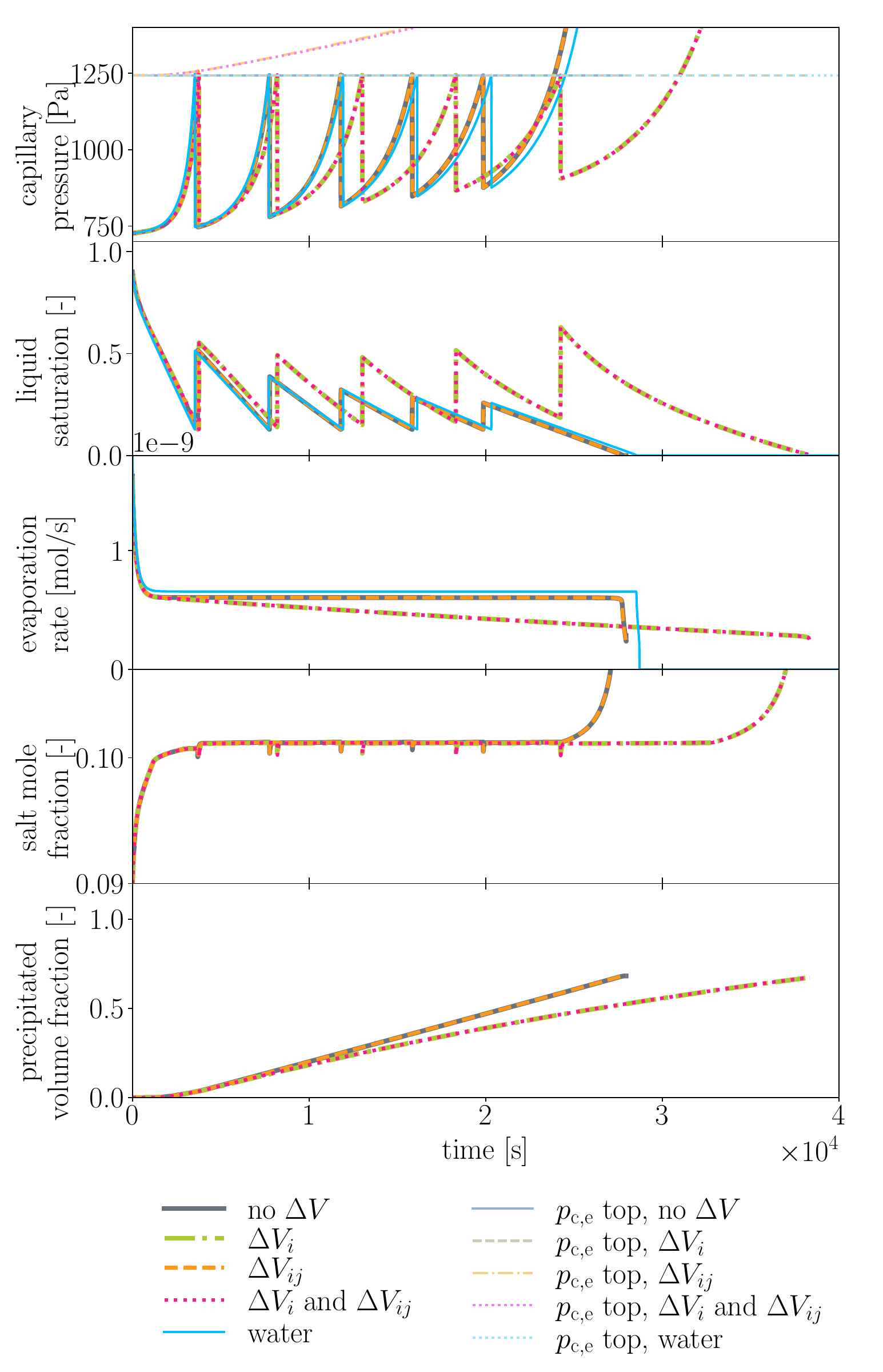}
\caption{Volume alteration for brine for square-shaped throat cross sections, considering an open (left) and closed system (right). Comparison of different cases: no volume alteration (no $\Delta V$), just pore-body alteration ($\Delta V_i$), just pore-throat volume alterations ($\Delta V_{ij}$), and both pore-body and pore-throat volume alterations ($\Delta V_{i}$ and $\Delta V_{ij}$)}
\label{fig:Brine_Square_VAlterations}
\end{figure}

In the open and the closed setup, high precipitated volume fractions occur. Thus, the decrease of pore volume is not neglectable as assumed in the previous section. In this section, we take pore space alterations into account, as described in Section~\ref{subsec:PoreSpaceAlterations}. Their influence is analyzed, first for pore bodies, then for pore throats and finally the interactions between them. The results of the top pb of all different cases are summarized in Figure~\ref{fig:Brine_Square_VAlterations}.
%\FloatBarrier
%
\subsubsection{Influence of Pore-Body Volume Alteration} \label{subsubsec:PoreBodyVolumeAlteration}
\begin{figure}[t]%[ht!]%
\centering
\begin{subfigure}[t]{0.48\textwidth}
\includegraphics[width=0.99\textwidth]{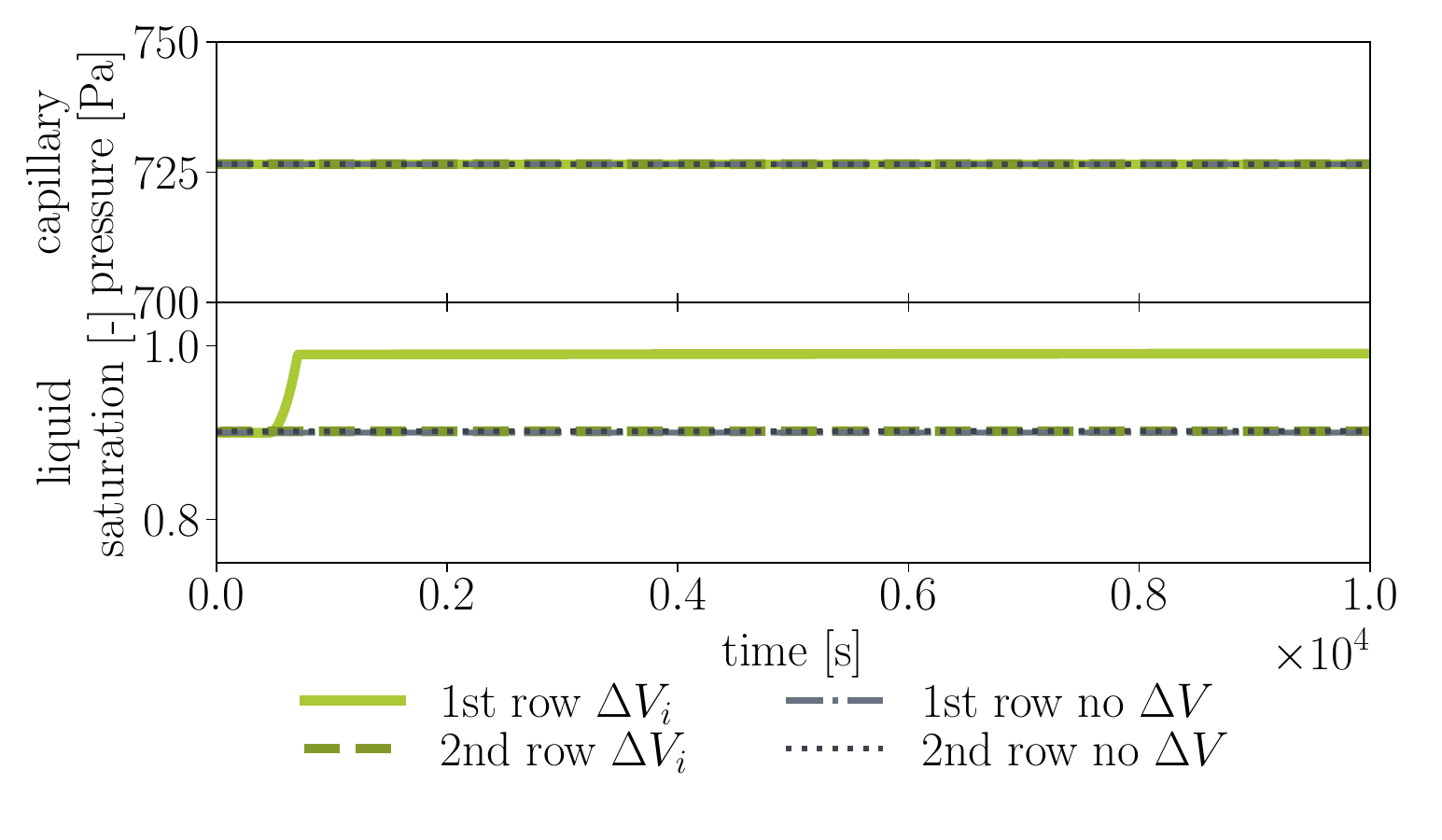}
\caption{Development of capillary pressure and liquid saturation}
\label{fig:Brine_Square_VBody_pcSl_open}
\end{subfigure}
%\end{figure}
%
\hfill
%\begin{figure}[ht!]%[ht!]%
\centering
\begin{subfigure}[t]{0.48\textwidth}
\includegraphics[width=0.99\textwidth]{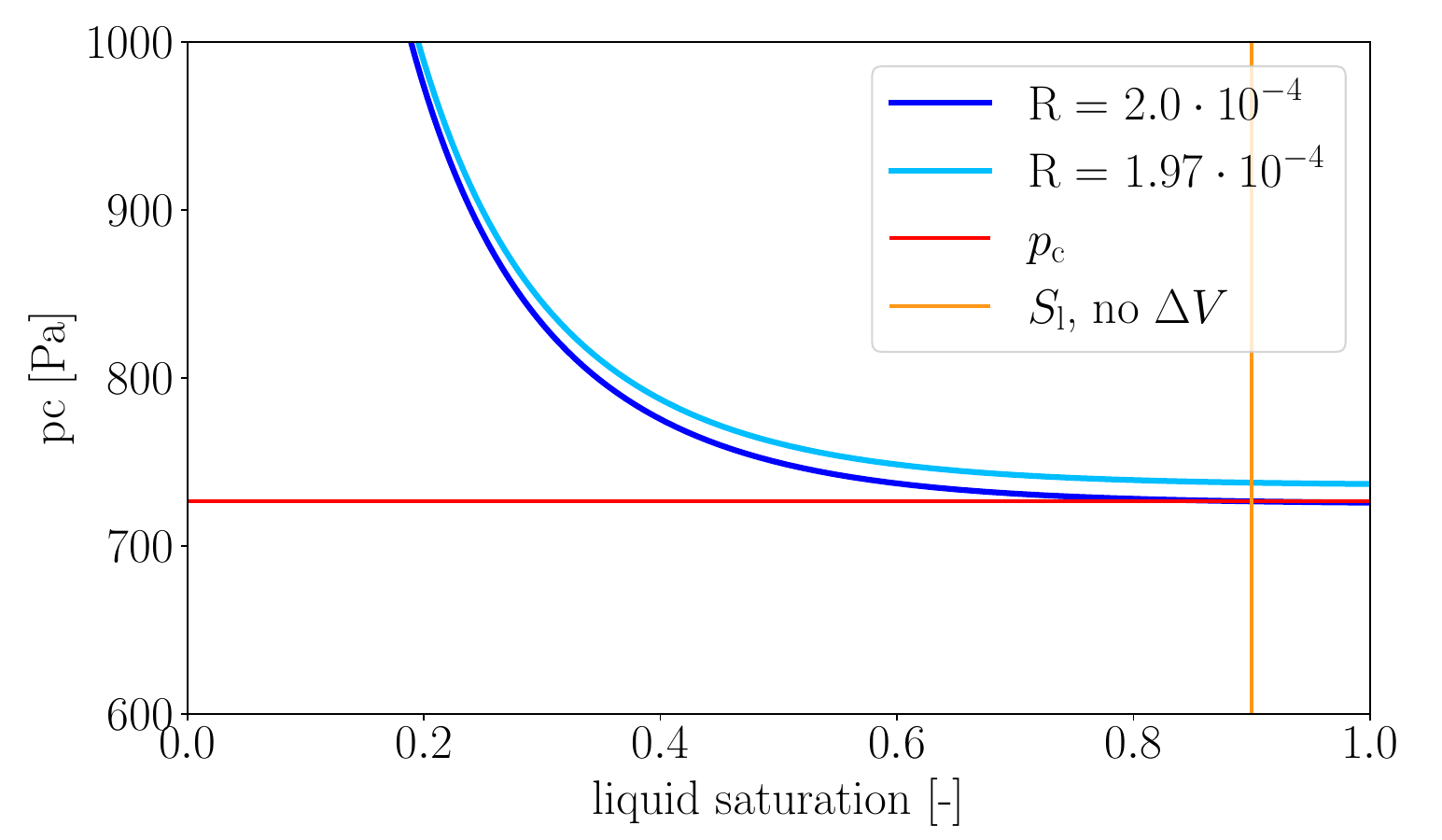}
\caption{Capillary pressure-saturation relationships initially and for top pb after $2000$s}
\label{fig:Brine_Square_VBody_pcSlRelationships_open}
\end{subfigure}
\caption{Comparison of case without volume alterations and with pore-body volume alterations for square throat cross sections and the open system}
\end{figure}

\begin{figure}[ht!]%
\centering
\begin{subfigure}[t]{0.48\textwidth}
\includegraphics[width=0.99\textwidth]{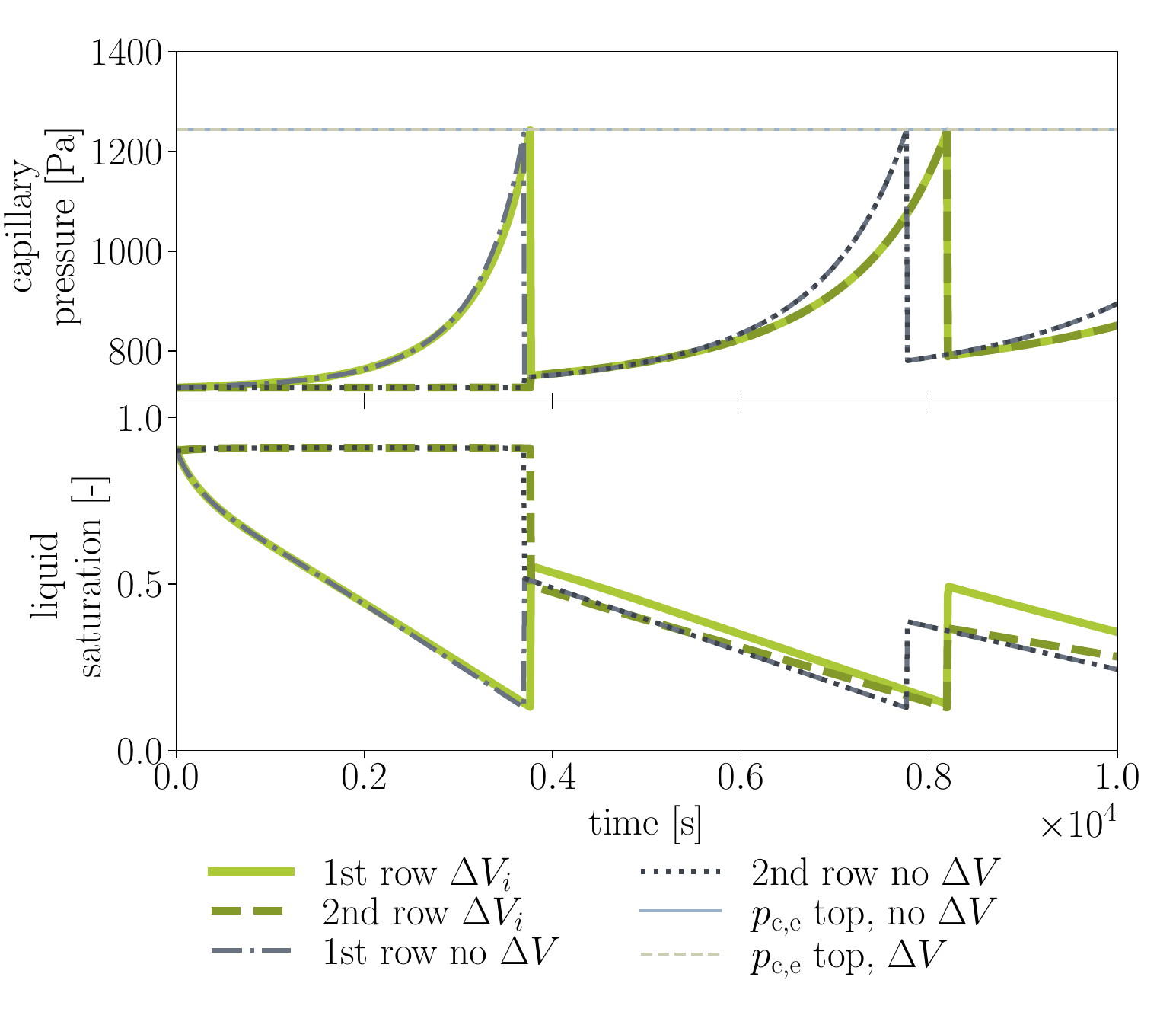}
\caption{Development of capillary pressure and liquid saturation}
\label{fig:Brine_Square_VBody_pcSl_closed}
\end{subfigure}
%\end{figure}
\hfill
%
%\begin{figure}[ht!]%[ht!]%
\centering
\begin{subfigure}[t]{0.48\textwidth}
\includegraphics[width=0.99\textwidth]{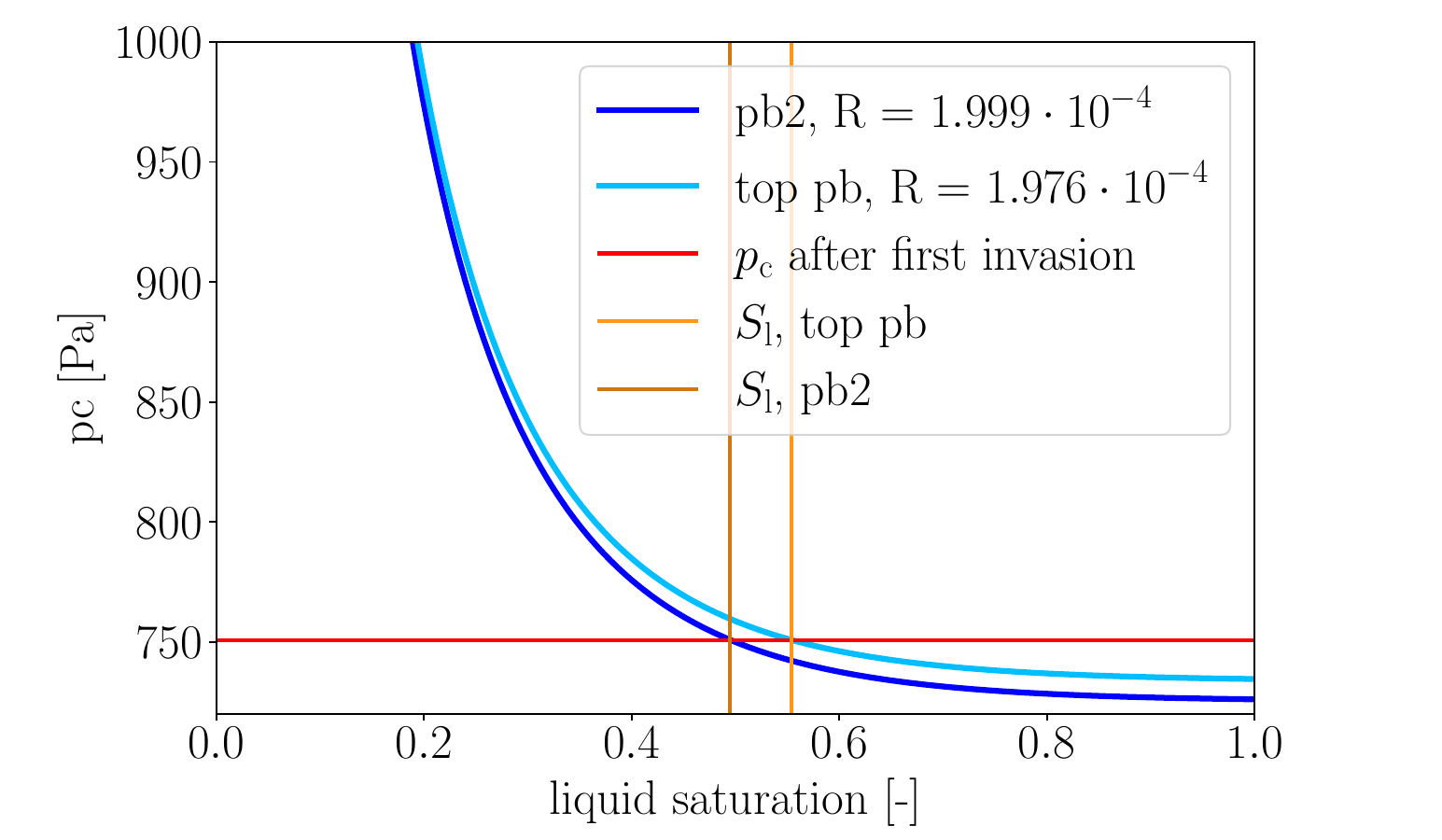}
\caption{Capillary pressure-saturation relationships after the first invasion for top pb and pb2}
\label{fig:Brine_Square_VBody_pcSlRelationships_closed}
\end{subfigure}
\caption{Comparison of case without volume alterations and with pore-body volume alterations for square throat cross sections and the closed system}
\end{figure}

In this section, we consider only the alteration of the pore-body volume as described in equation~\ref{eq:poreBodyVolumeAlterations}. In both setups, the precipitated volume fraction is limited now to $1.0$. 
%Open system
In the open system the salt precipitates in the top pb, consequently the pore-body radius of the top pb decreases with the increasing precipitated volume fraction. In Figure~\ref{fig:Brine_Square_VBody_pcSl_open} the $p_\mathrm{c}$ and $S_\mathrm{l}$ are shown for the initial $1 \cdot 10^4$s of the top pb and pb2. As reference the case for brine without volume changes, discussed in the previous section, is plotted. The $p_\mathrm{c}$ of both cases is constant and the same for both cases and pore bodies. However, the saturation of top pb increases as soon as salt precipitates. Due to the resulting decrease of volume of the top pb the local $p_\mathrm{c}$-$S_\mathrm{l}$ relationship for this pore body changes, which leads to different $S_\mathrm{l}$ for the same $p_\mathrm{c}$, see Figure~\ref{fig:Brine_Square_VBody_pcSlRelationships_open}. The $S_\mathrm{l}$ in the top pb increases to nearly 1.0.
%\FloatBarrier
\\
%Closed system
Similar effects can also be observed for the closed systems. 
After invasion of the top pt the same $p_\mathrm{c}$ in the top pb and pb2 leads to different saturations, see Figure~\ref{fig:Brine_Square_VBody_pcSl_closed}. The saturation in the top pb is higher due to the decreased pore-body radius and the resulting shift of the $p_\mathrm{c}$-$S_\mathrm{l}$ relationship, see Figure~\ref{fig:Brine_Square_VBody_pcSlRelationships_closed}. After each invasion, this difference in $S_\mathrm{l}$ gets bigger as the difference in pore-body volume increases due to the continuing precipitation in the top pb. Further, the $p_\mathrm{c}$ increases and $S_\mathrm{l}$ decreases slower than if volume alterations are neglected due to smaller evaporative area in the top pb. This results in later invasion of pore throats than without volume change and a slower precipitation rate, see Figure~\ref{fig:Brine_Square_VAlterations}. Additionally, the precipitation rate is lowered by the decrease of reactive surface in the source term of
precipitation.
\FloatBarrier
\subsubsection{Influence of Pore-Throat Volume Alteration} \label{subsubsec:PoreThroatVolumeAlteration}
\begin{figure}[t]%
\centering
\begin{subfigure}[t]{0.48\textwidth}
\includegraphics[width=0.99\textwidth]{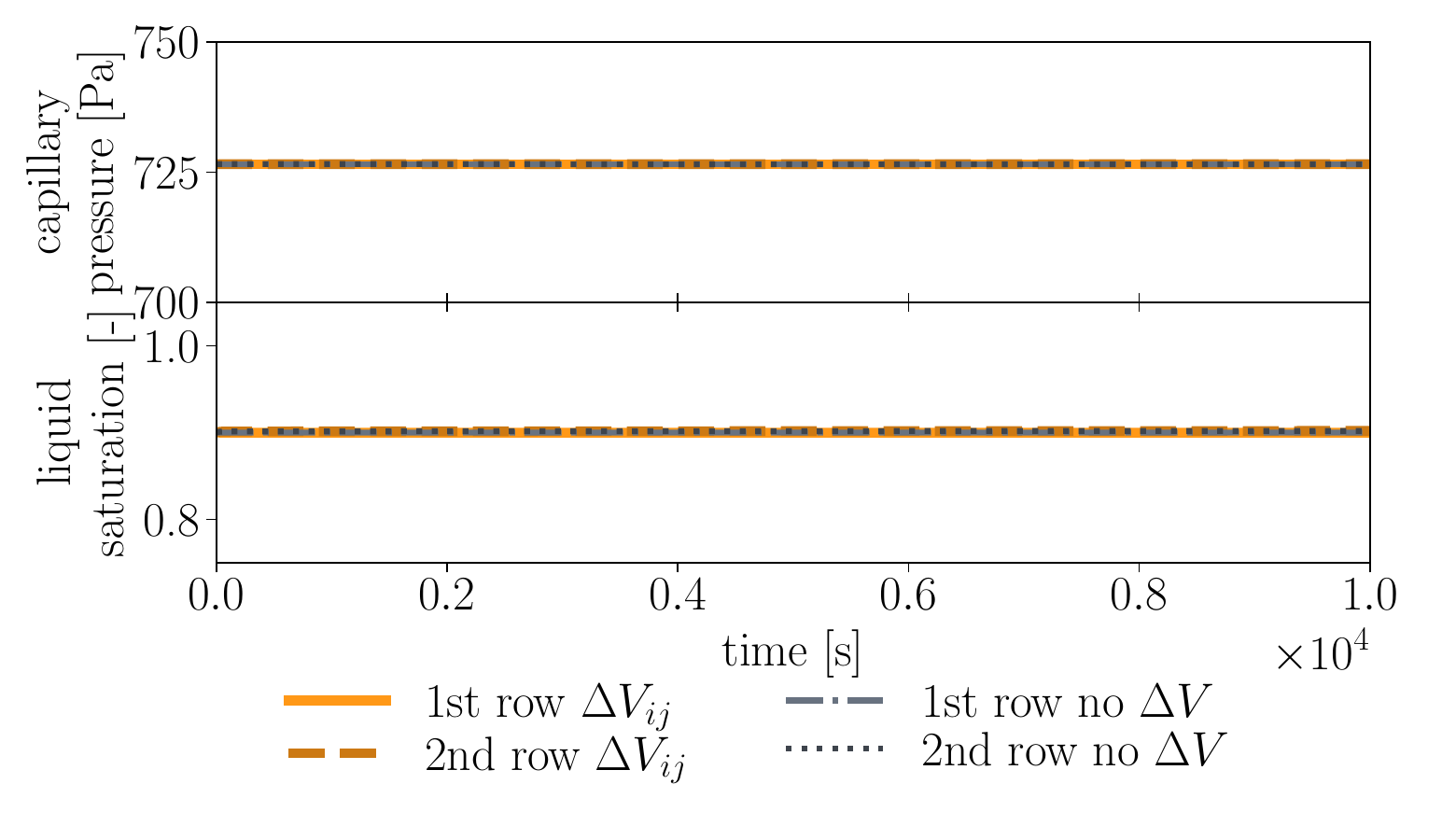}
\caption{Open system}
\label{fig:Brine_Square_VThroat_pcSl_open}
\end{subfigure}
\hfill
%\end{figure}
%\FloatBarrier
%\begin{figure}[h!]%
\begin{subfigure}[t]{0.48\textwidth}
\centering
\includegraphics[width=0.99\textwidth]{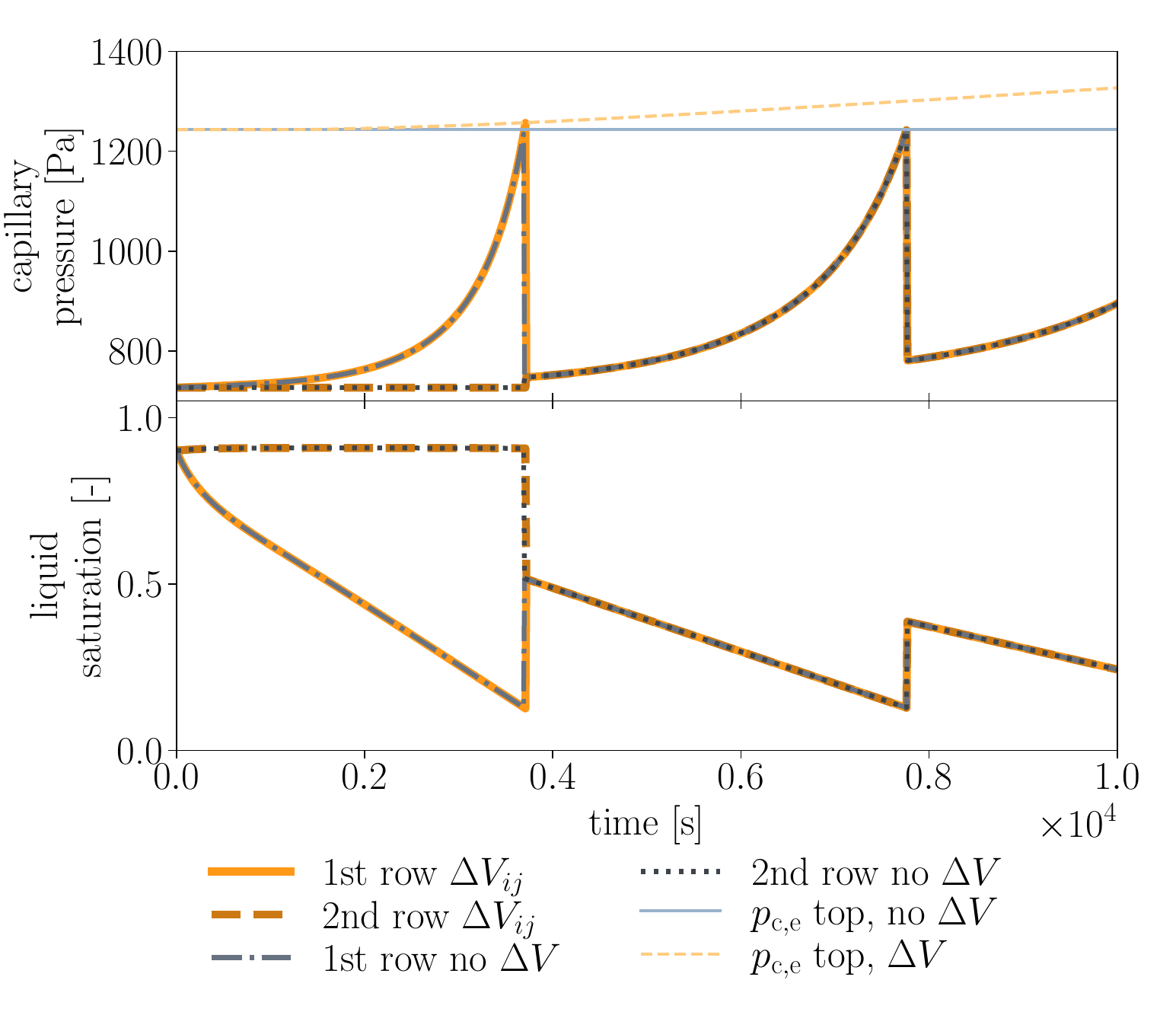}
\caption{Closed system}
\label{fig:Brine_Square_VThroat_pcSl_closed}
\end{subfigure}
\caption{Comparison of case without volume alterations and with pore-throat volume alterations for square throat cross sections}
\end{figure}

In this section, only alterations of the pore-throat volumes are considered (no alteration in pore-body volume). 
%Open system
For the open system the influences of smaller pore-throat radii are neglectable, see Figure~\ref{fig:Brine_Square_VThroat_pcSl_open}. As salt precipitates in the top pb, just the radius of the top pt decreases. As the throats get not invaded by air in the open system, the only influence of a decreasing throat radius is a decreasing liquid transmissibility for the throat, see Equation~\ref{eq:1pTransmissibilitySquare}. This is compensated by higher liquid flux velocities in the beginning, but later with a larger reduction in transmissibility leads to a slight reduction in the evaporation rate.
The diffusive flux decreases due to smaller cross-sectional area, which leads to smaller $x_\mathrm{l}^\mathrm{NaCl}$ in the lower pore bodies and slightly more precipitation in the top pb.
%\FloatBarrier
%\\
%Closed system
The width of the pore throat also effects its entry pressure $p_\mathrm{c,e}$, see Equation~\ref{eq:pceSquare}. This influences the invasion processes in the closed system. Due to precipitation in the top pb the radius of the top pt decreases. This leads to an increase of $p_\mathrm{c,e}$ for the top pt, see Figure~\ref{fig:Brine_Square_VThroat_pcSl_closed}. In the following, a higher $p_\mathrm{c}$ or a lower $S_\mathrm{l}$ is needed in the top pb to invade the top pt. As a result, the top pt is invaded later, as for the case without volume alterations. The effect of an increasing $p_\mathrm{c}$ is not visible for the lower pore throats, as here no salt precipitates before the invasion of the respective throat. Hence, the $p_\mathrm{c,e}$ needed for the second invasion process, representing the invasion of pt2, has not increased.
The liquid transmissibility of an invaded throat is independent of the radius, see Equation~\ref{eq:liquidTransmissibility}. The liquid occupied area is in the corners of the throat cross section and so does not change with the throat radius. The gas transmissibility, however, decreases with decreasing throat radius. As the top throat gets invaded short after the precipitation starts, the reducing radius of the top pt has no visible influence on the liquid transmissibility of the not-invaded throat. No difference of evaporation rate or precipitated volume fraction occurs compared to the case without volume alterations.
\subsubsection{Interaction of Pore-Body and Pore-Throat Volume Alteration} \label{subsubsec:PoreBodyThroatVolumeAlteration}
The processes based on the decreasing pore-body and pore-throat radii interact with each other. 
The decrease in pore-body radius leads to lower evaporation rates, which results in lower salt mole fractions and precipitated volume fraction in the pore body, see Section~\ref{subsubsec:PoreBodyVolumeAlteration}. This leads to less decrease of the pore-throat radius compared to the case where pore-body volume alterations are neglected, see Equation~\ref{eq:poreThroatVolumeAlterations}. 
\\
For decrease of the pore-throat radius, however, different effects are predominant for the open and closed system due to the different invasion states of the top pt. 
For the open system, the decrease in radius of the top pt leads to a reduced diffusive flux. This causes higher $x_\mathrm{l}^\mathrm{NaCl}$ and $\phi_\mathrm{s}$ in the top pb. Further, the liquid transmissibility decreases with the decrease of the top pt radius, which leads to a lower evaporation rate. In contrast to the reduced diffusive flux, this causes smaller $x_\mathrm{l}^\mathrm{NaCl}$ and $\phi_\mathrm{s}$ in the top pb. In case of the open system setup, the effect of the lower diffusive flux dominates, resulting in higher $\phi_\mathrm{s}$. As a result, the reduction in pore-throat radii enhances the reduction in pore-body radii and its effects. Further, the reduction of evaporation rate due to pore-body volume and due to pore-throat volume reduction add up. Here, the reduction due to pore-body volume alteration dominates. The reduction of the evaporation rate due to pore-body volume reduction gets enhanced in case of pore-throat volume reduction, whereas the reduction due to pore-throat volume reduction gets reduced in case of pore-body volume reduction.
\\
For the closed system, the reduction in pore-throat radius increases $p_\mathrm{c,e}$ of the top pt, which leads to a later invasion of the throat. This effect is reduced in combination with the reduction of pore-body volume, because the reduction in pore-body volume reduces the reduction of the pore-throat volume, as derived in the beginning of the section. In case of the closed system, the effects of the reduced pore-throat volume have no influence on the pore-body volume reduction. 
\subsection{Location of Interfaces} \label{subsec:InterfaceLocation}
As salt often precipitates at the gas-liquid interface, the location of the interface has a large influence on the location of precipitation. In the following, the location of interfaces and the resulting precipitation distribution is analyzed for different setups. 
\begin{figure}[ht!]%
\centering
\includegraphics[width=0.47\textwidth]{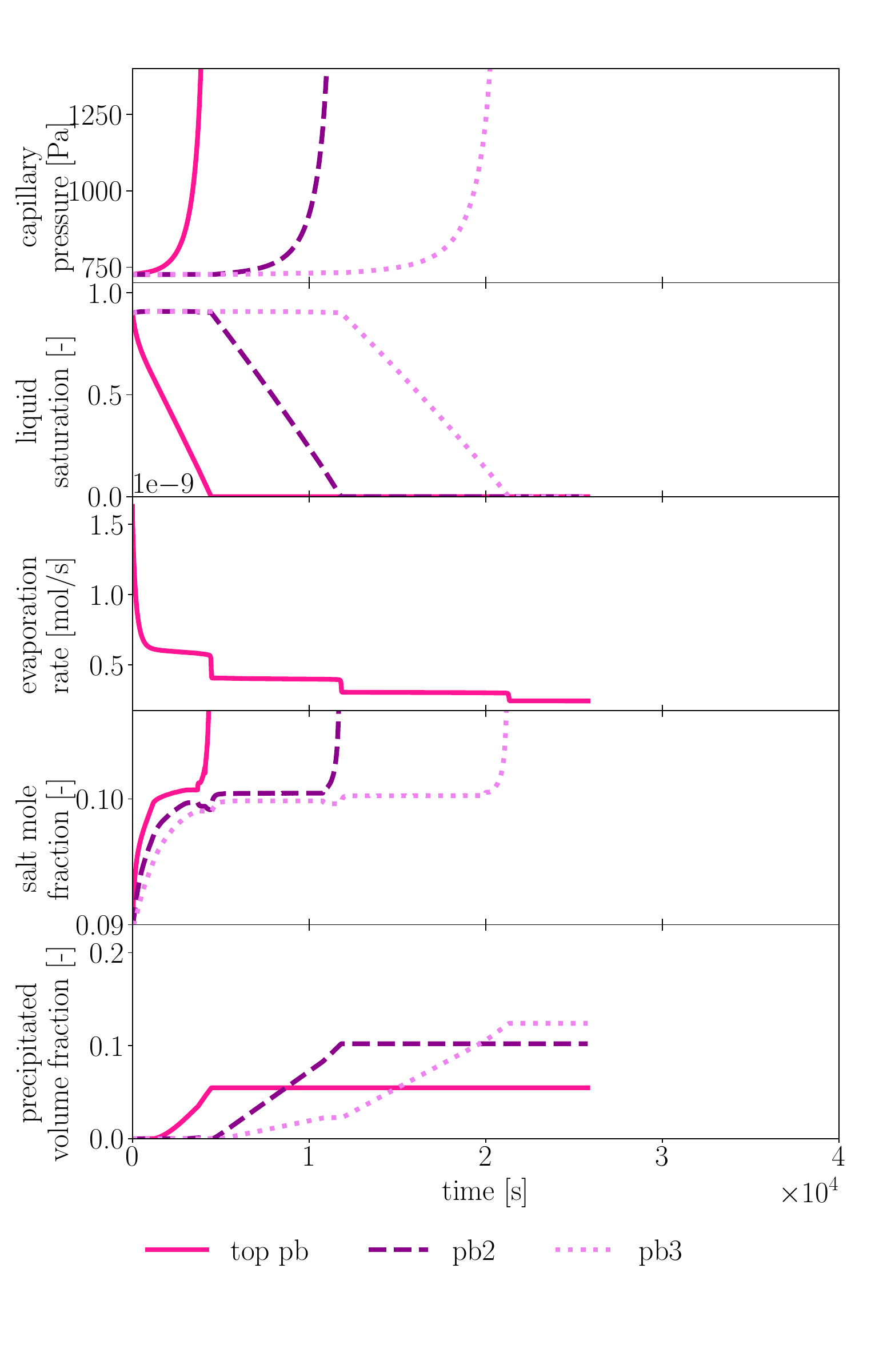}
\caption{Parameter development of the upper three pore bodies for the closed system with brine, circular-shaped throat cross sections, considering volume alterations in pore bodies and pore throats ($\Delta V_{i}$ and $\Delta V_{ij}$)}
\label{fig:Brine_Circular_Interfaces_closed}
\end{figure}
%\FloatBarrier
%
\subsubsection{Open and Closed System} \label{subsubsec:InterfacesOpenClosed}
In the previous sections, already system with different interface locations were discussed: the open and closed system, see Figure~\ref{fig:Brine_Square_VAlterations}. In the open system, the gas-liquid interface stays in the top pb. This leads to the exclusive precipitation of salt in the top pb. In the closed system an interface stays at the top due to liquid corner films but also recedes in the pore-network system. The major part of the salt precipitates in the top pb as the salt is transported to the top through liquid films and the major part of the evaporation takes place at the interface of the top pb. But salt also precipitates in the lower pore bodies as the solution in the system gets more and more concentrated. Further, there are small amounts of evaporation at the interface in e.g. pb2 after invasion, which leads to a slight increase of $x_\mathrm{l}^\mathrm{NaCl}$. Before the top pb is completely dried out in the end, the evaporation in pb2 and lower pbs is very low as in the top pb the gas is nearly saturated with water due to ongoing saturation. Consequently, the gradient of the mass fraction of water in the gas phase between pb2 and top pb is very small, which leads to small diffusive fluxes. This limits the transport of water vapor out of pb2 and the evaporation. 
\subsubsection{Influence of Liquid Films} \label{subsubsec:InterfacesLiquidFilms}
In the following, the influence of liquid corner films are discussed. There are also setups where no liquid corner films occur. This changes the locations of the interfaces and so has a huge influence on the salt transport and precipitation behavior.
In the pore-network, this is represented using pore-throats with circular throat cross sections. If circular throats get invaded, they are completely filled with gas phase and no corner liquid film can develop. In this section, circular cross sections are used with the same cross-sectional area as the square cross sections in the previous sections.
\\
%Open system
For the open system, there is nearly no difference to the square throat cross sections as all throats stay saturated. The small differences result from the little higher transmissibility of circular throats, due to smaller ratio of wall surface to cross-sectional area and thus less wall friction of circular throats. This has just minor influence on the precipitation.
\\
%Closed system
In the closed system, however, the pore bodies dry out completely one after each other, see Figure~\ref{fig:Brine_Circular_Interfaces_closed}. Initially, the $p_\mathrm{c}$ rises in the top pb. When $p_\mathrm{c,e}$ is reached, pt1 is invaded. In contrast to the use of square pore-throat cross sections, the invasion has no major influence on $S_\mathrm{l}$. In the top pb, $S_\mathrm{l}$ decreases continuously until it is completely dry. With the invasion of the pore-throat there is a liquid-gas interface in pb2, which leads to small evaporation and small decrease in $S_\mathrm{l}$ in pb2. After the top pb is completely dry, $S_\mathrm{l}$ in pb2 starts to drop noticeable. 
Initially, $x_\mathrm{l}^\mathrm{NaCl}$ increases until it reaches an equilibrium of advective, diffusive fluxes and precipitation and stays constant. After the invasion of pt1, however, $x_\mathrm{l}^\mathrm{NaCl}$ increases again in the top pb. The liquid phase in the top pb is now disconnected from the liquid phase in the lower pore bodies. The isolated salt solution gets more and more concentrated due to evaporation and lack of diffusive downward flux, faster than it precipitates. Salt precipitates until the top pb is completely dry. In pb2, the salt starts to precipitate after invasion of pt1 when at the interface small amounts of water evaporates and $x_\mathrm{l}^\mathrm{NaCl}$ starts to increase slightly. The precipitation rate increases when the top pb is completely dry and pb2 is the main pore body from which water evaporates. In lower pore bodies (pb3 - pb5) salt also precipitates before invasion of the respective pore-throat above as the total concentration in the remaining solution is above the solubility limit. In the end, this leads to precipitation in the whole pore network not just at the top. The precipitation in the dry system is even higher in the lower pore bodies as here the concentration was already more concentrated when they started to dry out and concentrated solution started to precipitate already before the invasion of pore throat above.
The evaporation rate decreases each time the main evaporative interface recedes in a lower pore body. Then the diffusive transport of the water vapor has to overcome the resistance of one further pore throat.
\subsection{Local Heterogeneities} \label{subsec:LocalHeterogeneities}
\begin{figure}[t]%
\centering
\includegraphics[width=0.47\textwidth]{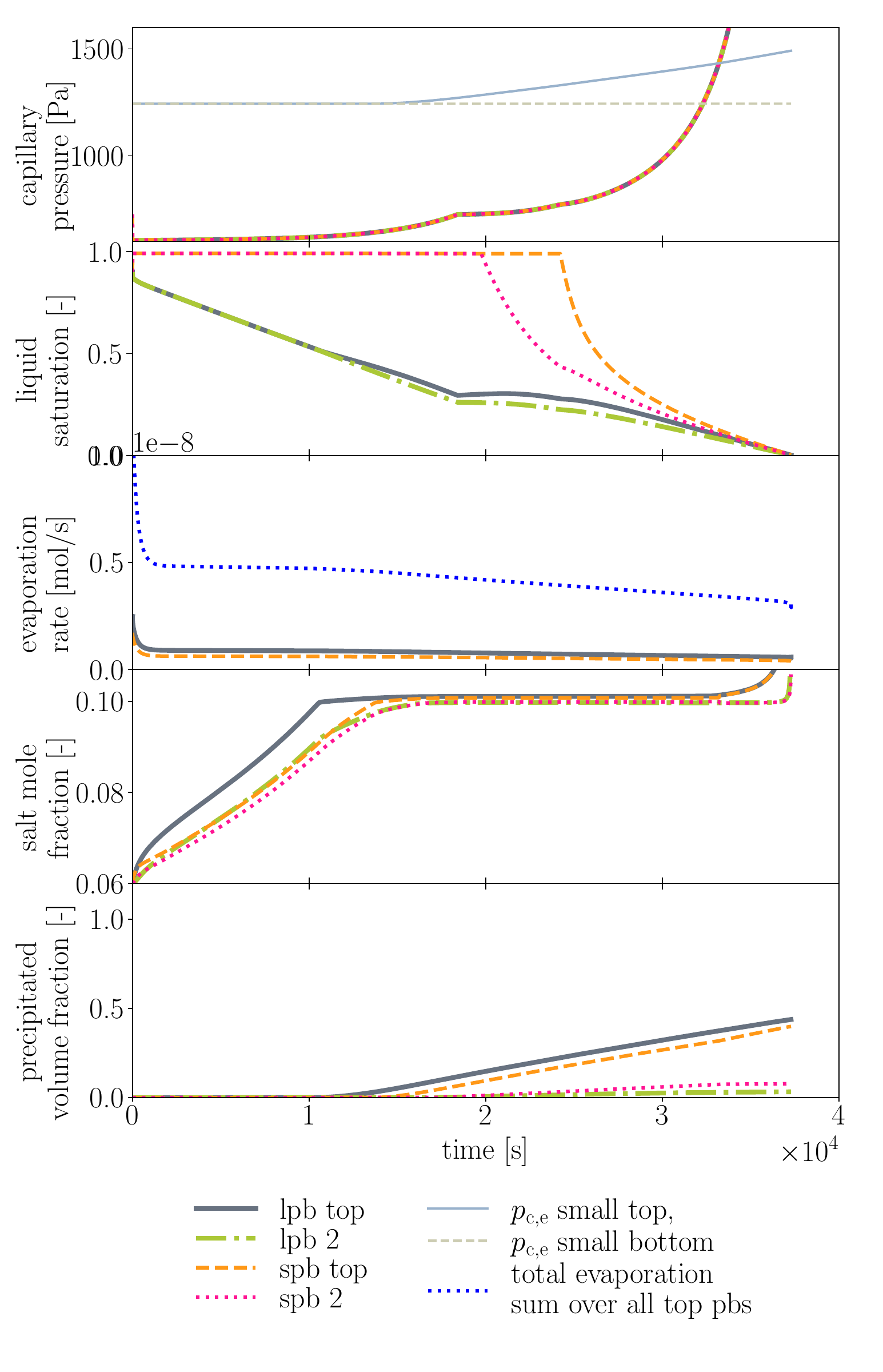}
\caption{Parameter development in the two-dimensional setup of two  small pore bodies (spb) and two large pore bodies (lpb) in the top and second row. In this setup brine and square-shaped throat cross sections are used. Further, volume alterations in pore bodies and pore throats ($\Delta V_{i}$ and $\Delta V_{ij}$) are considered.}
\label{fig:2d_smallMiddle}
\end{figure}
%\FloatBarrier
%
\begin{figure}[ht]%[ht!]%
\centering
\includegraphics[width=0.47\textwidth]{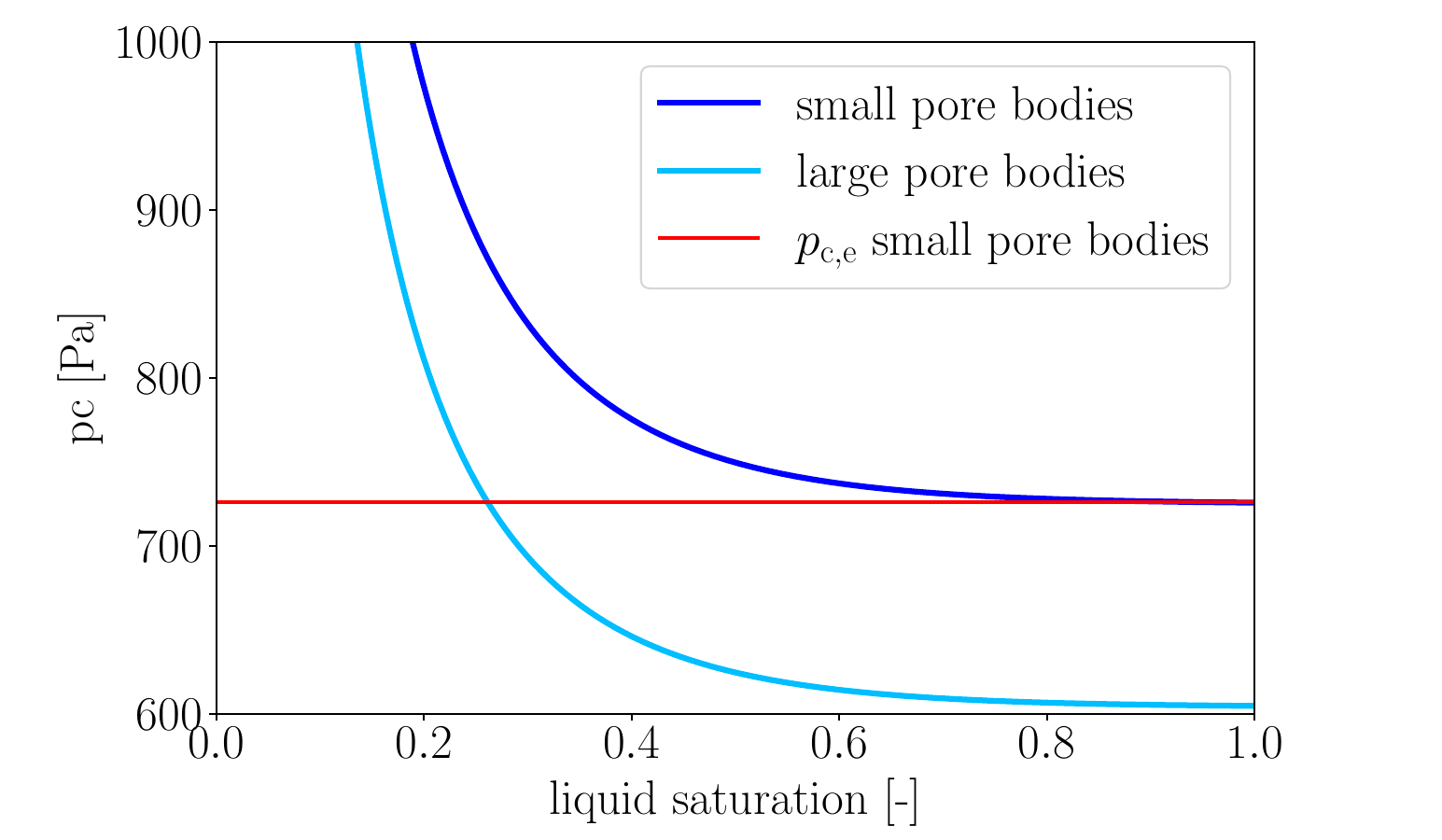}
\caption{Capillary pressure-saturation relationships for heterogeneous two-dimensional system for the initial dimensions of the small and large pore bodies}
\label{fig:2D_pcSlRelationships}
\end{figure}
\begin{figure}[ht!]%
\centering
\begin{subfigure}[t]{0.32\textwidth}
\includegraphics[width=\textwidth]{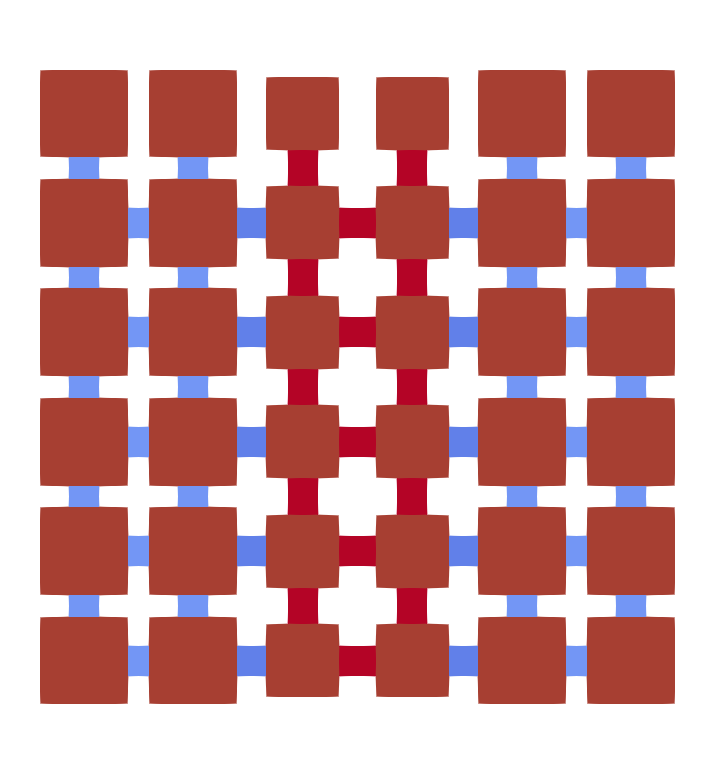}
\subcaption{Initial state at 0s}
\label{fig:2DPicturesInitial}
\end{subfigure}
\begin{subfigure}[t]{0.32\textwidth}
\includegraphics[width=\textwidth]{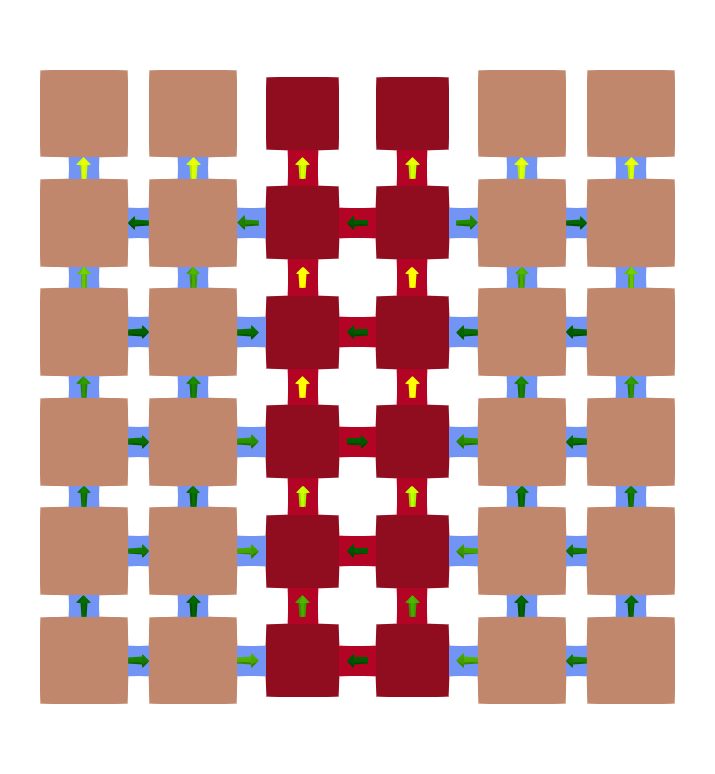}
\subcaption{Phase 1 at 5000s}
\label{fig:2DPicturesPhase1}
\end{subfigure}
\begin{subfigure}[t]{0.32\textwidth}
\includegraphics[width=\textwidth]{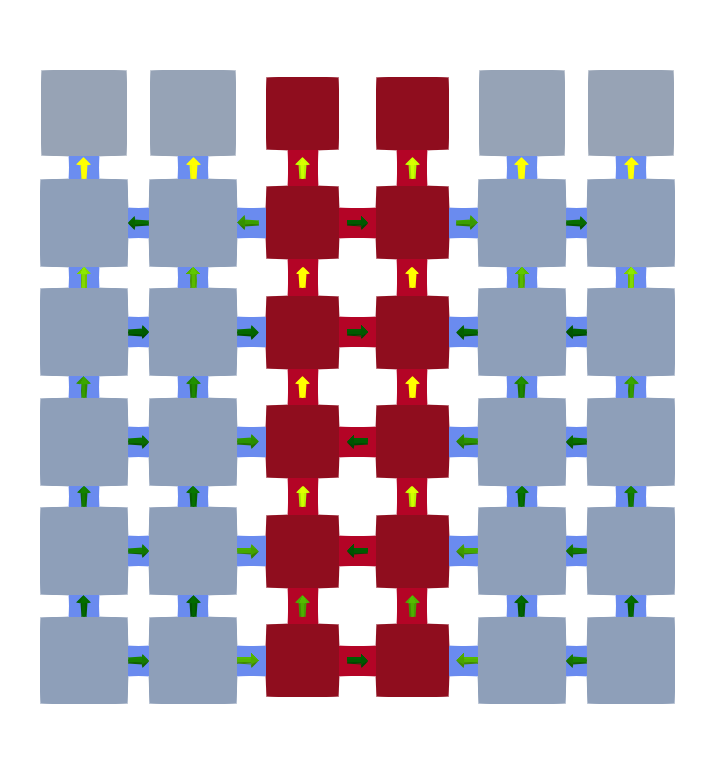}
\subcaption{Phase 2 at 15000s}
\label{fig:2DPicturesPhase2}
\end{subfigure}
\begin{subfigure}[t]{0.32\textwidth}
\includegraphics[width=\textwidth]{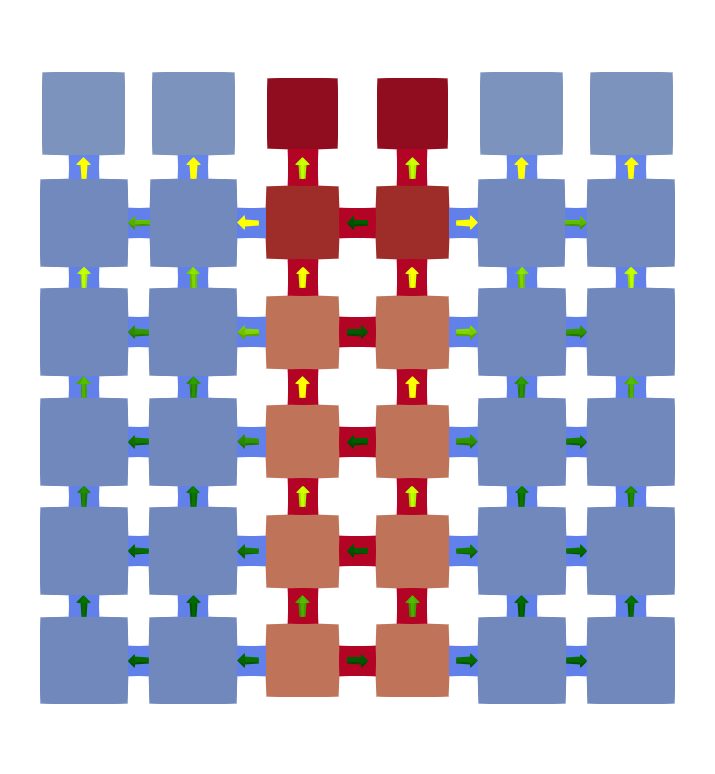}
\subcaption{Phase 3 at 20000s}
\label{fig:2DPicturesPhase3}
\end{subfigure}
\begin{subfigure}[t]{0.32\textwidth}
\includegraphics[width=\textwidth]{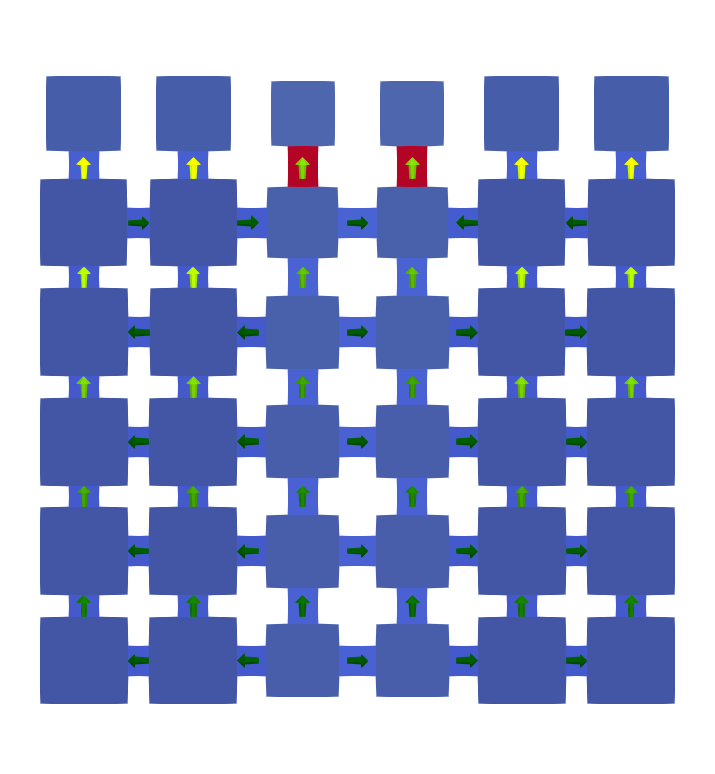}
\subcaption{Phase 4 at 33000s}
\label{fig:2DPicturesPhase4}
\end{subfigure}
\begin{subfigure}[t]{0.32\textwidth}
\includegraphics[width=\textwidth] {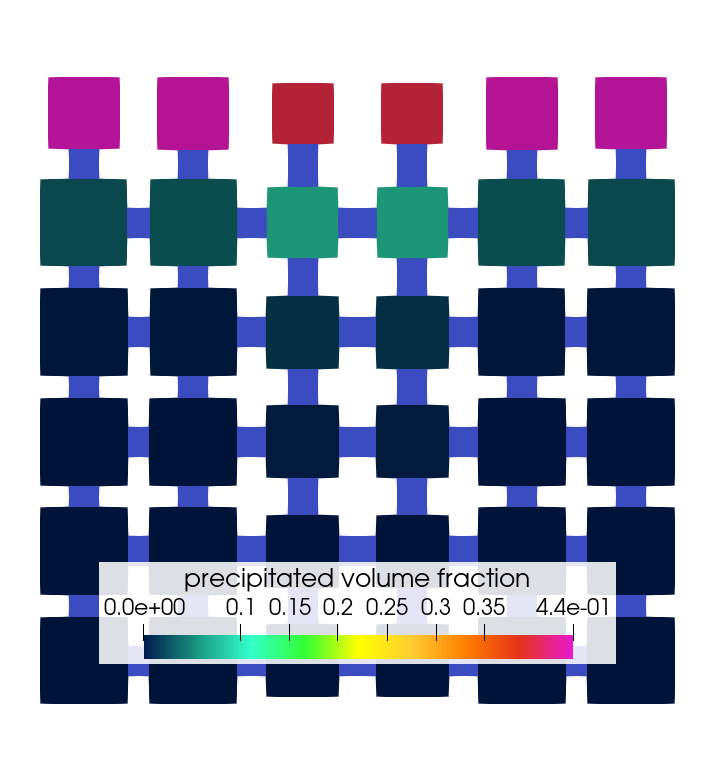}
\subcaption{End state at 37290s}
\label{fig:2DPicturesEnd}
\end{subfigure}
%
%legend
\includegraphics[width=0.32\textwidth, trim={0cm, 1.0cm, 0cm, 1.0cm},clip]{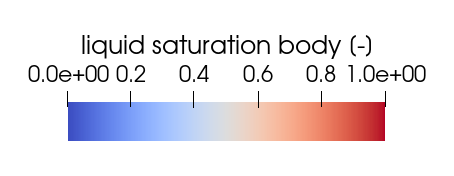}
\includegraphics[width=0.32\textwidth, trim={0cm, 1.0cm, 0cm, 1.0cm},clip]{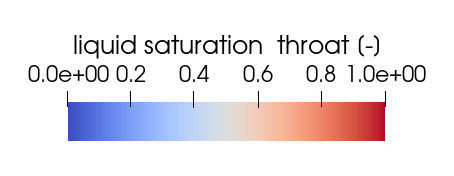}
\includegraphics[width=0.32\textwidth, trim={0cm, 1.0cm, 0cm, 1.0cm},clip]{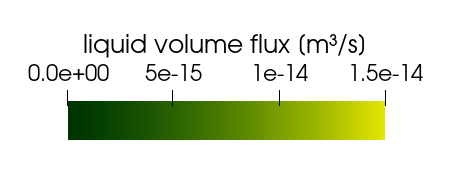}
\caption{Development of the saturation in pore bodies and throats in the two-dimensional setup. For the throats the liquid saturation is defined as the ratio of the liquid filled cross-sectional area to the total cross-sectional area. }
\label{fig:2DPictures}
\end{figure}
%\FloatBarrier
%
The influence of local heterogeneities is investigated using the two-dimensional setup, see Section~\ref{subsec:2DSetup}. 
Here, a region with smaller pores and a region with larger pores interact and generate a two-dimensional flow field, where different pore-scale processes interact.
%A combination of the individual effects analyzed for the one-dimensional case occure. 
\\
To analyze the system, we consider the development of parameters in specific pore bodies in detail: a large pore body at the top (top lpb) and in the second row (lpb 2) and also a small pore body at the top (top spb) and in the second row (spb 2), see Figure~\ref{fig:2d_smallMiddle}. Here, exemplary, the values of pore-body column 2 and 3 are used, but the values in the pore bodies of equal size are similar. Further, the saturation and volume flux of the whole system is shown at different times as well as the end state of $\phi_\mathrm{s}$ in Figure~\ref{fig:2DPictures}.
\\
In this system the gas phase in every pore body is connected to the ambient air as the larger pore throats are invaded initially. Further, the liquid phase in the pore bodies is connected through liquid corner film in the square-shaped throats. This results in equal gas and liquid pressures in the whole system as small pressure differences get balanced out quickly through advective fluxes. Consequently, the capillary pressure in every pore body is the same, see Figure~\ref{fig:2d_smallMiddle}. 
\\
Different phases can be determined during the period of evaporation-driven drying of the initially nearly saturated system until it is completely dry. In the beginning, the saturation starts to decrease in the large pore bodies, whereas $S_\mathrm{l}$ in the small pore bodies increases to $1.0$ due to the different $p_\mathrm{c}$-$S_\mathrm{l}$ relationship of the pore bodies. Here, the liquid phase gets sucked into the smaller pores due to capillary forces, see figure~\ref{fig:2DPicturesPhase1}. Consequently, the liquid flux flows from the large towards the smaller pore bodies, but only in the lower part of the pore network. In the top horizontal throats the liquid phase flows towards the large pore bodies.  In the small throats more liquid phase is transported towards the top as can evaporate. This leads to the vertical flux towards the lpbs, where the evaporation rate is bigger due to a larger cross-sectional area. This first phase is also characterized by the increase of $x_\mathrm{l}^\mathrm{NaCl}$ in the top pbs. Here, the increase is faster in the lpbs as in the spbs. The end of phase 1 is defined when the solubility limit is reached in the first pore body after $10530$s.
\\
In phase 2 salt starts to precipitate in the top pbs. The first salt precipitates in the top lpb followed by the top spb, due to higher evaporation rates in the lpbs.
Later small amounts of salt precipitates also in the second row of pbs. 
With the precipitation and following decrease of pb volume the $p_\mathrm{c}$-$S_\mathrm{l}$ relationship changes. This leads to a higher $S_\mathrm{l}$ of the top lpb as for the lpb 2.
During this phase, the lpbs in the pore network dry further as in phase 1.
\\
In phase 3, after 18410 s, $S_\mathrm{l}$ in the spbs starts to drop. At this time the $p_\mathrm{c}$ reaches the entry pressure $p_\mathrm{c,e} \approx 726~\mathrm{Pa}$ of the spbs, see Figure~\ref{fig:2D_pcSlRelationships}. The drying of the spbs starts at the bottom as here the pb volume is not reduced by precipitated salt. One row of pbs after each other is invaded towards the top when the respective entry pressure is reached. 
The saturation in the spbs decreases faster as in the lpb as the derivation $\frac{dS_\mathrm{l}}{dp_\mathrm{c}} $ of the $p_\mathrm{c}$-$S_\mathrm{l}$ relationship is higher in the region of high liquid saturation, as is the case in the spbs. This leads to a change in direction of the horizontal liquid flux which now flows from the spbs to the lpbs. Also, the decrease of $S_\mathrm{l}$ in the lpb slows down as the evaporative flux is now also supported by the liquid in the spbs. After some drying of the spbs the horizontal flux direction changes back from lpbs to spbs. 
\\
After 32310 s phase 4 starts when the first small pore throats gets invaded by the gas phase. The entry pressure of the thin throat is reached first for the spts at the bottom and then row by row for the upper spts. In the upper spts the entry pressure is increased due to precipitation and consequently reached later. The invasion shows no influence on the liquid saturation in the pbs.
\section{Conclusions} \label{sec:conclusion}
%Summary
In this paper, a dynamic, two-phase, non-isothermal pore-network model was presented for evaporation-driven precipitation processes, which includes the alteration of pore space due to precipitation. A kinetic precipitation-reaction is solved in every pore body, which results in a precipitated volume reducing the pore-body volume. The reduction of the pore-throat volume is estimated based on the oversaturations in the adjacent pore bodies. The model is able to represent the reduction of the evaporation rate of a salt solution compared to the evaporation rate of pure water based on the reduction of the saturated vapor pressure. Evaporation at the top of the porous medium is considered by calculating advective, diffusive and conductive fluxes assuming constant parameters in the ambient air.%Also the influence of capillarity on the evaporation is considered. 
\\
The model is able to represent pore-scale processes, which on the REV scale can only be estimated by upscaled relations. 
In the pore-network model the influence of pore-space alterations on the transport and precipitation processes on the pore scale can be described. The following effects of reduced pore-space volume due to precipitation of salt are identified:
\begin{itemize}
    \item The decrease of pore-body volume shifts the local capillary pressure saturation relationship of the pore body towards higher capillary pressure values. With the same, capillary pressure in the whole system, this leads to higher liquid saturation in pore bodies with reduced pore-body volumes.
    \item Further, the decrease in pore-body volume reduces the evaporation rate. The diffusive flux of water vapor into the ambient air at the top, which determines the evaporation rate, is reduced based on the reduction of cross-sectional area of the top pore bodies. This slows down the drying of the porous medium.
    \item The decrease of the pore-throat volume leads to an increase in entry pressure. Consequently,  the throat gets invaded by the gas phase at higher capillary pressures or lower liquid saturation. In context of drying, so the invasion of the throat happens later.
\end{itemize}
If the volume change of pore body and pore throat is combined, the different effects interact and amplify or attenuate. Further, the influence of the location of the liquid-gas interface can be represented with the presented model:
\begin{itemize}
    \item In a pore-network system, in which salt solution can enter at the bottom from a large reservoir, the interface stays at the top of the porous medium. This leads to exclusive precipitation of salt at the top.
    \item In a system, in which the bottom is closed, the system dries out and the interface recedes into the pore-network system. Here, the presence of liquid corner flow in invaded pore throats has a huge influence. Liquid films bring the main evaporative interface to the top of the porous medium, where the majority of salt precipitates. Without liquid corner film, the main evaporative surface recedes in the porous medium and precipitation occurs in the whole system. 
\end{itemize}
Further, the model is able to simulate the influence of pore-scale heterogeneities on the transport and precipitation. Effects like capillary pumping can be observed in regions with smaller pores, which act as the main region for transporting the salt solution upwards due to higher capillary forces.
Here, the different effects mentioned above interact and lead to a heterogeneous distribution of precipitated salt. 
\\
\\
The model is limited to precipitation within the porous medium, so called subflorescent precipitation. In experiments and field observations, the development of salt crusts on top of the porous medium (efflorescent precipitation) was observed~\cite{Nachshon2018b} for sodium chloride and other salts, which can not be represented by the current model. 
Further, \citeA{Rad2013} propose the porous properties of the precipitated salt itself. With comparable smaller pore sizes in the precipitated salt than in the porous medium, the liquid phase is sucked into the porous salt. The saturated salt solution is transported in the salt towards the top, where water evaporates from the wet salt. The porous structure of the salt and the crust formation on top of the porous medium lead to an enlargement of the evaporative area and the evaporation rate.
This is in contrast to the reduction of the evaporative area due to precipitation in the presented model. To incorporate the effects in the model, the salt should be considered as a porous medium using a double continuum model in the next step. 
\\
To validate the model, a comparison with experiments is necessary. 
Therefor it was already started to conduct micro-fluidic experiments with a setup comparable to the two-dimensional setup in this paper. The experimental and numerical results will be compared to validate and improve the presented model. 
\\
A comparison with REV-scale simulations can reveal the influence of the pore-scale effects. 
In the end, upscaled relations can be found with the pore-scale model which can be used in the REV-scale model to incorporate pore-scale effects.

%%%%%%%%%%%%%%%%%%%%%%%%%%%%%%%%%%%%%%%%%%%%%%%
%% Optional Appendices go here
%
% The \appendix command resets counters and redefines section heads
\newpage
\appendix
\section{Top Boundary Conditions}\label{app:TopBC}
At the top of the domain, described in Section~\ref{sec:Setup}, the evaporative fluxes are calculated and applied as a source term for each component. The values $p_\mathrm{g,amb}, S_\mathrm{g,amb}, x_\mathrm{g,amb}^\mathrm{w}, T_\mathrm{amb}$ are assumed to be constant in the ambient air and are set as boundary conditions, see Table~\ref{tab:Setups}. Based on the gradient of these parameters between the top pore bodies and the ambient air the fluxes are calculated based on the balance equations presented in Section~\ref{subsec:BalanceEquations}.
It is assumed that only components in the gaseous phase enter and leave the porous media at the top due to evaporation. The flux for the liquid and solid phase is zero as well as the flux of the component NaCl in the gas phase as it is just present in the liquid phase. 
\begin{linenomath*}
\begin{align}
j_{\mathrm{g},i}^\mathrm{w} &= j_{\mathrm{adv,g},i}^\mathrm{w} + j_{\mathrm{diff,g},i}^\mathrm{w}\\
j_{\mathrm{g},i}^\mathrm{a} &= j_{\mathrm{adv,g},i}^\mathrm{a} + j_{\mathrm{diff,g},i}^\mathrm{a}\\
j_{\mathrm{g},i}^\mathrm{NaCl} &= 0 ~&& \text{no NaCl in the gas phase}\\
j_{\mathrm{l},i}^\kappa &= 0 ~&& \text{no liquid phase leaves and enters the domain} \\
j_{\mathrm{s},i} &= 0 ~ &&\text{no solid phase leaves and enters the domain}
\end{align}
\end{linenomath*}
The advective fluxes of the components in the gas phase are calculated as follows:
\begin{linenomath*}
\begin{align}
j_{\mathrm{adv,g},i}^\mathrm{w} &=     
	\begin{cases}
       -  x_{\mathrm{g},i}^\mathrm{w} \frac{\rho_{\mathrm{m,g},i}}{\mu_{\mathrm{g},i}} \frac{ r_i}{\delta} \left(p_{\mathrm{g},i} - p_\mathrm{g,amb}\right)& \text{if $p_\mathrm{g} > p_\mathrm{g,amb} \rightarrow j<0$, outflow},\\
      -  x_{\mathrm{g,amb}}^\mathrm{w} \frac{\rho_{\mathrm{m,g,amb}}}{\mu_{\mathrm{g,amb}}} \frac{ r_i}{\delta} \left(p_{\mathrm{g},i} - p_\mathrm{g,amb}\right)& \text{if $p_\mathrm{g} \leq p_\mathrm{g,amb} \rightarrow j\geq 0$, inflow}
    \end{cases}.\\
j_{\mathrm{adv,g},i}^\mathrm{a} &=     
	\begin{cases}
       -  x_{\mathrm{g},i}^\mathrm{a} \frac{\rho_{\mathrm{m,g},i}}{\mu_{\mathrm{g},i}}  \frac{r_i }{\delta} \left(p_{\mathrm{g},i} - p_\mathrm{g,amb} \right)& \text{if $p_\mathrm{g} > p_\mathrm{g,amb} \rightarrow j<0$, outflow},\\
      -  \left( 1- x_{\mathrm{g,amb}}^\mathrm{w} \right) \frac{\rho_{\mathrm{m,g,amb}}}{\mu_{\mathrm{g,amb}}}  \frac{r_i }{\delta} \left(p_{\mathrm{g},i} - p_\mathrm{g,amb}\right) & \text{if $p_\mathrm{g} \leq p_\mathrm{g,amb} \rightarrow j\geq 0$, inflow}.
    \end{cases} 
\end{align}
\end{linenomath*}
with resistance factor $r_i = 1.0 ~\mathrm{m^4}$.\\
The diffusive fluxes of the components in the gas phase are calculated as follows:
\begin{linenomath*}
\begin{align}
j_{\mathrm{diff,g},i}^\mathrm{w} &= - \frac{\bar{\rho}_{\mathrm{g},i}}{M^\mathrm{w}} \bar{D}_{\mathrm{g},i}^\mathrm{w,a} A_{i} \frac{X_{\mathrm{g},i}^\mathrm{w} - X_\mathrm{g,amb}^\mathrm{w}}{\delta}, \\
j_{\mathrm{diff,g},i}^\mathrm{a} &= - \frac{\bar{\rho}_{\mathrm{g},i}}{M^\mathrm{a}} \bar{D}_{\mathrm{g},i}^\mathrm{w,a} A_{i} \frac{X_{\mathrm{g},i}^\mathrm{a} - \left(1-X_\mathrm{g,amb}^\mathrm{w}\right)}{\delta}, \\
\end{align}
\end{linenomath*}
with the averaged quantities:
\begin{linenomath*}
\begin{align}
\bar{D}_{\mathrm{g},i}^\mathrm{w,a} &= 0.5 \left(D_{\mathrm{g},i}^\mathrm{w,a} + D_\mathrm{g,amb}^\mathrm{w,a} \right),\\
\bar{\rho}_{\mathrm{g},i} &= 0.5 \left(\rho_{\mathrm{g},i} + \rho_\mathrm{g,amb} \right).
\end{align}
\end{linenomath*}

The energy fluxes due to evaporation are calculated based on the advective and diffusive fluxes and the conductive flux: %[$J/(m^3 s)$]
\begin{linenomath*}
\begin{align}
j_{\mathrm{e},i} &= j_{\mathrm{e,adv},i} + j_{\mathrm{e,diff},i} + j_{\mathrm{e,cond},i}. \\
j_{\mathrm{e,adv},i} &= 
\begin{cases}
j_{\mathrm{adv,g},i}^\mathrm{w} h_{\mathrm{g},i}^\mathrm{w} M^\mathrm{w} + j_{\mathrm{adv,g},i}^\mathrm{a} h_{\mathrm{g},i}^\mathrm{a} M^\mathrm{a} & \text{if $p_\mathrm{g} > p_\mathrm{amb} \rightarrow j<0$, outflow},\\
j_{\mathrm{adv,g},i}^\mathrm{w} h_{\mathrm{g,amb}}^\mathrm{w} M^\mathrm{w} + j_{\mathrm{adv,g},i}^\mathrm{a} h_{\mathrm{g,amb}}^\mathrm{a} M^\mathrm{a} & \text{if $p_\mathrm{g} \leq p_\mathrm{amb} \rightarrow j\geq 0$, inflow}.
\end{cases} \\
j_{\mathrm{e,diff},i} &= 
\begin{cases}
j_{\mathrm{diff,g},i}^\mathrm{w} h_{\mathrm{g},i}^\mathrm{w} M^\mathrm{w} + j_{\mathrm{diff,g},i}^\mathrm{a} h_{\mathrm{g,amb}}^\mathrm{a} M^\mathrm{a} & \text{if $X_\mathrm{g}^\mathrm{w} > X_\mathrm{g,amb}^\mathrm{w}$}\\
 & \text{outflow of water, inflow of air},\\
j_{\mathrm{diff,g},i}^\mathrm{w} h_{\mathrm{g,amb}}^\mathrm{w} M^\mathrm{w} + j_{\mathrm{diff,g},i}^\mathrm{a} h_{\mathrm{g},i}^\mathrm{a} M^\mathrm{a} & \text{if $X_\mathrm{g}^\mathrm{w} \leq X_\mathrm{g,amb}^\mathrm{w}$}\\
&\text{inflow of water, outflow of air}.
\end{cases} \\
j_{\mathrm{e,cond},i} &= \bar{\lambda}_{\mathrm{g},i} \frac{A_i}{V_i} \frac{T_\mathrm{amb} - T_i}{\delta},
\end{align}
\end{linenomath*}
with the averaged thermal conductivity
\begin{linenomath*}
\begin{align}
\bar{\lambda}_{\mathrm{g},i} = 0.5 \left( \lambda_{\mathrm{g},i} + \lambda_\mathrm{g,amb} \right).
\end{align}
\end{linenomath*}
%
%
%
%%%%%%%%%%%%%%%%%%%%%%%%%%%%%%%%%%%%%%%%%%%%%%%
%
% DATA SECTION and ACKNOWLEDGMENTS
%
%%%%%%%%%%%%%%%%%%%%%%%%%%%%%%%%%%%%%%%%%%%%%%%
\newpage
\section*{Open Research Section}
The numerical code developed for this publication and used to obtain all results and conclusions is available publicly in the following git repository \cite{schollenberger2025aPubModul}:\\
https://git.iws.uni-stuttgart.de/dumux-pub/schollenberger2025a.

\acknowledgments
We thank the Deutsche Forschungsgemeinschaft (DFG, German Research Foundation) for supporting this work by funding SFB 1313, Project Number 327154368.

%%%%%%%%%%%%%%%%%%%%%%%%%%%%%%%%%%%%%%%%%%%%%%%
% REFERENCES and BIBLIOGRAPHY
%
% \bibliography{<name of your .bib file>} don't specify the file extension
% don't specify bibliographystyle
%
%%%%%%%%%%%%%%%%%%%%%%%%%%%%%%%%%%%%%%%%%%%%%%%

\bibliography{literature}

\begin{thebibliography}{}

\bibitem [\protect \citeauthoryear {%
Bakke%
\ \BBA {} Øren%
}{%
Bakke%
\ \BBA {} Øren%
}{%
{\protect \APACyear {1997}}%
}]{%
Bakke1997}
\APACinsertmetastar {%
Bakke1997}%
\begin{APACrefauthors}%
Bakke, S.%
\BCBT {}\ \BBA {} Øren, P\BHBI E.%
\end{APACrefauthors}%
\unskip\
\newblock
\APACrefYearMonthDay{1997}{}{}.
\newblock
{\BBOQ}\APACrefatitle {3-D Pore-Scale Modelling of Sandstones and Flow
  Simulations in the Pore Networks} {3-d pore-scale modelling of sandstones and
  flow simulations in the pore networks}.{\BBCQ}
\newblock
\APACjournalVolNumPages{SPE Journal}{2}{02}{136-149}.
\newblock
\begin{APACrefDOI} \doi{10.2118/35479-PA} \end{APACrefDOI}
\PrintBackRefs{\CurrentBib}

\bibitem [\protect \citeauthoryear {%
Bastian%
\ \protect \BOthers {.}}{%
Bastian%
\ \protect \BOthers {.}}{%
{\protect \APACyear {2021}}%
}]{%
Bastian2021}
\APACinsertmetastar {%
Bastian2021}%
\begin{APACrefauthors}%
Bastian, P.%
, Blatt, M.%
, Dedner, A.%
, Dreier, N\BHBI A.%
, Engwer, C.%
, Fritze, R.%
\BDBL {}Sander, O.%
\end{APACrefauthors}%
\unskip\
\newblock
\APACrefYearMonthDay{2021}{}{}.
\newblock
{\BBOQ}\APACrefatitle {The Dune framework: Basic concepts and recent
  developments} {The dune framework: Basic concepts and recent
  developments}.{\BBCQ}
\newblock
\APACjournalVolNumPages{Computers and Mathematics with
  Applications}{81}{}{75-112}.
\newblock
\APACrefnote{Development and Application of Open-source Software for Problems
  with Numerical PDEs}
\newblock
\begin{APACrefDOI} \doi{10.1016/j.camwa.2020.06.007} \end{APACrefDOI}
\PrintBackRefs{\CurrentBib}

\bibitem [\protect \citeauthoryear {%
Batzle%
\ \BBA {} Wang%
}{%
Batzle%
\ \BBA {} Wang%
}{%
{\protect \APACyear {1992}}%
}]{%
Batzle1992}
\APACinsertmetastar {%
Batzle1992}%
\begin{APACrefauthors}%
Batzle, M.%
\BCBT {}\ \BBA {} Wang, Z\BPBI J.%
\end{APACrefauthors}%
\unskip\
\newblock
\APACrefYearMonthDay{1992}{}{}.
\newblock
{\BBOQ}\APACrefatitle {Seismic Properties of Pore fluids} {Seismic properties
  of pore fluids}.{\BBCQ}
\newblock
\APACjournalVolNumPages{Geophysics}{57}{11}{1396--1408}.
\newblock
\begin{APACrefDOI} \doi{10.1190/1.1443207} \end{APACrefDOI}
\PrintBackRefs{\CurrentBib}

\bibitem [\protect \citeauthoryear {%
Bernabé%
, Mok%
\BCBL {}\ \BBA {} Evans%
}{%
Bernabé%
\ \protect \BOthers {.}}{%
{\protect \APACyear {2003}}%
}]{%
Bernabe2003}
\APACinsertmetastar {%
Bernabe2003}%
\begin{APACrefauthors}%
Bernabé, Y.%
, Mok, U.%
\BCBL {}\ \BBA {} Evans, B.%
\end{APACrefauthors}%
\unskip\
\newblock
\APACrefYearMonthDay{2003}{}{}.
\newblock
{\BBOQ}\APACrefatitle {Permeability-porosity Relationships in Rocks Subjected
  to Various Evolution Processes} {Permeability-porosity relationships in rocks
  subjected to various evolution processes}.{\BBCQ}
\newblock
\APACjournalVolNumPages{Pure and Applied Geophysics}{160}{}{937-960}.
\newblock
\begin{APACrefDOI} \doi{10.1007/PL00012574} \end{APACrefDOI}
\PrintBackRefs{\CurrentBib}

\bibitem [\protect \citeauthoryear {%
Blunt%
}{%
Blunt%
}{%
{\protect \APACyear {2017}}%
}]{%
Blunt2017}
\APACinsertmetastar {%
Blunt2017}%
\begin{APACrefauthors}%
Blunt, M.%
\end{APACrefauthors}%
\unskip\
\newblock
\APACrefYear{2017}.
\newblock
\APACrefbtitle {Multiphase Flow in Permeable Media: A Pore-Scale Perspective}
  {Multiphase flow in permeable media: A pore-scale perspective}.
\newblock
\APACaddressPublisher{}{Cambridge University Press}.
\newblock
\begin{APACrefDOI} \doi{10.1017/9781316145098} \end{APACrefDOI}
\PrintBackRefs{\CurrentBib}

\bibitem [\protect \citeauthoryear {%
Bringedal%
, von Wolff%
\BCBL {}\ \BBA {} Pop%
}{%
Bringedal%
\ \protect \BOthers {.}}{%
{\protect \APACyear {2020}}%
}]{%
Bringedal2020}
\APACinsertmetastar {%
Bringedal2020}%
\begin{APACrefauthors}%
Bringedal, C.%
, von Wolff, L.%
\BCBL {}\ \BBA {} Pop, I\BPBI S.%
\end{APACrefauthors}%
\unskip\
\newblock
\APACrefYearMonthDay{2020}{}{}.
\newblock
{\BBOQ}\APACrefatitle {Phase Field Modeling of Precipitation and Dissolution
  Processes in Porous Media: Upscaling and Numerical Experiments} {Phase field
  modeling of precipitation and dissolution processes in porous media:
  Upscaling and numerical experiments}.{\BBCQ}
\newblock
\APACjournalVolNumPages{Multiscale Modeling \& Simulation}{18}{2}{1076-1112}.
\newblock
\begin{APACrefDOI} \doi{10.1137/19M1239003} \end{APACrefDOI}
\PrintBackRefs{\CurrentBib}

\bibitem [\protect \citeauthoryear {%
Bruus%
}{%
Bruus%
}{%
{\protect \APACyear {2011}}%
}]{%
Bruus2011}
\APACinsertmetastar {%
Bruus2011}%
\begin{APACrefauthors}%
Bruus, H.%
\end{APACrefauthors}%
\unskip\
\newblock
\APACrefYearMonthDay{2011}{}{}.
\newblock
{\BBOQ}\APACrefatitle {Acoustofluidics 1: Governing equations in microfluidics}
  {Acoustofluidics 1: Governing equations in microfluidics}.{\BBCQ}
\newblock
\APACjournalVolNumPages{Lab Chip}{11}{}{3742--3751}.
\newblock
\begin{APACrefDOI} \doi{10.1039/C1LC20658C} \end{APACrefDOI}
\PrintBackRefs{\CurrentBib}

\bibitem [\protect \citeauthoryear {%
Camassel%
, Sghaier%
, Prat%
\BCBL {}\ \BBA {} {Ben Nasrallah}%
}{%
Camassel%
\ \protect \BOthers {.}}{%
{\protect \APACyear {2005}}%
}]{%
Camassel2005}
\APACinsertmetastar {%
Camassel2005}%
\begin{APACrefauthors}%
Camassel, B.%
, Sghaier, N.%
, Prat, M.%
\BCBL {}\ \BBA {} {Ben Nasrallah}, S.%
\end{APACrefauthors}%
\unskip\
\newblock
\APACrefYearMonthDay{2005}{}{}.
\newblock
{\BBOQ}\APACrefatitle {Evaporation in a capillary tube of square cross-section:
  application to ion transport} {Evaporation in a capillary tube of square
  cross-section: application to ion transport}.{\BBCQ}
\newblock
\APACjournalVolNumPages{Chemical Engineering Science}{60}{3}{815-826}.
\newblock
\begin{APACrefDOI} \doi{10.1016/j.ces.2004.09.044} \end{APACrefDOI}
\PrintBackRefs{\CurrentBib}

\bibitem [\protect \citeauthoryear {%
Carman%
}{%
Carman%
}{%
{\protect \APACyear {1937}}%
}]{%
Carman1937}
\APACinsertmetastar {%
Carman1937}%
\begin{APACrefauthors}%
Carman, P.%
\end{APACrefauthors}%
\unskip\
\newblock
\APACrefYearMonthDay{1937}{}{}.
\newblock
{\BBOQ}\APACrefatitle {Fluid flow through granular beds} {Fluid flow through
  granular beds}.{\BBCQ}
\newblock
\APACjournalVolNumPages{Trans. Inst. Chem. Eng.}{15}{}{}.
\PrintBackRefs{\CurrentBib}

\bibitem [\protect \citeauthoryear {%
Daliakopoulos%
\ \protect \BOthers {.}}{%
Daliakopoulos%
\ \protect \BOthers {.}}{%
{\protect \APACyear {2016}}%
}]{%
Daliakopoulos2016}
\APACinsertmetastar {%
Daliakopoulos2016}%
\begin{APACrefauthors}%
Daliakopoulos, I.%
, Tsanis, I.%
, Koutroulis, A.%
, Kourgialas, N.%
, Varouchakis, A.%
, Karatzas, G.%
\BCBL {}\ \BBA {} Ritsema, C.%
\end{APACrefauthors}%
\unskip\
\newblock
\APACrefYearMonthDay{2016}{}{}.
\newblock
{\BBOQ}\APACrefatitle {The threat of soil salinity: A European scale review}
  {The threat of soil salinity: A european scale review}.{\BBCQ}
\newblock
\APACjournalVolNumPages{Science of The Total Environment}{573}{}{727-739}.
\newblock
\begin{APACrefDOI} \doi{10.1016/j.scitotenv.2016.08.177} \end{APACrefDOI}
\PrintBackRefs{\CurrentBib}

\bibitem [\protect \citeauthoryear {%
Dashtian%
, Shokri%
\BCBL {}\ \BBA {} Sahimi%
}{%
Dashtian%
\ \protect \BOthers {.}}{%
{\protect \APACyear {2018}}%
}]{%
Dashtian2018}
\APACinsertmetastar {%
Dashtian2018}%
\begin{APACrefauthors}%
Dashtian, H.%
, Shokri, N.%
\BCBL {}\ \BBA {} Sahimi, M.%
\end{APACrefauthors}%
\unskip\
\newblock
\APACrefYearMonthDay{2018}{}{}.
\newblock
{\BBOQ}\APACrefatitle {Pore-network model of evaporation-induced salt
  precipitation in porous media: The effect of correlations and heterogeneity}
  {Pore-network model of evaporation-induced salt precipitation in porous
  media: The effect of correlations and heterogeneity}.{\BBCQ}
\newblock
\APACjournalVolNumPages{Advances in Water Resources}{112}{}{59-71}.
\newblock
\begin{APACrefDOI} \doi{10.1016/j.advwatres.2017.12.004} \end{APACrefDOI}
\PrintBackRefs{\CurrentBib}

\bibitem [\protect \citeauthoryear {%
Daubert%
\ \BBA {} Danner%
}{%
Daubert%
\ \BBA {} Danner%
}{%
{\protect \APACyear {1989}}%
}]{%
Daubert1989}
\APACinsertmetastar {%
Daubert1989}%
\begin{APACrefauthors}%
Daubert, T\BPBI E.%
\BCBT {}\ \BBA {} Danner, R\BPBI P.%
\end{APACrefauthors}%
\unskip\
\newblock
\APACrefYear{1989}.
\newblock
\APACrefbtitle {{Physical and Thermodynamic Properties of Pure Chemicals:
  Design institute for physical property data, American institute of chemical
  engineers. vp}} {{Physical and Thermodynamic Properties of Pure Chemicals:
  Design institute for physical property data, American institute of chemical
  engineers. vp}}.
\newblock
\APACaddressPublisher{}{Hemisphere Publishing Corporation}.
\PrintBackRefs{\CurrentBib}

\bibitem [\protect \citeauthoryear {%
Derluyn%
, Moonen%
\BCBL {}\ \BBA {} Carmeliet%
}{%
Derluyn%
\ \protect \BOthers {.}}{%
{\protect \APACyear {2014}}%
}]{%
Derluyn2014}
\APACinsertmetastar {%
Derluyn2014}%
\begin{APACrefauthors}%
Derluyn, H.%
, Moonen, P.%
\BCBL {}\ \BBA {} Carmeliet, J.%
\end{APACrefauthors}%
\unskip\
\newblock
\APACrefYearMonthDay{2014}{}{}.
\newblock
{\BBOQ}\APACrefatitle {Deformation and damage due to drying-induced salt
  crystallization in porous limestone} {Deformation and damage due to
  drying-induced salt crystallization in porous limestone}.{\BBCQ}
\newblock
\APACjournalVolNumPages{Journal of the Mechanics and Physics of
  Solids}{63}{}{242-255}.
\newblock
\begin{APACrefDOI} \doi{10.1016/j.jmps.2013.09.005} \end{APACrefDOI}
\PrintBackRefs{\CurrentBib}

\bibitem [\protect \citeauthoryear {%
Dong%
\ \protect \BOthers {.}}{%
Dong%
\ \protect \BOthers {.}}{%
{\protect \APACyear {2021}}%
}]{%
Dong2021}
\APACinsertmetastar {%
Dong2021}%
\begin{APACrefauthors}%
Dong, L.%
, Xiong, Y.%
, Huang, Q.%
, Xu, X.%
, Huo, Z.%
\BCBL {}\ \BBA {} Huang, G.%
\end{APACrefauthors}%
\unskip\
\newblock
\APACrefYearMonthDay{2021}{}{}.
\newblock
{\BBOQ}\APACrefatitle {Evaporation-induced salt crystallization and feedback on
  hydrological functions in porous media with different grain morphologies}
  {Evaporation-induced salt crystallization and feedback on hydrological
  functions in porous media with different grain morphologies}.{\BBCQ}
\newblock
\APACjournalVolNumPages{Journal of Hydrology}{598}{}{126427}.
\newblock
\begin{APACrefDOI} \doi{10.1016/j.jhydrol.2021.126427} \end{APACrefDOI}
\PrintBackRefs{\CurrentBib}

\bibitem [\protect \citeauthoryear {%
Espinosa-Marzal%
\ \BBA {} Scherer%
}{%
Espinosa-Marzal%
\ \BBA {} Scherer%
}{%
{\protect \APACyear {2010}}%
}]{%
Espinosa2010}
\APACinsertmetastar {%
Espinosa2010}%
\begin{APACrefauthors}%
Espinosa-Marzal, R\BPBI M.%
\BCBT {}\ \BBA {} Scherer, G\BPBI W.%
\end{APACrefauthors}%
\unskip\
\newblock
\APACrefYearMonthDay{2010}{}{}.
\newblock
{\BBOQ}\APACrefatitle {Mechanisms of damage by salt} {Mechanisms of damage by
  salt}.{\BBCQ}
\newblock
\APACjournalVolNumPages{Geological Society, London, Special
  Publications}{331}{}{61--77}.
\newblock
\begin{APACrefDOI} \doi{10.1144/SP331.5} \end{APACrefDOI}
\PrintBackRefs{\CurrentBib}

\bibitem [\protect \citeauthoryear {%
Fekih%
, Sghaier%
, Amara%
\BCBL {}\ \BBA {} Prat%
}{%
Fekih%
\ \protect \BOthers {.}}{%
{\protect \APACyear {2024}}%
}]{%
Fekih2024}
\APACinsertmetastar {%
Fekih2024}%
\begin{APACrefauthors}%
Fekih, O.%
, Sghaier, N.%
, Amara, M\BPBI E\BPBI A\BPBI B.%
\BCBL {}\ \BBA {} Prat, M.%
\end{APACrefauthors}%
\unskip\
\newblock
\APACrefYearMonthDay{2024}{}{}.
\newblock
{\BBOQ}\APACrefatitle {Sodium chloride crystallization in a model porous medium
  during drying with a receding sharp front} {Sodium chloride crystallization
  in a model porous medium during drying with a receding sharp front}.{\BBCQ}
\newblock
\APACjournalVolNumPages{Physics of Fluids}{36}{4}{043301}.
\newblock
\begin{APACrefDOI} \doi{10.1063/5.0198793} \end{APACrefDOI}
\PrintBackRefs{\CurrentBib}

\bibitem [\protect \citeauthoryear {%
Ferrell%
\ \BBA {} Himmelblau%
}{%
Ferrell%
\ \BBA {} Himmelblau%
}{%
{\protect \APACyear {1967}}%
}]{%
Ferrell1967}
\APACinsertmetastar {%
Ferrell1967}%
\begin{APACrefauthors}%
Ferrell, R\BPBI T.%
\BCBT {}\ \BBA {} Himmelblau, D\BPBI M.%
\end{APACrefauthors}%
\unskip\
\newblock
\APACrefYearMonthDay{1967}{}{}.
\newblock
{\BBOQ}\APACrefatitle {Diffusion coefficients of nitrogen and oxygen in water}
  {Diffusion coefficients of nitrogen and oxygen in water}.{\BBCQ}
\newblock
\APACjournalVolNumPages{Journal of chemical and engineering
  data}{12}{1}{111--115}.
\newblock
\begin{APACrefDOI} \doi{10.1021/je60032a036} \end{APACrefDOI}
\PrintBackRefs{\CurrentBib}

\bibitem [\protect \citeauthoryear {%
Gholami%
}{%
Gholami%
}{%
{\protect \APACyear {2023}}%
}]{%
Gholami2023}
\APACinsertmetastar {%
Gholami2023}%
\begin{APACrefauthors}%
Gholami, R.%
\end{APACrefauthors}%
\unskip\
\newblock
\APACrefYearMonthDay{2023}{}{}.
\newblock
{\BBOQ}\APACrefatitle {Hydrogen storage in geological porous media: Solubility,
  mineral trapping, H2S generation and salt precipitation} {Hydrogen storage in
  geological porous media: Solubility, mineral trapping, h2s generation and
  salt precipitation}.{\BBCQ}
\newblock
\APACjournalVolNumPages{Journal of Energy Storage}{59}{}{106576}.
\newblock
\begin{APACrefDOI} \doi{10.1016/j.est.2022.106576} \end{APACrefDOI}
\PrintBackRefs{\CurrentBib}

\bibitem [\protect \citeauthoryear {%
Hommel%
, Coltman%
\BCBL {}\ \BBA {} Class%
}{%
Hommel%
\ \protect \BOthers {.}}{%
{\protect \APACyear {2018}}%
}]{%
Hommel2018}
\APACinsertmetastar {%
Hommel2018}%
\begin{APACrefauthors}%
Hommel, J.%
, Coltman, E.%
\BCBL {}\ \BBA {} Class, H.%
\end{APACrefauthors}%
\unskip\
\newblock
\APACrefYearMonthDay{2018}{}{}.
\newblock
{\BBOQ}\APACrefatitle {Porosity–Permeability Relations for Evolving Pore
  Space: A Review with a Focus on (Bio-)geochemically Altered Porous Media}
  {Porosity–permeability relations for evolving pore space: A review with a
  focus on (bio-)geochemically altered porous media}.{\BBCQ}
\newblock
\APACjournalVolNumPages{Transport in Porous Media}{124}{2}{589-629}.
\newblock
\begin{APACrefDOI} \doi{10.1007/s11242-018-1086-2} \end{APACrefDOI}
\PrintBackRefs{\CurrentBib}

\bibitem [\protect \citeauthoryear {%
Jambhekar%
, Helmig%
, Schr{\"o}der%
\BCBL {}\ \BBA {} Shokri%
}{%
Jambhekar%
\ \protect \BOthers {.}}{%
{\protect \APACyear {2015}}%
}]{%
Jambhekar2015}
\APACinsertmetastar {%
Jambhekar2015}%
\begin{APACrefauthors}%
Jambhekar, V\BPBI A.%
, Helmig, R.%
, Schr{\"o}der, N.%
\BCBL {}\ \BBA {} Shokri, N.%
\end{APACrefauthors}%
\unskip\
\newblock
\APACrefYearMonthDay{2015}{}{}.
\newblock
{\BBOQ}\APACrefatitle {Free-Flow--Porous-Media Coupling for Evaporation-Driven
  Transport and Precipitation of Salt in Soil} {Free-flow--porous-media
  coupling for evaporation-driven transport and precipitation of salt in
  soil}.{\BBCQ}
\newblock
\APACjournalVolNumPages{Transport in Porous Media}{110}{2}{251--280}.
\newblock
\begin{APACrefDOI} \doi{10.1007/s11242-015-0516-7} \end{APACrefDOI}
\PrintBackRefs{\CurrentBib}

\bibitem [\protect \citeauthoryear {%
Jambhekar%
, Mejri%
, Schr{\"o}der%
, Helmig%
\BCBL {}\ \BBA {} Shokri%
}{%
Jambhekar%
\ \protect \BOthers {.}}{%
{\protect \APACyear {2016}}%
}]{%
Jambhekar2016}
\APACinsertmetastar {%
Jambhekar2016}%
\begin{APACrefauthors}%
Jambhekar, V\BPBI A.%
, Mejri, E.%
, Schr{\"o}der, N.%
, Helmig, R.%
\BCBL {}\ \BBA {} Shokri, N.%
\end{APACrefauthors}%
\unskip\
\newblock
\APACrefYearMonthDay{2016}{}{}.
\newblock
{\BBOQ}\APACrefatitle {Kinetic Approach to Model Reactive Transport and Mixed
  Salt Precipitation in a Coupled Free-Flow--Porous-Media System} {Kinetic
  approach to model reactive transport and mixed salt precipitation in a
  coupled free-flow--porous-media system}.{\BBCQ}
\newblock
\APACjournalVolNumPages{Transport in Porous Media}{114}{2}{341--369}.
\newblock
\begin{APACrefDOI} \doi{10.1007/s11242-016-0665-3} \end{APACrefDOI}
\PrintBackRefs{\CurrentBib}

\bibitem [\protect \citeauthoryear {%
Joekar-Niasar%
, Hassanizadeh%
\BCBL {}\ \BBA {} Dahle%
}{%
Joekar-Niasar%
\ \protect \BOthers {.}}{%
{\protect \APACyear {2010}}%
}]{%
Joekar2010}
\APACinsertmetastar {%
Joekar2010}%
\begin{APACrefauthors}%
Joekar-Niasar, V.%
, Hassanizadeh, S\BPBI M.%
\BCBL {}\ \BBA {} Dahle, H\BPBI K.%
\end{APACrefauthors}%
\unskip\
\newblock
\APACrefYearMonthDay{2010}{}{}.
\newblock
{\BBOQ}\APACrefatitle {Non-equilibrium effects in capillarity and interfacial
  area in two-phase flow: dynamic pore-network modelling} {Non-equilibrium
  effects in capillarity and interfacial area in two-phase flow: dynamic
  pore-network modelling}.{\BBCQ}
\newblock
\APACjournalVolNumPages{Journal of Fluid Mechanics}{655}{}{38–71}.
\newblock
\begin{APACrefDOI} \doi{10.1017/S0022112010000704} \end{APACrefDOI}
\PrintBackRefs{\CurrentBib}

\bibitem [\protect \citeauthoryear {%
Kang%
, Zhang%
\BCBL {}\ \BBA {} Chen%
}{%
Kang%
\ \protect \BOthers {.}}{%
{\protect \APACyear {2003}}%
}]{%
Kang2003}
\APACinsertmetastar {%
Kang2003}%
\begin{APACrefauthors}%
Kang, Q.%
, Zhang, D.%
\BCBL {}\ \BBA {} Chen, S.%
\end{APACrefauthors}%
\unskip\
\newblock
\APACrefYearMonthDay{2003}{}{}.
\newblock
{\BBOQ}\APACrefatitle {Simulation of dissolution and precipitation in porous
  media} {Simulation of dissolution and precipitation in porous media}.{\BBCQ}
\newblock
\APACjournalVolNumPages{Journal of Geophysical Research: Solid
  Earth}{108}{B10}{}.
\newblock
\begin{APACrefDOI} \doi{10.1029/2003JB002504} \end{APACrefDOI}
\PrintBackRefs{\CurrentBib}

\bibitem [\protect \citeauthoryear {%
Koch%
\ \protect \BOthers {.}}{%
Koch%
\ \protect \BOthers {.}}{%
{\protect \APACyear {2020}}%
}]{%
Koch2020}
\APACinsertmetastar {%
Koch2020}%
\begin{APACrefauthors}%
Koch, T.%
, Gl{\"a}ser, D.%
, Weishaupt, K.%
, Ackermann, S.%
, Beck, M.%
, Becker, B.%
\BDBL {}Flemisch, B.%
\end{APACrefauthors}%
\unskip\
\newblock
\APACrefYearMonthDay{2020}{}{}.
\newblock
{\BBOQ}\APACrefatitle {DuMuX 3 - an open-source simulator for solving flow and
  transport problems in porous media with a focus on model coupling} {Dumux 3 -
  an open-source simulator for solving flow and transport problems in porous
  media with a focus on model coupling}.{\BBCQ}
\newblock
\APACjournalVolNumPages{Computers and Mathematics with Applications}{}{}{}.
\newblock
\begin{APACrefDOI} \doi{10.1016/j.camwa.2020.02.012} \end{APACrefDOI}
\PrintBackRefs{\CurrentBib}

\bibitem [\protect \citeauthoryear {%
Koniorczyk%
}{%
Koniorczyk%
}{%
{\protect \APACyear {2012}}%
}]{%
Koniorczyk2012}
\APACinsertmetastar {%
Koniorczyk2012}%
\begin{APACrefauthors}%
Koniorczyk, M.%
\end{APACrefauthors}%
\unskip\
\newblock
\APACrefYearMonthDay{2012}{}{}.
\newblock
{\BBOQ}\APACrefatitle {Salt transport and crystallization in non-isothermal,
  partially saturated porous materials considering ions interaction model}
  {Salt transport and crystallization in non-isothermal, partially saturated
  porous materials considering ions interaction model}.{\BBCQ}
\newblock
\APACjournalVolNumPages{International Journal of Heat and Mass
  Transfer}{55}{4}{665-679}.
\newblock
\begin{APACrefDOI} \doi{10.1016/j.ijheatmasstransfer.2011.10.043}
  \end{APACrefDOI}
\PrintBackRefs{\CurrentBib}

\bibitem [\protect \citeauthoryear {%
Kozeny%
}{%
Kozeny%
}{%
{\protect \APACyear {1927}}%
}]{%
Kozeny1927}
\APACinsertmetastar {%
Kozeny1927}%
\begin{APACrefauthors}%
Kozeny, J.%
\end{APACrefauthors}%
\unskip\
\newblock
\APACrefYearMonthDay{1927}{}{}.
\newblock
{\BBOQ}\APACrefatitle {Über kapillare {L}eitung des {W}assers im {B}oden}
  {Über kapillare {L}eitung des {W}assers im {B}oden}.{\BBCQ}
\newblock
\APACjournalVolNumPages{Sitzungsber Akad Wiss Wien}{136}{}{271-306}.
\PrintBackRefs{\CurrentBib}

\bibitem [\protect \citeauthoryear {%
Leverett%
}{%
Leverett%
}{%
{\protect \APACyear {1941}}%
}]{%
Leverett1941}
\APACinsertmetastar {%
Leverett1941}%
\begin{APACrefauthors}%
Leverett, M.%
\end{APACrefauthors}%
\unskip\
\newblock
\APACrefYearMonthDay{1941}{}{}.
\newblock
{\BBOQ}\APACrefatitle {Capillary Behavior in Porous Solids} {Capillary behavior
  in porous solids}.{\BBCQ}
\newblock
\APACjournalVolNumPages{Transactions of the AIME}{142}{01}{152-169}.
\newblock
\begin{APACrefDOI} \doi{10.2118/941152-G} \end{APACrefDOI}
\PrintBackRefs{\CurrentBib}

\bibitem [\protect \citeauthoryear {%
Ma%
, Mason%
\BCBL {}\ \BBA {} Morrow%
}{%
Ma%
\ \protect \BOthers {.}}{%
{\protect \APACyear {1996}}%
}]{%
Ma1996}
\APACinsertmetastar {%
Ma1996}%
\begin{APACrefauthors}%
Ma, S.%
, Mason, G.%
\BCBL {}\ \BBA {} Morrow, N\BPBI R.%
\end{APACrefauthors}%
\unskip\
\newblock
\APACrefYearMonthDay{1996}{}{}.
\newblock
{\BBOQ}\APACrefatitle {Effect of contact angle on drainage and imbibition in
  regular polygonal tubes} {Effect of contact angle on drainage and imbibition
  in regular polygonal tubes}.{\BBCQ}
\newblock
\APACjournalVolNumPages{Colloids and Surfaces A: Physicochemical and
  Engineering Aspects}{117}{3}{273-291}.
\newblock
\begin{APACrefDOI} \doi{10.1016/0927-7757(96)03702-8} \end{APACrefDOI}
\PrintBackRefs{\CurrentBib}

\bibitem [\protect \citeauthoryear {%
Mason%
\ \BBA {} Morrow%
}{%
Mason%
\ \BBA {} Morrow%
}{%
{\protect \APACyear {1991}}%
}]{%
Mason1991}
\APACinsertmetastar {%
Mason1991}%
\begin{APACrefauthors}%
Mason, G.%
\BCBT {}\ \BBA {} Morrow, N\BPBI R.%
\end{APACrefauthors}%
\unskip\
\newblock
\APACrefYearMonthDay{1991}{}{}.
\newblock
{\BBOQ}\APACrefatitle {Capillary behavior of a perfectly wetting liquid in
  irregular triangular tubes} {Capillary behavior of a perfectly wetting liquid
  in irregular triangular tubes}.{\BBCQ}
\newblock
\APACjournalVolNumPages{Journal of Colloid and Interface
  Science}{141}{1}{262-274}.
\newblock
\begin{APACrefDOI} \doi{10.1016/0021-9797(91)90321-X} \end{APACrefDOI}
\PrintBackRefs{\CurrentBib}

\bibitem [\protect \citeauthoryear {%
Mejri%
, Bouhlila%
\BCBL {}\ \BBA {} Helmig%
}{%
Mejri%
\ \protect \BOthers {.}}{%
{\protect \APACyear {2017}}%
}]{%
Mejri2017}
\APACinsertmetastar {%
Mejri2017}%
\begin{APACrefauthors}%
Mejri, E.%
, Bouhlila, R.%
\BCBL {}\ \BBA {} Helmig, R.%
\end{APACrefauthors}%
\unskip\
\newblock
\APACrefYearMonthDay{2017}{}{}.
\newblock
{\BBOQ}\APACrefatitle {Heterogeneity Effects on Evaporation-Induced Halite and
  Gypsum Co-precipitation in Porous Media} {Heterogeneity effects on
  evaporation-induced halite and gypsum co-precipitation in porous
  media}.{\BBCQ}
\newblock
\APACjournalVolNumPages{Transport in Porous Media}{118}{}{39-64}.
\newblock
\begin{APACrefDOI} \doi{10.1007/s11242-017-0846-8} \end{APACrefDOI}
\PrintBackRefs{\CurrentBib}

\bibitem [\protect \citeauthoryear {%
Mejri%
, Helmig%
\BCBL {}\ \BBA {} Bouhlila%
}{%
Mejri%
\ \protect \BOthers {.}}{%
{\protect \APACyear {2020}}%
}]{%
Mejri2020}
\APACinsertmetastar {%
Mejri2020}%
\begin{APACrefauthors}%
Mejri, E.%
, Helmig, R.%
\BCBL {}\ \BBA {} Bouhlila, R.%
\end{APACrefauthors}%
\unskip\
\newblock
\APACrefYearMonthDay{2020}{}{}.
\newblock
{\BBOQ}\APACrefatitle {Modeling of Evaporation-Driven Multiple Salt
  Precipitation in Porous Media with a Real Field Application} {Modeling of
  evaporation-driven multiple salt precipitation in porous media with a real
  field application}.{\BBCQ}
\newblock
\APACjournalVolNumPages{Geosciences}{10}{10}{}.
\newblock
\begin{APACrefDOI} \doi{10.3390/geosciences10100395} \end{APACrefDOI}
\PrintBackRefs{\CurrentBib}

\bibitem [\protect \citeauthoryear {%
Michaelides%
}{%
Michaelides%
}{%
{\protect \APACyear {1981}}%
}]{%
Michaelides1981}
\APACinsertmetastar {%
Michaelides1981}%
\begin{APACrefauthors}%
Michaelides, E\BPBI E.%
\end{APACrefauthors}%
\unskip\
\newblock
\APACrefYearMonthDay{1981}{}{}.
\newblock
{\BBOQ}\APACrefatitle {Thermodynamic properties of geothermal fluids}
  {Thermodynamic properties of geothermal fluids}.{\BBCQ}
\newblock
\APACjournalVolNumPages{Trans. - Geotherm. Resour. Counc.; (United
  States)}{5}{}{}.
\newblock
\begin{APACrefURL} \url{https://www.osti.gov/biblio/6760030} \end{APACrefURL}
\PrintBackRefs{\CurrentBib}

\bibitem [\protect \citeauthoryear {%
Miri%
, {van Noort}%
, Aagaard%
\BCBL {}\ \BBA {} Hellevang%
}{%
Miri%
\ \protect \BOthers {.}}{%
{\protect \APACyear {2015}}%
}]{%
Miri2015}
\APACinsertmetastar {%
Miri2015}%
\begin{APACrefauthors}%
Miri, R.%
, {van Noort}, R.%
, Aagaard, P.%
\BCBL {}\ \BBA {} Hellevang, H.%
\end{APACrefauthors}%
\unskip\
\newblock
\APACrefYearMonthDay{2015}{}{}.
\newblock
{\BBOQ}\APACrefatitle {New insights on the physics of salt precipitation during
  injection of CO2 into saline aquifers} {New insights on the physics of salt
  precipitation during injection of co2 into saline aquifers}.{\BBCQ}
\newblock
\APACjournalVolNumPages{International Journal of Greenhouse Gas
  Control}{43}{}{10-21}.
\newblock
\begin{APACrefDOI} \doi{10.1016/j.ijggc.2015.10.004} \end{APACrefDOI}
\PrintBackRefs{\CurrentBib}

\bibitem [\protect \citeauthoryear {%
Molins%
}{%
Molins%
}{%
{\protect \APACyear {2015}}%
}]{%
Molins2015}
\APACinsertmetastar {%
Molins2015}%
\begin{APACrefauthors}%
Molins, S.%
\end{APACrefauthors}%
\unskip\
\newblock
\APACrefYearMonthDay{2015}{}{}.
\newblock
{\BBOQ}\APACrefatitle {Reactive Interfaces in Direct Numerical Simulation of
  Pore-Scale Processes} {Reactive interfaces in direct numerical simulation of
  pore-scale processes}.{\BBCQ}
\newblock
\APACjournalVolNumPages{Reviews in Mineralogy and
  Geochemistry}{80}{1}{461-481}.
\newblock
\begin{APACrefDOI} \doi{10.2138/rmg.2015.80.14} \end{APACrefDOI}
\PrintBackRefs{\CurrentBib}

\bibitem [\protect \citeauthoryear {%
Molins%
, Trebotich%
, Steefel%
\BCBL {}\ \BBA {} Shen%
}{%
Molins%
\ \protect \BOthers {.}}{%
{\protect \APACyear {2012}}%
}]{%
Molins2012}
\APACinsertmetastar {%
Molins2012}%
\begin{APACrefauthors}%
Molins, S.%
, Trebotich, D.%
, Steefel, C\BPBI I.%
\BCBL {}\ \BBA {} Shen, C.%
\end{APACrefauthors}%
\unskip\
\newblock
\APACrefYearMonthDay{2012}{}{}.
\newblock
{\BBOQ}\APACrefatitle {An investigation of the effect of pore scale flow on
  average geochemical reaction rates using direct numerical simulation} {An
  investigation of the effect of pore scale flow on average geochemical
  reaction rates using direct numerical simulation}.{\BBCQ}
\newblock
\APACjournalVolNumPages{Water Resources Research}{48}{3}{}.
\newblock
\begin{APACrefDOI} \doi{10.1029/2011WR011404} \end{APACrefDOI}
\PrintBackRefs{\CurrentBib}

\bibitem [\protect \citeauthoryear {%
Nachshon%
}{%
Nachshon%
}{%
{\protect \APACyear {2018}}%
}]{%
Nachshon2018a}
\APACinsertmetastar {%
Nachshon2018a}%
\begin{APACrefauthors}%
Nachshon, U.%
\end{APACrefauthors}%
\unskip\
\newblock
\APACrefYearMonthDay{2018}{}{}.
\newblock
{\BBOQ}\APACrefatitle {Cropland Soil Salinization and Associated Hydrology:
  Trends, Processes and Examples} {Cropland soil salinization and associated
  hydrology: Trends, processes and examples}.{\BBCQ}
\newblock
\APACjournalVolNumPages{Water}{10}{8}{}.
\newblock
\begin{APACrefDOI} \doi{10.3390/w10081030} \end{APACrefDOI}
\PrintBackRefs{\CurrentBib}

\bibitem [\protect \citeauthoryear {%
Nachshon%
, Weisbrod%
, Katzir%
\BCBL {}\ \BBA {} Nasser%
}{%
Nachshon%
\ \protect \BOthers {.}}{%
{\protect \APACyear {2018}}%
}]{%
Nachshon2018b}
\APACinsertmetastar {%
Nachshon2018b}%
\begin{APACrefauthors}%
Nachshon, U.%
, Weisbrod, N.%
, Katzir, R.%
\BCBL {}\ \BBA {} Nasser, A.%
\end{APACrefauthors}%
\unskip\
\newblock
\APACrefYearMonthDay{2018}{}{}.
\newblock
{\BBOQ}\APACrefatitle {NaCl Crust Architecture and Its Impact on Evaporation:
  Three-Dimensional Insights} {Nacl crust architecture and its impact on
  evaporation: Three-dimensional insights}.{\BBCQ}
\newblock
\APACjournalVolNumPages{Geophysical Research Letters}{45}{12}{6100-6108}.
\newblock
\begin{APACrefDOI} \doi{10.1029/2018GL078363} \end{APACrefDOI}
\PrintBackRefs{\CurrentBib}

\bibitem [\protect \citeauthoryear {%
Nicolai%
, Grunewald%
\BCBL {}\ \BBA {} Zhang~PhD%
}{%
Nicolai%
\ \protect \BOthers {.}}{%
{\protect \APACyear {2007}}%
}]{%
Nicolai2007}
\APACinsertmetastar {%
Nicolai2007}%
\begin{APACrefauthors}%
Nicolai, A.%
, Grunewald, J.%
\BCBL {}\ \BBA {} Zhang~PhD, J\BPBI S.%
\end{APACrefauthors}%
\unskip\
\newblock
\APACrefYearMonthDay{2007}{}{}.
\newblock
{\BBOQ}\APACrefatitle {Salztransport und Phasenumwandlung – Modellierung und
  numerische Lösung im Simulationsprogramm Delphin 5} {Salztransport und
  phasenumwandlung – modellierung und numerische lösung im
  simulationsprogramm delphin 5}.{\BBCQ}
\newblock
\APACjournalVolNumPages{Bauphysik}{29}{3}{231-239}.
\newblock
\begin{APACrefDOI} \doi{10.1002/bapi.200710032} \end{APACrefDOI}
\PrintBackRefs{\CurrentBib}

\bibitem [\protect \citeauthoryear {%
Nogues%
, Fitts%
, Celia%
\BCBL {}\ \BBA {} Peters%
}{%
Nogues%
\ \protect \BOthers {.}}{%
{\protect \APACyear {2013}}%
}]{%
Nogues2013}
\APACinsertmetastar {%
Nogues2013}%
\begin{APACrefauthors}%
Nogues, J\BPBI P.%
, Fitts, J\BPBI P.%
, Celia, M\BPBI A.%
\BCBL {}\ \BBA {} Peters, C\BPBI A.%
\end{APACrefauthors}%
\unskip\
\newblock
\APACrefYearMonthDay{2013}{}{}.
\newblock
{\BBOQ}\APACrefatitle {Permeability evolution due to dissolution and
  precipitation of carbonates using reactive transport modeling in pore
  networks} {Permeability evolution due to dissolution and precipitation of
  carbonates using reactive transport modeling in pore networks}.{\BBCQ}
\newblock
\APACjournalVolNumPages{Water Resources Research}{49}{9}{6006--6021}.
\newblock
\begin{APACrefDOI} \doi{10.1002/wrcr.20486} \end{APACrefDOI}
\PrintBackRefs{\CurrentBib}

\bibitem [\protect \citeauthoryear {%
{Norouzi Rad}%
, Shokri%
\BCBL {}\ \BBA {} Sahimi%
}{%
{Norouzi Rad}%
\ \protect \BOthers {.}}{%
{\protect \APACyear {2013}}%
}]{%
Rad2013}
\APACinsertmetastar {%
Rad2013}%
\begin{APACrefauthors}%
{Norouzi Rad}, M.%
, Shokri, N.%
\BCBL {}\ \BBA {} Sahimi, M.%
\end{APACrefauthors}%
\unskip\
\newblock
\APACrefYearMonthDay{2013}{}{}.
\newblock
{\BBOQ}\APACrefatitle {Pore-scale dynamics of salt precipitation in drying
  porous media} {Pore-scale dynamics of salt precipitation in drying porous
  media}.{\BBCQ}
\newblock
\APACjournalVolNumPages{Phys. Rev. E}{88}{}{032404}.
\newblock
\begin{APACrefDOI} \doi{10.1103/PhysRevE.88.032404} \end{APACrefDOI}
\PrintBackRefs{\CurrentBib}

\bibitem [\protect \citeauthoryear {%
Ondrasek%
, Rengel%
\BCBL {}\ \BBA {} Veres%
}{%
Ondrasek%
\ \protect \BOthers {.}}{%
{\protect \APACyear {2011}}%
}]{%
Ondrasek2011}
\APACinsertmetastar {%
Ondrasek2011}%
\begin{APACrefauthors}%
Ondrasek, G.%
, Rengel, Z.%
\BCBL {}\ \BBA {} Veres, S.%
\end{APACrefauthors}%
\unskip\
\newblock
\APACrefYearMonthDay{2011}{}{}.
\newblock
{\BBOQ}\APACrefatitle {Soil Salinisation and Salt Stress in Crop Production}
  {Soil salinisation and salt stress in crop production}.{\BBCQ}
\newblock
\BIn{} A\BPBI K.~Shanker\ \BBA {} B.~Venkateswarlu\ (\BEDS), \APACrefbtitle
  {Abiotic Stress in Plants - Mechanisms and Adaptations} {Abiotic stress in
  plants - mechanisms and adaptations}\ (\BPG~171-190).
\newblock
\begin{APACrefDOI} \doi{10.5772/22248} \end{APACrefDOI}
\PrintBackRefs{\CurrentBib}

\bibitem [\protect \citeauthoryear {%
Pérez-Villaseñor%
, Iglesias-Silva%
\BCBL {}\ \BBA {} Hall%
}{%
Pérez-Villaseñor%
\ \protect \BOthers {.}}{%
{\protect \APACyear {2002}}%
}]{%
Perez2002}
\APACinsertmetastar {%
Perez2002}%
\begin{APACrefauthors}%
Pérez-Villaseñor, F.%
, Iglesias-Silva, G\BPBI A.%
\BCBL {}\ \BBA {} Hall, K\BPBI R.%
\end{APACrefauthors}%
\unskip\
\newblock
\APACrefYearMonthDay{2002}{}{}.
\newblock
{\BBOQ}\APACrefatitle {Osmotic and Activity Coefficients Using a Modified
  Pitzer Equation for Strong Electrolytes 1:1 and 1:2 at 298.15 K} {Osmotic and
  activity coefficients using a modified pitzer equation for strong
  electrolytes 1:1 and 1:2 at 298.15 k}.{\BBCQ}
\newblock
\APACjournalVolNumPages{Industrial \& Engineering Chemistry
  Research}{41}{5}{1031-1037}.
\newblock
\begin{APACrefDOI} \doi{10.1021/ie0103153} \end{APACrefDOI}
\PrintBackRefs{\CurrentBib}

\bibitem [\protect \citeauthoryear {%
Qadir%
\ \protect \BOthers {.}}{%
Qadir%
\ \protect \BOthers {.}}{%
{\protect \APACyear {2014}}%
}]{%
Qadir2014}
\APACinsertmetastar {%
Qadir2014}%
\begin{APACrefauthors}%
Qadir, M.%
, Quill{\'e}rou, E.%
, Nangia, V.%
, Murtaza, G.%
, Singh, M.%
, Thomas, R.%
\BDBL {}Noble, A.%
\end{APACrefauthors}%
\unskip\
\newblock
\APACrefYearMonthDay{2014}{}{}.
\newblock
{\BBOQ}\APACrefatitle {Economics of salt-induced land degradation and
  restoration} {Economics of salt-induced land degradation and
  restoration}.{\BBCQ}
\newblock
\APACjournalVolNumPages{Natural Resources Forum}{38}{4}{282--295}.
\newblock
\begin{APACrefDOI} \doi{10.1111/1477-8947.12054} \end{APACrefDOI}
\PrintBackRefs{\CurrentBib}

\bibitem [\protect \citeauthoryear {%
Rad%
, Shokri%
, Keshmiri%
\BCBL {}\ \BBA {} Withers%
}{%
Rad%
\ \protect \BOthers {.}}{%
{\protect \APACyear {2015}}%
}]{%
Rad2015}
\APACinsertmetastar {%
Rad2015}%
\begin{APACrefauthors}%
Rad, M\BPBI N.%
, Shokri, N.%
, Keshmiri, A.%
\BCBL {}\ \BBA {} Withers, P\BPBI J.%
\end{APACrefauthors}%
\unskip\
\newblock
\APACrefYearMonthDay{2015}{}{}.
\newblock
{\BBOQ}\APACrefatitle {Effects of Grain and Pore Size on Salt Precipitation
  During Evaporation from Porous Media} {Effects of grain and pore size on salt
  precipitation during evaporation from porous media}.{\BBCQ}
\newblock
\APACjournalVolNumPages{Transport in Porous Media}{110}{2}{281--294}.
\newblock
\begin{APACrefDOI} \doi{10.1007/s11242-015-0515-8} \end{APACrefDOI}
\PrintBackRefs{\CurrentBib}

\bibitem [\protect \citeauthoryear {%
Ransohoff%
\ \BBA {} Radke%
}{%
Ransohoff%
\ \BBA {} Radke%
}{%
{\protect \APACyear {1988}}%
}]{%
Ransohoff1988}
\APACinsertmetastar {%
Ransohoff1988}%
\begin{APACrefauthors}%
Ransohoff, T.%
\BCBT {}\ \BBA {} Radke, C.%
\end{APACrefauthors}%
\unskip\
\newblock
\APACrefYearMonthDay{1988}{}{}.
\newblock
{\BBOQ}\APACrefatitle {Laminar flow of a wetting liquid along the corners of a
  predominantly gas-occupied noncircular pore} {Laminar flow of a wetting
  liquid along the corners of a predominantly gas-occupied noncircular
  pore}.{\BBCQ}
\newblock
\APACjournalVolNumPages{Journal of Colloid and Interface
  Science}{121}{2}{392-401}.
\newblock
\begin{APACrefDOI} \doi{10.1016/0021-9797(88)90442-0} \end{APACrefDOI}
\PrintBackRefs{\CurrentBib}

\bibitem [\protect \citeauthoryear {%
Rard%
\ \BBA {} Miller%
}{%
Rard%
\ \BBA {} Miller%
}{%
{\protect \APACyear {1979}}%
}]{%
Rard1979}
\APACinsertmetastar {%
Rard1979}%
\begin{APACrefauthors}%
Rard, J\BPBI A.%
\BCBT {}\ \BBA {} Miller, D\BPBI G.%
\end{APACrefauthors}%
\unskip\
\newblock
\APACrefYearMonthDay{1979}{}{}.
\newblock
{\BBOQ}\APACrefatitle {The mutual diffusion coefficients of NaCl-H2O and
  CaCl2-H2O at 25°C from Rayleigh interferometry} {The mutual diffusion
  coefficients of nacl-h2o and cacl2-h2o at 25°c from rayleigh
  interferometry}.{\BBCQ}
\newblock
\APACjournalVolNumPages{Journal of Solution Chemistry}{8}{10}{701-716}.
\newblock
\begin{APACrefDOI} \doi{10.1007/BF00648776} \end{APACrefDOI}
\PrintBackRefs{\CurrentBib}

\bibitem [\protect \citeauthoryear {%
Rengasamy%
}{%
Rengasamy%
}{%
{\protect \APACyear {2006}}%
}]{%
Rengasamy2006}
\APACinsertmetastar {%
Rengasamy2006}%
\begin{APACrefauthors}%
Rengasamy, P.%
\end{APACrefauthors}%
\unskip\
\newblock
\APACrefYearMonthDay{2006}{}{}.
\newblock
{\BBOQ}\APACrefatitle {World salinization with emphasis on Australia} {World
  salinization with emphasis on australia}.{\BBCQ}
\newblock
\APACjournalVolNumPages{Journal of Experimental Botany}{57}{5}{1017--1023}.
\newblock
\begin{APACrefDOI} \doi{10.1093/jxb/erj108} \end{APACrefDOI}
\PrintBackRefs{\CurrentBib}

\bibitem [\protect \citeauthoryear {%
Rohde%
\ \BBA {} von Wolff%
}{%
Rohde%
\ \BBA {} von Wolff%
}{%
{\protect \APACyear {2021}}%
}]{%
Rohde2021}
\APACinsertmetastar {%
Rohde2021}%
\begin{APACrefauthors}%
Rohde, C.%
\BCBT {}\ \BBA {} von Wolff, L.%
\end{APACrefauthors}%
\unskip\
\newblock
\APACrefYearMonthDay{2021}{}{}.
\newblock
{\BBOQ}\APACrefatitle {A ternary Cahn–Hilliard–Navier–Stokes model for
  two-phase flow with precipitation and dissolution} {A ternary
  cahn–hilliard–navier–stokes model for two-phase flow with precipitation
  and dissolution}.{\BBCQ}
\newblock
\APACjournalVolNumPages{Mathematical Models and Methods in Applied
  Sciences}{31}{01}{1-35}.
\newblock
\begin{APACrefDOI} \doi{10.1142/S0218202521500019} \end{APACrefDOI}
\PrintBackRefs{\CurrentBib}

\bibitem [\protect \citeauthoryear {%
Roy%
, Weibel%
\BCBL {}\ \BBA {} Garimella%
}{%
Roy%
\ \protect \BOthers {.}}{%
{\protect \APACyear {2022}}%
}]{%
Roy2022}
\APACinsertmetastar {%
Roy2022}%
\begin{APACrefauthors}%
Roy, R.%
, Weibel, J\BPBI A.%
\BCBL {}\ \BBA {} Garimella, S\BPBI V.%
\end{APACrefauthors}%
\unskip\
\newblock
\APACrefYearMonthDay{2022}{}{}.
\newblock
{\BBOQ}\APACrefatitle {Modeling the formation of efflorescence and
  subflorescence caused by salt solution evaporation from porous media}
  {Modeling the formation of efflorescence and subflorescence caused by salt
  solution evaporation from porous media}.{\BBCQ}
\newblock
\APACjournalVolNumPages{International Journal of Heat and Mass
  Transfer}{189}{}{122645}.
\newblock
\begin{APACrefDOI} \doi{10.1016/j.ijheatmasstransfer.2022.122645}
  \end{APACrefDOI}
\PrintBackRefs{\CurrentBib}

\bibitem [\protect \citeauthoryear {%
Scherer%
}{%
Scherer%
}{%
{\protect \APACyear {2004}}%
}]{%
Scherer2004}
\APACinsertmetastar {%
Scherer2004}%
\begin{APACrefauthors}%
Scherer, G\BPBI W.%
\end{APACrefauthors}%
\unskip\
\newblock
\APACrefYearMonthDay{2004}{}{}.
\newblock
{\BBOQ}\APACrefatitle {Stress from crystallization of salt} {Stress from
  crystallization of salt}.{\BBCQ}
\newblock
\APACjournalVolNumPages{Cement and Concrete Research}{34}{9}{1613--1624}.
\newblock
\APACrefnote{H. F. W. Taylor Commemorative Issue}
\newblock
\begin{APACrefDOI} \doi{10.1016/j.cemconres.2003.12.034} \end{APACrefDOI}
\PrintBackRefs{\CurrentBib}

\bibitem [\protect \citeauthoryear {%
Schollenbegrer%
}{%
Schollenbegrer%
}{%
{\protect \APACyear {2025}}%
}]{%
schollenberger2025aPubModul}
\APACinsertmetastar {%
schollenberger2025aPubModul}%
\begin{APACrefauthors}%
Schollenbegrer, T.%
\end{APACrefauthors}%
\unskip\
\newblock
\APACrefYearMonthDay{2025}{}{}.
\newblock
\APACrefbtitle {Schollenberger2025a [source code].} {Schollenberger2025a
  [source code].}
\newblock
\begin{APACrefURL}
  \url{https://git.iws.uni-stuttgart.de/dumux-pub/schollenberger2025a}
  \end{APACrefURL}
\PrintBackRefs{\CurrentBib}

\bibitem [\protect \citeauthoryear {%
Schollenberger%
\ \protect \BOthers {.}}{%
Schollenberger%
\ \protect \BOthers {.}}{%
{\protect \APACyear {2024}}%
}]{%
Schollenberger2024}
\APACinsertmetastar {%
Schollenberger2024}%
\begin{APACrefauthors}%
Schollenberger, T.%
, von Wolff, L.%
, Bringedal, C.%
, Pop, I\BPBI S.%
, Rohde, C.%
\BCBL {}\ \BBA {} Helmig, R.%
\end{APACrefauthors}%
\unskip\
\newblock
\APACrefYearMonthDay{2024}{}{}.
\newblock
{\BBOQ}\APACrefatitle {Investigation of Different Throat Concepts for
  Precipitation Processes in Saturated Pore-Network Models} {Investigation of
  different throat concepts for precipitation processes in saturated
  pore-network models}.{\BBCQ}
\newblock
\APACjournalVolNumPages{Transport in Porous Media}{151}{14}{2647-2692}.
\newblock
\begin{APACrefDOI} \doi{10.1007/s11242-024-02125-5} \end{APACrefDOI}
\PrintBackRefs{\CurrentBib}

\bibitem [\protect \citeauthoryear {%
Shokri%
}{%
Shokri%
}{%
{\protect \APACyear {2014}}%
}]{%
Shokri2014}
\APACinsertmetastar {%
Shokri2014}%
\begin{APACrefauthors}%
Shokri, N.%
\end{APACrefauthors}%
\unskip\
\newblock
\APACrefYearMonthDay{2014}{}{}.
\newblock
{\BBOQ}\APACrefatitle {Pore-scale dynamics of salt transport and distribution
  in drying porous media} {Pore-scale dynamics of salt transport and
  distribution in drying porous media}.{\BBCQ}
\newblock
\APACjournalVolNumPages{Physics of Fluids}{26}{1}{012106}.
\newblock
\begin{APACrefDOI} \doi{10.1063/1.4861755} \end{APACrefDOI}
\PrintBackRefs{\CurrentBib}

\bibitem [\protect \citeauthoryear {%
Singh%
}{%
Singh%
}{%
{\protect \APACyear {2015}}%
}]{%
Singh2015}
\APACinsertmetastar {%
Singh2015}%
\begin{APACrefauthors}%
Singh, A.%
\end{APACrefauthors}%
\unskip\
\newblock
\APACrefYearMonthDay{2015}{}{}.
\newblock
{\BBOQ}\APACrefatitle {Soil salinization and waterlogging: A threat to
  environment and agricultural sustainability} {Soil salinization and
  waterlogging: A threat to environment and agricultural
  sustainability}.{\BBCQ}
\newblock
\APACjournalVolNumPages{Ecological Indicators}{57}{}{128--130}.
\newblock
\begin{APACrefDOI} \doi{10.1016/j.ecolind.2015.04.027} \end{APACrefDOI}
\PrintBackRefs{\CurrentBib}

\bibitem [\protect \citeauthoryear {%
Thomson%
}{%
Thomson%
}{%
{\protect \APACyear {1871}}%
}]{%
Kelvin1871}
\APACinsertmetastar {%
Kelvin1871}%
\begin{APACrefauthors}%
Thomson, W.%
\end{APACrefauthors}%
\unskip\
\newblock
\APACrefYearMonthDay{1871}{}{}.
\newblock
{\BBOQ}\APACrefatitle {On the equilibrium of vapour at a curved surface of
  liquid} {On the equilibrium of vapour at a curved surface of liquid}.{\BBCQ}
\newblock
\APACjournalVolNumPages{The London, Edinburgh, and Dublin Philosophical
  Magazine and Journal of Science}{42}{282}{448--452}.
\newblock
\begin{APACrefDOI} \doi{10.1080/14786447108640606} \end{APACrefDOI}
\PrintBackRefs{\CurrentBib}

\bibitem [\protect \citeauthoryear {%
Verma%
\ \BBA {} Pruess%
}{%
Verma%
\ \BBA {} Pruess%
}{%
{\protect \APACyear {1988}}%
}]{%
Verma1988}
\APACinsertmetastar {%
Verma1988}%
\begin{APACrefauthors}%
Verma, A.%
\BCBT {}\ \BBA {} Pruess, K.%
\end{APACrefauthors}%
\unskip\
\newblock
\APACrefYearMonthDay{1988}{}{}.
\newblock
{\BBOQ}\APACrefatitle {Thermohydrological conditions and silica redistribution
  near high-level nuclear wastes emplaced in saturated geological formations}
  {Thermohydrological conditions and silica redistribution near high-level
  nuclear wastes emplaced in saturated geological formations}.{\BBCQ}
\newblock
\APACjournalVolNumPages{Journal of Geophysical Research: Solid
  Earth}{93}{B2}{1159-1173}.
\newblock
\begin{APACrefDOI} \doi{10.1029/JB093iB02p01159} \end{APACrefDOI}
\PrintBackRefs{\CurrentBib}

\bibitem [\protect \citeauthoryear {%
von Wolff%
\ \BBA {} Pop%
}{%
von Wolff%
\ \BBA {} Pop%
}{%
{\protect \APACyear {2022}}%
}]{%
vonWolff2022}
\APACinsertmetastar {%
vonWolff2022}%
\begin{APACrefauthors}%
von Wolff, L.%
\BCBT {}\ \BBA {} Pop, I\BPBI S.%
\end{APACrefauthors}%
\unskip\
\newblock
\APACrefYearMonthDay{2022}{}{}.
\newblock
{\BBOQ}\APACrefatitle {Upscaling of a Cahn–Hilliard Navier–Stokes model
  with precipitation and dissolution in a thin strip} {Upscaling of a
  cahn–hilliard navier–stokes model with precipitation and dissolution in a
  thin strip}.{\BBCQ}
\newblock
\APACjournalVolNumPages{Journal of Fluid Mechanics}{941}{}{A49}.
\newblock
\begin{APACrefDOI} \doi{10.1017/jfm.2022.308} \end{APACrefDOI}
\PrintBackRefs{\CurrentBib}

\bibitem [\protect \citeauthoryear {%
Wagner%
\ \BBA {} Kretzschmar%
}{%
Wagner%
\ \BBA {} Kretzschmar%
}{%
{\protect \APACyear {2008}}%
}]{%
IAPWS2008}
\APACinsertmetastar {%
IAPWS2008}%
\begin{APACrefauthors}%
Wagner, W.%
\BCBT {}\ \BBA {} Kretzschmar, H\BHBI J.%
\end{APACrefauthors}%
\unskip\
\newblock
\APACrefYear{2008}.
\newblock
\APACrefbtitle {IAPWS Industrial Formulation 1997for the Thermodynamic
  Properties of Water and Steam} {Iapws industrial formulation 1997for the
  thermodynamic properties of water and steam}\ (\PrintOrdinal{2}\ \BEd).
\newblock
\APACaddressPublisher{}{Springer Verlag, Berlin}.
\newblock
\begin{APACrefDOI} \doi{10.1007/978-3-540-74234-0_3} \end{APACrefDOI}
\PrintBackRefs{\CurrentBib}

\bibitem [\protect \citeauthoryear {%
Weishaupt%
, Koch%
\BCBL {}\ \BBA {} Helmig%
}{%
Weishaupt%
\ \protect \BOthers {.}}{%
{\protect \APACyear {2022}}%
}]{%
Weishaupt2022}
\APACinsertmetastar {%
Weishaupt2022}%
\begin{APACrefauthors}%
Weishaupt, K.%
, Koch, T.%
\BCBL {}\ \BBA {} Helmig, R.%
\end{APACrefauthors}%
\unskip\
\newblock
\APACrefYearMonthDay{2022}{}{}.
\newblock
{\BBOQ}\APACrefatitle {A fully implicit coupled pore-network/free-flow model
  for the pore-scale simulation of drying processes} {A fully implicit coupled
  pore-network/free-flow model for the pore-scale simulation of drying
  processes}.{\BBCQ}
\newblock
\APACjournalVolNumPages{Drying Technology}{40}{4}{697--718}.
\newblock
\begin{APACrefDOI} \doi{10.1080/07373937.2021.1955706} \end{APACrefDOI}
\PrintBackRefs{\CurrentBib}

\bibitem [\protect \citeauthoryear {%
Wicke%
\ \protect \BOthers {.}}{%
Wicke%
\ \protect \BOthers {.}}{%
{\protect \APACyear {2011}}%
}]{%
Wicke2011}
\APACinsertmetastar {%
Wicke2011}%
\begin{APACrefauthors}%
Wicke, B.%
, Smeets, E.%
, Dornburg, V.%
, Vashev, B.%
, Gaiser, T.%
, Turkenburg, W.%
\BCBL {}\ \BBA {} Faaij, A.%
\end{APACrefauthors}%
\unskip\
\newblock
\APACrefYearMonthDay{2011}{}{}.
\newblock
{\BBOQ}\APACrefatitle {The global technical and economic potential of bioenergy
  from salt-affected soils} {The global technical and economic potential of
  bioenergy from salt-affected soils}.{\BBCQ}
\newblock
\APACjournalVolNumPages{Energy Environ. Sci.}{4}{}{2669--2681}.
\newblock
\begin{APACrefDOI} \doi{10.1039/C1EE01029H} \end{APACrefDOI}
\PrintBackRefs{\CurrentBib}

\bibitem [\protect \citeauthoryear {%
H.~Wu%
, Veyskarami%
, Schneider%
\BCBL {}\ \BBA {} Helmig%
}{%
H.~Wu%
\ \protect \BOthers {.}}{%
{\protect \APACyear {2024}}%
}]{%
Wu2024}
\APACinsertmetastar {%
Wu2024}%
\begin{APACrefauthors}%
Wu, H.%
, Veyskarami, M.%
, Schneider, M.%
\BCBL {}\ \BBA {} Helmig, R.%
\end{APACrefauthors}%
\unskip\
\newblock
\APACrefYearMonthDay{2024}{}{}.
\newblock
{\BBOQ}\APACrefatitle {A New Fully Implicit Two-Phase Pore-Network Model by
  Utilizing Regularization Strategies} {A new fully implicit two-phase
  pore-network model by utilizing regularization strategies}.{\BBCQ}
\newblock
\APACjournalVolNumPages{Transport in Porous Media}{151}{}{1-26}.
\newblock
\begin{APACrefDOI} \doi{10.1007/s11242-023-02031-2} \end{APACrefDOI}
\PrintBackRefs{\CurrentBib}

\bibitem [\protect \citeauthoryear {%
R.~Wu%
\ \BBA {} Chen%
}{%
R.~Wu%
\ \BBA {} Chen%
}{%
{\protect \APACyear {2023}}%
}]{%
Wu2023}
\APACinsertmetastar {%
Wu2023}%
\begin{APACrefauthors}%
Wu, R.%
\BCBT {}\ \BBA {} Chen, F.%
\end{APACrefauthors}%
\unskip\
\newblock
\APACrefYearMonthDay{2023}{}{}.
\newblock
{\BBOQ}\APACrefatitle {Interplay between salt precipitation, corner liquid film
  flow, and gas–liquid displacement during evaporation in microfluidic pore
  networks} {Interplay between salt precipitation, corner liquid film flow, and
  gas–liquid displacement during evaporation in microfluidic pore
  networks}.{\BBCQ}
\newblock
\APACjournalVolNumPages{Journal of Applied Physics}{133}{7}{074701}.
\newblock
\begin{APACrefDOI} \doi{10.1063/5.0135135} \end{APACrefDOI}
\PrintBackRefs{\CurrentBib}

\bibitem [\protect \citeauthoryear {%
Xu%
\ \protect \BOthers {.}}{%
Xu%
\ \protect \BOthers {.}}{%
{\protect \APACyear {2004}}%
}]{%
Xu2004}
\APACinsertmetastar {%
Xu2004}%
\begin{APACrefauthors}%
Xu, T.%
, Ontoy, Y.%
, Molling, P.%
, Spycher, N.%
, Parini, M.%
\BCBL {}\ \BBA {} Pruess, K.%
\end{APACrefauthors}%
\unskip\
\newblock
\APACrefYearMonthDay{2004}{}{}.
\newblock
{\BBOQ}\APACrefatitle {Reactive transport modeling of injection well scaling
  and acidizing at Tiwi field, Philippines} {Reactive transport modeling of
  injection well scaling and acidizing at tiwi field, philippines}.{\BBCQ}
\newblock
\APACjournalVolNumPages{Geothermics}{33}{4}{477-491}.
\newblock
\APACrefnote{Selected papers from the TOUGH Symposium 2003}
\newblock
\begin{APACrefDOI} \doi{10.1016/j.geothermics.2003.09.012} \end{APACrefDOI}
\PrintBackRefs{\CurrentBib}

\bibitem [\protect \citeauthoryear {%
Yang%
\ \protect \BOthers {.}}{%
Yang%
\ \protect \BOthers {.}}{%
{\protect \APACyear {2023}}%
}]{%
Yang2023}
\APACinsertmetastar {%
Yang2023}%
\begin{APACrefauthors}%
Yang, J.%
, Lei, T.%
, Wang, G.%
, Xu, Q.%
, Chen, J.%
\BCBL {}\ \BBA {} Luo, K\BPBI H.%
\end{APACrefauthors}%
\unskip\
\newblock
\APACrefYearMonthDay{2023}{}{}.
\newblock
{\BBOQ}\APACrefatitle {Lattice Boltzmann modelling of salt precipitation during
  brine evaporation} {Lattice boltzmann modelling of salt precipitation during
  brine evaporation}.{\BBCQ}
\newblock
\APACjournalVolNumPages{Advances in Water Resources}{180}{}{104542}.
\newblock
\begin{APACrefDOI} \doi{10.1016/j.advwatres.2023.104542} \end{APACrefDOI}
\PrintBackRefs{\CurrentBib}

\bibitem [\protect \citeauthoryear {%
Yoon%
, Kang%
\BCBL {}\ \BBA {} Valocchi%
}{%
Yoon%
\ \protect \BOthers {.}}{%
{\protect \APACyear {2015}}%
}]{%
Yoon2015}
\APACinsertmetastar {%
Yoon2015}%
\begin{APACrefauthors}%
Yoon, H.%
, Kang, Q.%
\BCBL {}\ \BBA {} Valocchi, A\BPBI J.%
\end{APACrefauthors}%
\unskip\
\newblock
\APACrefYearMonthDay{2015}{}{}.
\newblock
{\BBOQ}\APACrefatitle {Lattice Boltzmann-Based Approaches for Pore-Scale
  Reactive Transport} {Lattice boltzmann-based approaches for pore-scale
  reactive transport}.{\BBCQ}
\newblock
\APACjournalVolNumPages{Reviews in Mineralogy and
  Geochemistry}{80}{1}{393-431}.
\newblock
\begin{APACrefDOI} \doi{10.2138/rmg.2015.80.12} \end{APACrefDOI}
\PrintBackRefs{\CurrentBib}

\bibitem [\protect \citeauthoryear {%
Yusufova%
, Pepinov%
, Nikolaev%
\BCBL {}\ \BBA {} Guseinov%
}{%
Yusufova%
\ \protect \BOthers {.}}{%
{\protect \APACyear {1975}}%
}]{%
Yusufova1975}
\APACinsertmetastar {%
Yusufova1975}%
\begin{APACrefauthors}%
Yusufova, V.%
, Pepinov, R.%
, Nikolaev, V.%
\BCBL {}\ \BBA {} Guseinov, G.%
\end{APACrefauthors}%
\unskip\
\newblock
\APACrefYearMonthDay{1975}{}{}.
\newblock
{\BBOQ}\APACrefatitle {Thermal conductivity of aqueous solutions of NaCl}
  {Thermal conductivity of aqueous solutions of nacl}.{\BBCQ}
\newblock
\APACjournalVolNumPages{Journal of engineering physics}{29}{4}{1225--1229}.
\newblock
\begin{APACrefDOI} \doi{10.1007/BF00867119} \end{APACrefDOI}
\PrintBackRefs{\CurrentBib}

\bibitem [\protect \citeauthoryear {%
Zaretskiy%
, Geiger%
, Sorbie%
\BCBL {}\ \BBA {} Förster%
}{%
Zaretskiy%
\ \protect \BOthers {.}}{%
{\protect \APACyear {2010}}%
}]{%
Zaretskiy2010}
\APACinsertmetastar {%
Zaretskiy2010}%
\begin{APACrefauthors}%
Zaretskiy, Y.%
, Geiger, S.%
, Sorbie, K.%
\BCBL {}\ \BBA {} Förster, M.%
\end{APACrefauthors}%
\unskip\
\newblock
\APACrefYearMonthDay{2010}{}{}.
\newblock
{\BBOQ}\APACrefatitle {Efficient flow and transport simulations in
  reconstructed 3D pore geometries} {Efficient flow and transport simulations
  in reconstructed 3d pore geometries}.{\BBCQ}
\newblock
\APACjournalVolNumPages{Advances in Water Resources}{33}{12}{1508-1516}.
\newblock
\begin{APACrefDOI} \doi{10.1016/j.advwatres.2010.08.008} \end{APACrefDOI}
\PrintBackRefs{\CurrentBib}

\bibitem [\protect \citeauthoryear {%
Zhou%
, Blunt%
\BCBL {}\ \BBA {} Orr%
}{%
Zhou%
\ \protect \BOthers {.}}{%
{\protect \APACyear {1997}}%
}]{%
Zhou1997}
\APACinsertmetastar {%
Zhou1997}%
\begin{APACrefauthors}%
Zhou, D.%
, Blunt, M.%
\BCBL {}\ \BBA {} Orr, F.%
\end{APACrefauthors}%
\unskip\
\newblock
\APACrefYearMonthDay{1997}{}{}.
\newblock
{\BBOQ}\APACrefatitle {Hydrocarbon Drainage along Corners of Noncircular
  Capillaries} {Hydrocarbon drainage along corners of noncircular
  capillaries}.{\BBCQ}
\newblock
\APACjournalVolNumPages{Journal of Colloid and Interface
  Science}{187}{1}{11-21}.
\newblock
\begin{APACrefDOI} \doi{10.1006/jcis.1996.4699} \end{APACrefDOI}
\PrintBackRefs{\CurrentBib}

\bibitem [\protect \citeauthoryear {%
Øren%
, Bakke%
\BCBL {}\ \BBA {} Arntzen%
}{%
Øren%
\ \protect \BOthers {.}}{%
{\protect \APACyear {1998}}%
}]{%
Oren1998}
\APACinsertmetastar {%
Oren1998}%
\begin{APACrefauthors}%
Øren, P\BHBI E.%
, Bakke, S.%
\BCBL {}\ \BBA {} Arntzen, O\BPBI J.%
\end{APACrefauthors}%
\unskip\
\newblock
\APACrefYearMonthDay{1998}{}{}.
\newblock
{\BBOQ}\APACrefatitle {Extending Predictive Capabilities to Network Models}
  {Extending predictive capabilities to network models}.{\BBCQ}
\newblock
\APACjournalVolNumPages{SPE Journal}{3}{04}{324-336}.
\newblock
\begin{APACrefDOI} \doi{10.2118/52052-PA} \end{APACrefDOI}
\PrintBackRefs{\CurrentBib}

\end{thebibliography}
%\bibliography{paper}

\end{document}